# Probing local coordination and halide miscibility in single-, double-, and triple-halide perovskites using EXAFS

*Sonia S. Mulgund,[1†] Esther Y.-H. Hung,[2] Leslie Bostwick,[3] Ashley Galbraith,[1] Owen M. Romberg,[4] Justus Just,[2*] Rebecca A. Belisle[1*]*

[1] Department of Physics and Astronomy, Wellesley College, Wellesley, Massachusetts 02481, United States

[2] MAX IV Laboratory, Lund University, 22484 Lund, Sweden

[3] Olin College of Engineering, Needham, Massachusetts 02492, United States

[4] Department of Chemistry, Wellesley College, Wellesley, Massachusetts 02471 United States

Abstract

Lead-halide perovskites are a promising material platform as semiconductors in next-generation solar cells because of their solution processability, defect tolerance, and tunable optoelectronic properties. Mixed iodide-bromide perovskite compositions are attractive as wide bandgap absorbers but suffer from significant operational instabilities. Incorporation of chloride to form triple-halide perovskites has been shown to improve both stability and performance; however, the extent of halide miscibility and chloride incorporation remain poorly understood. While bulk metrics such as diffraction-derived lattice parameters and optical bandgap can confirm single-phase behavior, they do not establish homogeneous mixing on the halide site. Here, we use cryogenic X-ray absorption spectroscopy (XAS) to directly probe local lead-halide coordination across single-, double-, and triple-halide perovskite compositions. We show formation of a single-phase triple-halide perovskite whose miscibility is mediated by bromide content. We identify signatures of halide mixing from the Pb $L_3$-edge EXAFS of mixed double- and triple-halide perovskites using both quantitative fits and Cauchy wavelet transforms. Finally, using wavelet transforms of the Br K-edge EXAFS, we demonstrate halide intermixing on the single $PbX_6$ octahedron level, exploiting forward-scattering amplified third-shell halide-bromide interactions. These results are a step forward in the understanding of local structure that is required to fully describe and optimize halide incorporation for novel perovskite compositions.

## Introduction

Lead-halide perovskites have rapidly established themselves as leading semiconductor candidates for next-generation photovoltaics, where reported power conversion efficiencies (PCEs) have exceeded 27% for single-junction devices[1] and are approaching 35% for perovskite-silicon-tandem devices.[2] The greater efficiencies achievable in tandem and multijunction devices are enabled by the facile tunability of perovskites through halide composition control.[3–7] That tunability enables development of wide-bandgap perovskites ($E_g$ > 2 eV), which have been operationalized in all-perovskite and perovskite-on-silicon multijunction devices.[2,5,8,9] However, many wide-bandgap perovskites, typically combining iodide and bromide to access bandgaps up to approximately 2.4 eV, suffer from halide segregation that results in structural and electronic disorder and significant nonradiative losses.[7,10–13] Photoinduced halide segregation is a reversible phenomenon where, upon illumination, halide migration results in the formation of temporary iodide-rich and bromide-rich regions of distinct bandgaps and lattice spacings.[14,15] The exact mechanism underlying photoinduced halide segregation is debated, but whether induced by thermodynamic demixing,[16] polaron-related lattice deformation,[17,18] or photochemical halide oxidation,[19] this detrimental phenomenon relies on halide migration through the perovskite lattice, and it is as such accelerated by preexisting defects and structural disorder.[20–22]

Thus, in order to mitigate photoinduced halide segregation and improve the performance of current wide-bandgap perovskites, a variety of approaches have been proposed, including chloride addition to direct crystallization pathway and promote increased grain size and quality.[6,11,23–25] Chloride processing has been extensively studied for its role in improving crystallinity,[6,23–27] photostability,[27] and electronic performance[6,23,28] of iodide-bromide perovskites by directing crystallization,[25] filling lattice vacancies,[11] and deactivating trap states.[11] When chloride fraction is further increased, in addition to these dopant effects, chloride can incorporate into the bulk lattice alongside iodide and bromide to form a triple-halide perovskite phase, as first demonstrated by McGehee and coworkers – a new compositional variable that further raises the accessible bandgap to 3 eV while maintaining the beneficial structural and electronic properties from lower fractions of chloride incorporation.[5,7,19,29] In multijunction devices, such triple-halide perovskites have the benefit of both a widened accessible bandgap range and improved stability to halide segregation.[5,30–32]

While bulk characterization metrics such as optical bandgap and the average crystal structure point towards near-stoichiometric chloride incorporation,[5] the question remains whether chloride is truly incorporating into the perovskite, and if so to what extent and how homogenously. In mixed iodide-bromide perovskites, the solid-state miscibility has been documented by X-ray diffraction (XRD) and spectroscopic methods as well as theoretical modeling, showing lattice spacing and bandgap to change proportionally with halide ratio.[3,33,34] Signatures of phase segregation in these materials are also well-understood, both from the same structural and optical methods, and by a suite of nanoscale imaging and elemental analysis techniques.[10,15,35–40] For example, a significant miscibility gap has also been documented for $MAPb(Cl_xI_{1-x})_3$ with chloride content as low as 3%, and many studies of chloride processing in iodide-rich perovskites suggest that it largely leaves the film during annealing.[25,41] These results make quantification of chloride incorporation challenging. The Goldschmidt tolerance factor, *t*, is the classical geometric descriptor used to predict perovskite formability and degree of framework distortion for a given $ABX_3$ composition, based solely on the constituent ionic radii.[42] While a useful indicator of average symmetry, it treats the ions as hard spheres and so does not capture the covalent character of the Pb-X bond (which varies with halide identity), or the octahedral tilting and local distortions induced by hydrogen bonding between the organic cation and the halide sublattice – effects that are often unresolvable by XRD.[43] These complexities mean a detailed understanding of miscibility in mixed-halide perovskites is still missing.

Thus, in order to fully describe miscibility and halide incorporation in triple-halide perovskites, both bulk metrics (i.e., lattice spacing and bandgap) and more local descriptions of the coordination environment are required. Such a description should not only show chloride content in the perovskite film, but also that the chloride is incorporated into the perovskite lattice itself, and that a single homogeneous phase has formed. X-ray absorption spectroscopy (XAS) is a powerful technique for this task, providing a view of the average coordination environment (up to 10 Å) of a given element in a material. When combined with XRD lattice spacing, both local geometry and long-range structure are illuminated. XAS has already been applied to study lead-halide perovskites toward a variety of goals: bond dynamics and temperature/pressure induced phase transitions,[44–49] precursor evolution and film crystallization,[25,50–54] and local coordination of formed materials.[24,55–60] Here, we apply XAS to study a range of perovskite compositions, from single- to triple-halide, to confirm halide incorporation and determine EXAFS signatures of mixing. Our study moves toward a deeper structural understanding of triple-halide perovskites, thus enabling a more informed design of wide bandgap absorbers in tandem solar cells.

## Experimental Methods

*Perovskite fabrication*. Perovskite samples were solution-processed on UV-ozone-cleaned polyamide substrates in a dry $N_2$ environment. 1M solutions of $CH_3NH_3Pb(I_xBr_yCl_{1-x-y})_3$ were prepared by first making 1M solutions of $CH_3NH_3PbI_3$, $CH_3NH_3PbBr_3$, and $CH_3NH_3PbCl_3$ ($PbI_2$, TCI; $PbBr_2$, TCI; $PbCl_2$, TCI; $CH_3NH_3I$, Greatcell Chemicals; $CH_3NH_3Br$, Greatcell

Chemicals; $CH_3NH_3Cl$, Greatcell Chemicals) in anhydrous DMSO solvent (Sigma Aldrich) and then combining these solutions in the appropriate volumetric ratios. All solutions were prepared in an $N_2$ glovebox 24 hours before spin-coating and filtered through a 0.2 um PVDF filter before spin coating. For each sample, 30 μL of solution was converted into perovskite using the following process: 40 s of spinning at 6000 rpm; 30 s of immersion in anhydrous anisole with slight agitation; and 30 s of spinning at 6000 rpm to dry the film. All samples were then annealed at 100°C for 30 min. Separate samples were fabricated for XRD and XAS measurements.

*X-ray diffraction (XRD) measurements*. XRD data were collected in the grazing incidence geometry at beam line 11-3 of the Stanford Synchrotron Radiation Lightsource. Two-dimensional scattering was collected with monochromatic 12.7 keV X-Rays and recorded on a Rayonix MX-225 detector measuring 225 × 225 $mm^2$. Samples were measured in a chamber under flowing helium at room temperature. All 2D images were calibrated with a $LaB_6$ standard to determine experimental geometry. GSAS-II,[61] pyFAI[62] and pygix[63] python packages were used for data analysis (complete methods included in the SI).

*X-ray absorption spectroscopy (XAS) measurements*. XAS data were collected in transmission mode at the Balder beamline of the MAX IV Laboratory 3 GeV synchrotron using a Si(111) double crystal monochromator with fixed exit motion and vertical beam stabilization. To achieve an edge step allowing for high quality transmission mode measurements of thin-films, samples (25x25$mm^2$) coated on polyimide substrate were cut into strips and stacked on top of each other. Samples were combined into stacks of 8, except for $MAPb(Br_{0.6}Cl_{0.4})_3$, which was a stack of 6. Transmission was measured through this stack under a shallow incidence angle of approximately 5-10°. The beam energy was calibrated to the selenium K-edge at 12654 eV (maximum of first derivative of μd). Signals of both incident ($I_0$) and transmitted ($I_t$) X-rays were detected using ionization chambers. During measurement, samples were cooled in a liquid helium cryostat to a temperature of 15-20 K. For each sample, 10-20 spectra were recorded and subsequently merged, using an acquisition time of 30s per EXAFS spectrum. Background subtraction and normalization of raw data was performed in the ATHENA software, and the ARTEMIS software with FEFF6 and ATOMS was used to model the processed EXAFS spectra.[64,65] The wavelet transforms (Figs. 3 and 5) were computed using the Cauchy method, implemented in LARCH.[66,67]

## Results and Discussion

There are many existing examples of mixed-halide perovskite systems, spanning bromide-chloride,[68] bromide-iodide,[3] and mixed-cation triple-halide perovskites.[5,29–31] However, solution-processed single-cation triple-halide perovskites with significant chloride content remain largely unexplored. To reduce the complexity associated with mixed-cation disorder and enable clearer insight into halide miscibility, we focus on single-cation methylammonium lead halide perovskites, $MAPb(I_xBr_yCl_{1-x-y})_3$.

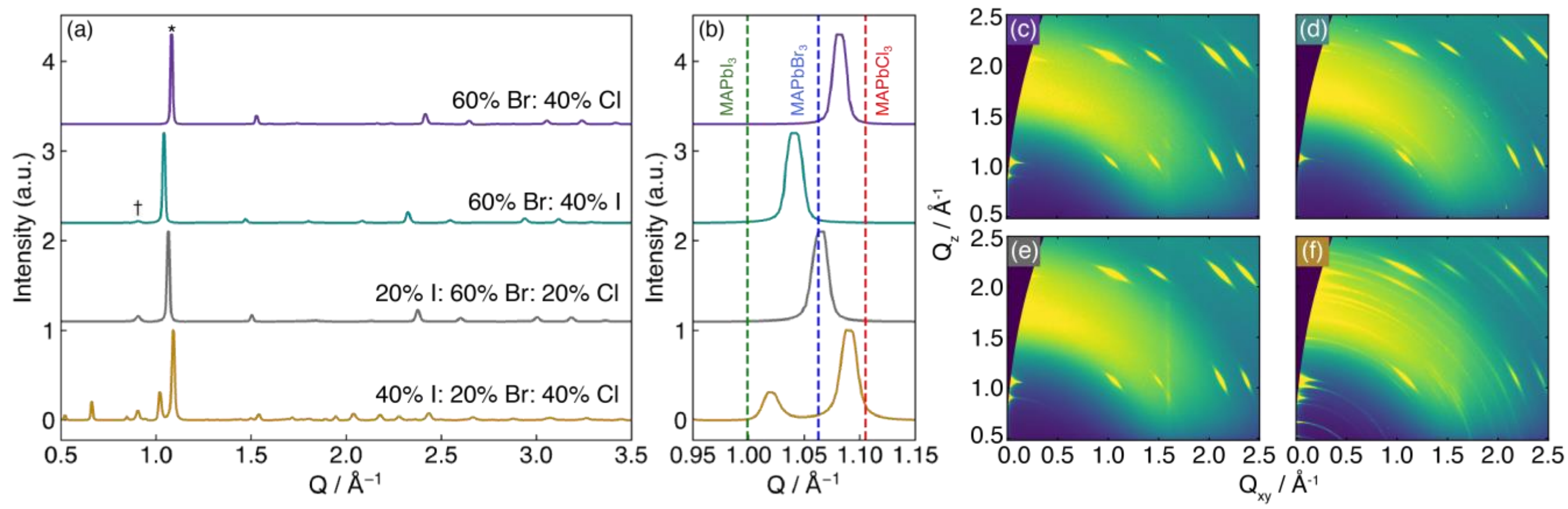

**Figure 1**. XRD patterns of mixed-halide perovskites. (a) Azimuthally integrated diffraction patterns of the mixed-halide perovskite thin-films, highlighting the cubic (100) (*) and $PbI_2$ (†) reflections. (b) Enlargement of the region of the first perovskite reflection, showing single phase formation in all but the low Br content triple-halide perovskite. Dashed lines indicate the (100) peak center of the pure compositions $MAPbCl_3$,[69] $MAPbBr_3$,[70] and $MAPbI_3$.[69] (c)-(f) 2D XRD patterns in linear scale of the four mixed-halide samples with halide ratios: (c) 60% Br: 40% Cl, (d) 60% Br: 40% I, (e) 20% I: 60% Br: 20% Cl, (f) 40% I: 20% Br: 40% Cl.

To probe the crystallographic structure and phase purity of the mixed-halide thin films, we perform X-ray diffraction (XRD) measurements on as-cast thin films (**Figure 1a**). All films exhibit a reflection at $Q \sim 1$Å with a strong preferential orientation in the out-of-plane direction (**Figure 1c-f**), assigned to the (100) cubic perovskite reflection. Relative to $MAPbBr_3$ (dashed blue line), the 60% Br: 40% I and the 60% Br:40% Cl films show systematic shifts to lower and higher $Q$, indicating larger and smaller unit cells, respectively ($a$ = 6.04 Å and $a$ = 5.81 Å, **Table S1**). This trend is consistent with the larger ionic radius of $I^-$ and the smaller ionic radius of $Br^-$ /$Cl^-$ (**Table S11**), suggesting substantial halide mixing inside the perovskite lattice. The lack of substantial competing crystalline phases (e.g. $PbI_2$) supports the existence of a crystalline majority (>90 %) mixed-halide perovskite phase whose long-range compositional heterogeneity is within the width of the diffraction peaks.

XRD patterns for two triple-halide compositions, with low (20%) and high (60%) target $Br^-$ content, highlight the role of composition in phase stability (**Figure 1a**, lower). At high $Br^-$ content, a single (100) reflection with a peak width similar to the binary mixtures is observed (**Table S1**). The high Br content apparent triple-halide perovskite, $MAPb(I_{0.2}Br_{0.6}Cl_{0.2})_3$, shows a lattice spacing intermediate to that of the $Br^-/Cl^-$ and $Br^-/I^-$ perovskites ($a$ = 5.91 Å), supporting the incorporation of all three halides into the bulk lattice and the formation of a single-phase perovskite. The similarity in peak center for the 60% Br triple-halide and $MAPbBr_3$ highlights the challenge of inferring stoichiometry and compositional homogeneity from diffraction data alone for triple-halide compositions. In contrast to the 60% $Br^-$ composition, the triple-halide film prepared from the 40% I: 20% Br: 40% Cl stoichiometry exhibits a clear splitting of the (100) reflection, consistent with phase segregation into two perovskite-like domains – one $Cl^-$ -rich and one $I^-$-rich. This suggests that the higher $Br^-$ content promotes halide miscibility in the perovskite lattice, consistent with prior reported trends reported for mixed-cation and mechanochemically synthesized triple-halide perovskites.[5,29] We also observe additional reflections attributable to unknown, non-3D-perovskite crystalline phases in the nominal $MAPb(I_{0.4}Br_{0.2}Cl_{0.4})_3$ composition.

The corresponding 2D-XRD patterns for each film (**Figure 1c-f**) reveal a perovskite structure that is highly textured independent of halide ratio, with all compositions showing a preferential orientation of the (n00) reflections along the $Q_z$ direction. This consistently out-of-plane alignment suggests that the crystalline orientation is not strongly influenced by as-cast halide composition. The highest, single-phase $Cl^-$-containing $MAPb(Br_{0.6}Cl_{0.4})_3$ (**Figure 1c**) exhibits well-defined Bragg reflections, and has the lowest fitted perovskite peak width (**Table S1**) — consistent with known effects of $Cl^-$ on perovskite crystallinity.[6] In contrast, the phase-segregated nominal $MAPb(I_{0.4}Br_{0.2}Cl_{0.4})_3$ composition (**Figure 1f**) appears less sharply textured, with perovskite reflections which are more smeared radially.

These XRD results represent the long-range, average perovskite crystal structure. While they can verify single or segregated perovskite phase formation, they do not describe how the three halides interact with and accommodate one another, leaving open the possibility of compositional heterogeneity at a very local length scale. To describe that, local coordination as probed by XAS is a more appropriate tool.

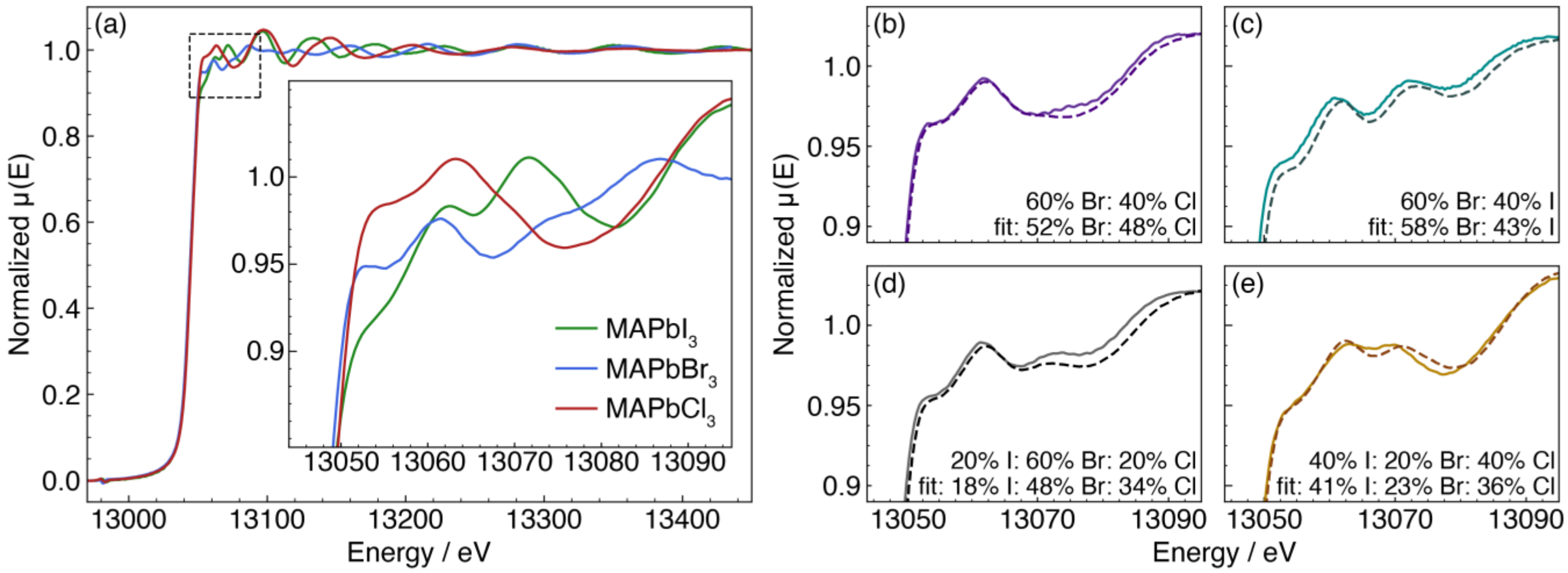

**Figure 2.** Normalized Pb $L_3$-edge X-ray absorption near edge structure (XANES) spectra for all compositions, measured at cryogenic temperatures. (a) Normalized $\mu(E)$ with detail inset for single halide perovskites. (b) – (e) Normalized XANES $\mu(E)$ for mixed-halide perovskites together with linear combination analysis (LCA) fits (dashed lines) for a quantitative estimate of halide coordination around Pb in comparison with the halide fractions for solution compositions.

Perovskite chemical composition influences long-range crystalline phase, short-range local coordination environment, and electronic orbital structure, so each of these should contain evidence of halide mixing. To probe these effects, we first employ XANES to investigate how halide composition modifies the electronic structure arising from the interactions between lead and halide electronic orbitals. Single-halide perovskite XANES $\mu(E)$ spectra (**Figure 2a**) reveal distinct spectral features, indicating differences in electronic structure that reflect unique local chemical environments. Notably, the overall EXAFS oscillation amplitude is low despite measurement at 15-20 K, reflecting the significant static disorder in lead-halide perovskites compared to most crystalline materials at cryogenic temperatures where thermal/dynamic disorder is minimized. We study mixed-halide compositions using linear combination analysis (LCA) (**Figure 2b-e**), which approximates the Pb local environment as a superposition of that in $MAPbI_3$, $MAPbBr_3$, and $MAPbCl_3$, consistent with a linear combination of atomic orbitals.[55] However, a mixed-halide lattice is significantly distorted to accommodate the varying ionic radii.[3,34,71] These changes in geometry influence how atomic orbitals combine to form the band structure of a mixed-halide perovskite; thus, the LCA fits (**Figure 2b-e**) provide only a crude indication of halide content in the absence of describing the intricacies of Pb-X interactions.

Despite that limitation, linear combination of the single-halide perovskite XANES can reproduce all major spectral features for the mixed-halide compositions very well, demonstrating the presence of all three Pb-X bonds at significant fractions. There is no indication, for example, that either of the triple-halide samples completely lack $Cl^-$ incorporation into the perovskite lattice. Moreover, the best-fit fractions resemble, within expected experimental uncertainties of approximately 5-10%, the solution stoichiometry. The largest discrepancy is for $MAPb(I_{0.2}Br_{0.6}Cl_{0.2})_3$, where the LCA fit indicates an effective stoichiometry of $MAPb(I_{0.18}Br_{0.48}Cl_{0.34})_3$, with substantially increased apparent $Cl^-$ content and moderately reduced $Br^-$ content compared with the solution composition. The LCA model is limited in its ability to capture the electronic structure of mixed-halide perovskite compositions, which deviate from a simple linear combination of single-halide orbitals. As a result, the elevated $Cl^-$ fraction is unlikely to reflect a true increase in $Cl^-$ content but instead represents an effective parameter in the linear combination fit that attempts to account for these more complex electronic and structural contributions. On the other hand, the LCA fits for the phase-segregated, nominal 20% $Br^-$ triple-halide composition yield LCA fractions more closely resembling the solution stoichiometry. As identified in XRD, this composition likely forms distinct $I^-$-rich and $Cl^-$-rich crystalline phases, likely to each be less complex electronically than the single-phase triple-halide. The phase separation allows most bond distances and angles to relax closer to their single-halide values --- thus making the linear combination assumptions more appropriate and enabling more accurate comparison of its XANES to the single-halide standards.[5,55]

Overall, the LCA results show that all halides contribute significantly to the measured XANES spectra at levels broadly consistent with the input solution stoichiometry. Thus, despite not providing a specific structural model, the Pb $L_3$-edge XANES LCA demonstrates that all halides coordinate to Pb. This result, coupled with our observations of single perovskite phases in XRD, suggests that if this lead-halide coordination occurs in a crystalline phase, it must be in a mixed-halide perovskite phase. While small quantities of $Cl^-$ have been shown to direct perovskite crystallization without incorporation into the bulk lattice,[25] we now show based on the local environment of the Pb atom, instead of bulk metrics, that at higher $Cl^-$ fractions it coordinates in the perovskite.

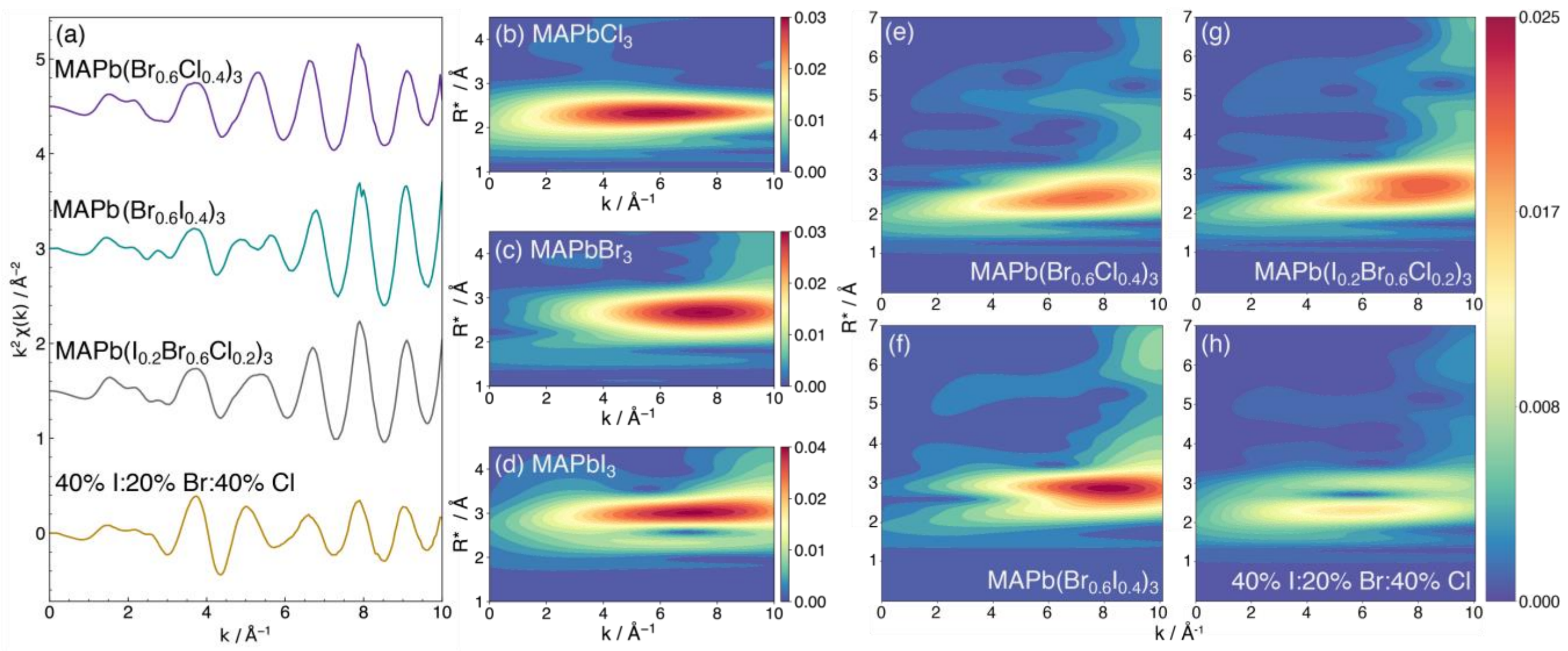

**Figure 3.** $k^2$-weighted Pb $L_3$ –edge EXAFS and wavelet transforms thereof. (a) Pb $L_3$ EXAFS $\chi(k)$ for the mixed-halide perovskite samples. Data is offset for visualization. (b-d) Cauchy wavelet transforms (WTs) W($k$,$R^*$) for single halide perovskites in the range of the first Pb-X shell. (e-h) Cauchy WTs for the mixed-halide compositions.

While XRD and XANES LCA show evidence of a triple-halide perovskite phase, a more direct local probe is needed to understand the extent of halide mixing and stoichiometric incorporation into the perovskite lattice. In contrast to XANES, EXAFS provides element-specific, local structural information by quantifying the identity, distances, and coordination numbers of atoms surrounding the absorber, typically out to approximately 8-10 Å. In lead halide perovskites, this means that EXAFS can resolve the average $PbX_6$ octahedral environment by identifying Pb-X nearest-neighbors along with tracking subtle changes in disorder. To minimize thermal disorder contributions to the EXAFS signal, we employ cryogenic EXAFS at 15 K at the Pb $L_3$-edge (**Figure 3a**). All mixed, Br-containing halide data is cut off at the Br K-edge at 14374 eV, limiting $\chi(k)$ to 10 $Å^{-1}$. The mixed-halide EXAFS $\chi(k)$ spectra (**Figure 3a**) reveal the presence of multiple frequency (*k*) components. In EXAFS, *k* is the wavevector of the photoelectron ejected from the absorbing atom (here, Pb). The backscattering amplitude and phase are characteristic of the neighboring atom and vary with *k* according to atomic number: lighter atoms generally backscatter at lower *k* and heavier atoms tend to contribute at higher *k*. Distinct backscattering atoms therefore contribute to $\chi(k)$ with different *k*-dependence (calculated amplitudes and phases for Pb-X single scattering paths shown in **Figure S10**). Accordingly, the mixed-halide data contain several superimposed components, indicating the presence of multiple distinct backscattering atoms (Cl, Br, I) around the absorbing Pb. To clarify these overlapping contributions to the EXAFS, we compute the Cauchy wavelet transform (WT). The WT describes the EXAFS oscillations as a superposition of wavelets, localized at a specific frequency and photoelectron wavenumber *k*, to yield a two-dimensional representation of the EXAFS signal, W(*k*,*R**), that simultaneously resolves contributions in *k* and *R** (**Figure 3b–h**).[66] As *R** incorporates both the radial distance *R* and the scattering phase shift, which is individual for specific absorber-backscatterer combinations, its absolute value cannot be directly interpreted as a distance in real space. The distribution along the *k*-axis, as mentioned above, depends on the atomic number of neighboring atoms, allowing differentiation of atomic species at distances *R** based on the *k*-dependent scattering amplitude. This allows direct visualization of how scattering contributions evolve in both domains. For the single-halide $MAPbX_3$ compositions (**Figure 3b–d**), the main wavelet intensity feature, corresponding to the first shell halides, shifts toward higher *R** and higher *k* as X varies from Cl to Br to I. This trend is consistent with the longer Pb-X bond distances associated with the increasing ionic radius going from lighter ($Cl^-$) to heavier ($I^-$) ions (**Table S11**) as well as the expected behavior that the *k*-dependent scattering amplitude favors higher *k* for heavier atoms.

A similar trend is observed in $MAPb(Br_{0.6}X_{0.4})_3$, where the maximum wavelet intensity shifts towards higher $R^*$ and higher $k$ when comparing X=Cl to X=I, indicating the presence of significant amounts of $Cl^-$ in the structure in agreement to XANES. Across all mixed-halide compositions (**Figure 3e–h**), the high-intensity regions appear more broad along $R^*$ than for the single-halides, indicating an increased distribution of local Pb-X bond distances. In the triple-halide compositions, these high-intensity regions appear as distinct lobes, which overlap in $MAPb(I_{0.2}Br_{0.6}Cl_{0.2})_3$ (**Figure 3g**) but are separated in nominal $MAPb(I_{0.4}Br_{0.2}Cl_{0.4})_3$ (**Figure 3h**). This separation can reflect the reduced presence of intermediate-length Pb-Br bonds at lower Br content, consistent with the observed crystallographic phase-segregation. We note that these two-dimensional representations of the first Pb $L_3$ coordination shell do not allow unequivocal distinction between single-phase and phase-segregated perovskite compositions, because the average Pb is still coordinated to all halides in both cases. However, they provide evidence that all three halides are present and coordinated to Pb in the perovskite lattice, consistent with XANES LCA.

To resolve distinct Pb-X contributions to the EXAFS, we next perform quantitative fitting of the EXAFS spectra. First shell fits are carried out in back-Fourier-transformed $q$-space using a $PbX_6$ octahedral model, establishing benchmark Pb-X bond distances and disorder parameters for the single-halide compounds (**Table S3**). These fits show a systematic contraction of the Pb-X bond with decreasing halide ionic radius (**Figure S8**, **Table S3**), with best-fit Pb-X distances ranging from 3.12 (X=I) to 2.73 Å (X=Cl), in agreement with published crystal structure data (**Table S2**). [3,33,34] Of note, $MAPbCl_3$ has a lower symmetry structure with four distinct Pb-Cl distances, which was taken into account in the fit. For $MAPbI_3$, we additionally extended the fit to successfully include second shell Pb-Pb-Pb and multiple scattering contributions (**Figure S9**), validating the appropriateness of the orthorhombic perovskite input model.

Among the single-halide compositions, the Pb-X scattering path in $MAPbBr_3$ exhibits the largest fitted mean-squared relative displacement (MRSD $\sigma^2$, which describes the distribution of distances between the absorbing atom and a specific back-scatterer: here, in Pb-X), indicating greater disorder than in the other single-halide compositions. At the measurement temperature (15-20 K), this disorder is predominantly static (i.e., structural rather than thermal). In single-halide $MAPbX_3$, this static disorder encompasses octahedral tilting and $MA^+$ orientational effects, which are coupled, especially at low temperature. The Pb $L_3$-edge EXAFS does not directly probe $MA^+$ orientation - the organic cation contains no heavy backscattering atoms – but it is sensitive to the correlated $PbX_6$ framework response, including octahedral tilt and distortions. The larger $\sigma^2$ for $MAPbBr_3$ compared with the other single-halide compositions is consistent with reports elsewhere,[45,48,58,71,72] which indicates more local static disorder compared with the iodide- or chloride-based perovskites. However, the limited energy range of the data due to the nearby Br K edge at 13474 eV limits the accuracy of this value.

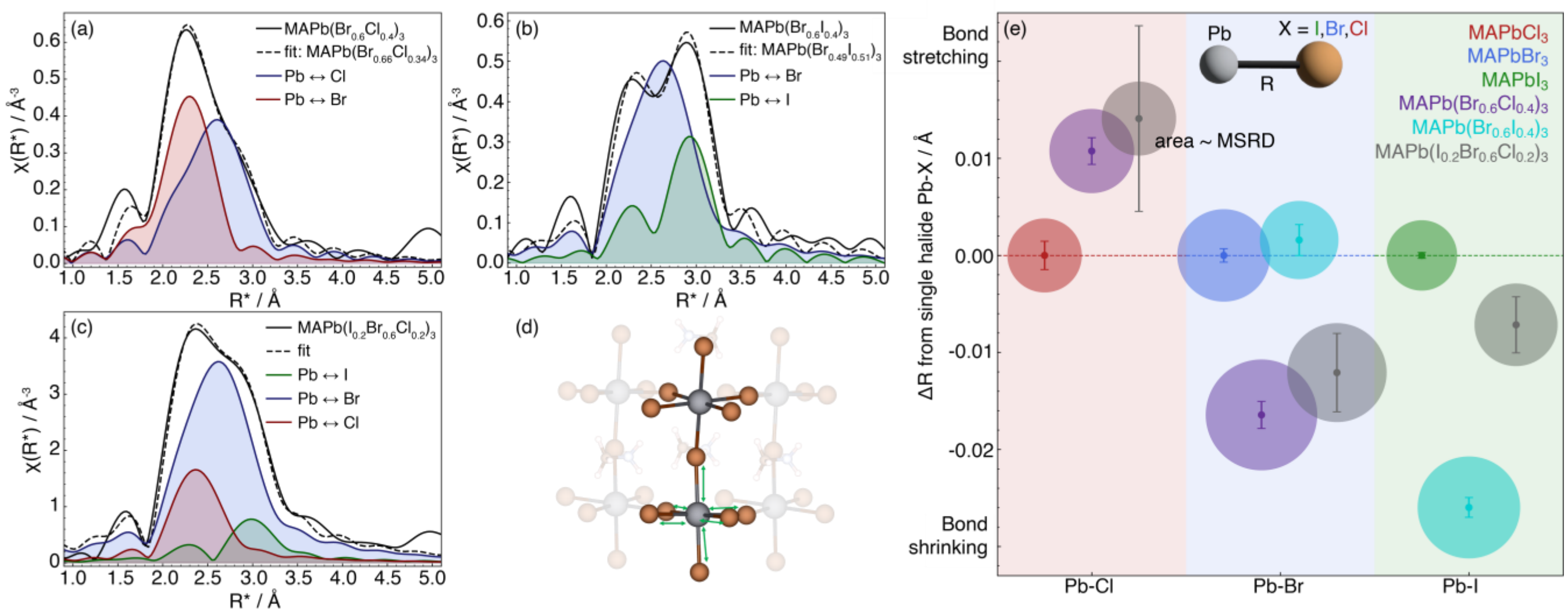

**Figure 4.** Local structure of $PbX_6$ octahedra as derived from quantitative EXAFS analysis. (a-c) Magnitudes of the Fourier transform of Pb $L_3$-edge EXAFS fits into $R^*$-space for the mixed-halide samples. Solid black lines show the magnitude of the experimental spectrum, dashed black lines show the total magnitude of the fit, and colored lines show the magnitude of the contribution of individual scattering path contributions; note that these path magnitudes sum coherently and are not directly additive. (d) Schematic of a $MAPbX_3$ perovskite structure. The arrows show the first shell Pb-X scattering paths that were considered in the analysis. (e) Deviation of Pb-X bond lengths, $\Delta R$, in the mixed-halide perovskites from the corresponding single-halide reference values. The area of each circle is 1/10 of the $\sigma^2$ (MSRD), and error bars reflect discrepancies obtained by fitting under different conditions and k-weights.

Extending this analysis to mixed-halide compositions, quantitative EXAFS fits are performed in *q*-space using *ab-initio*-calculated scattering amplitudes and phases, with the corresponding *R*-space results shown in **Figure 4** (fitted parameters in **Table S4**). The fit (dashed black lines) represents the coherent complex sum of the individual scattering path contributions (colored curves) for distinct backscattering halide atoms within the first coordination shell of Pb, here the $PbX_6$ octahedron (**Figure 4d**). The first shell model uses a single average Pb-X distance for each halide species and thus does not explicitly resolve the distinct X environments potentially present in the low-temperature orthorhombic structure. However, dual-halide perovskites have been shown experimentally to favor cubic structures at lower temperatures than their single-halide parent compositions,[3,73,74] with our mixed-halide perovskites all being cubic at room temperature as shown in XRD (**Figure 1**). Likewise, computational studies of halide mixing suggest that compositional disorder and strain can reduce octahedral tilting, rendering the local structure in these materials more cubic-like than a single-halide perovskite.[71,74] Under these conditions, the first shell environment is likely well-modeled by single Pb-X bond lengths for each X.

For double-halide samples, halide composition is incorporated as a fitting variable in addition to bond distances and mean-squared relative displacements (MSRD) $\sigma^2$; the relative amplitude of each path is constrained to the total expected coordination number of 6, enabling extraction of the average $PbX_6$ octahedral composition (**Figure 4a-b**, paths shown in **Figure 4d**, *q*-space fits shown in **Figures S13-S14**). Notably, there is significant uncertainty in the exact values for relative coordination number due to correlations with $\sigma^2$. The amplitude reduction factor ($S_0^2$), whose values for each path are fixed to those from the single-halide fits, also carries over uncertainty. To assess the numerical robustness of the extracted compositions, we repeat fits under variation of data extraction and fitting parameters as minimum *k* values and *k*-weightings (**Tables S8-S9**). The repeated fits yield an average Br:I ratio of 0.51:0.49 (± 0.03), and an average Br:Cl ratio of 0.67:0.33 (± 0.02). Note that these EXAFS-derived halide ratios describe local coordination around Pb in the perovskite structure, not the bulk elemental concentration. However, the limited variation across fitting conditions combined with the agreement between XANES, EXAFS, and input chemistry indicate that the film compositions closely match those of the precursor solutions.

Next, [3,33,34]we compare the distribution of Pb-X distances with MSRD $\sigma^2$, to validate the fitting protocol and examine trends in materials believed to be miscible.[3,33,34] In both double-halide materials, both the MSRD and the magnitude of bond-length deviation from the single-halide reference are larger for the longer Pb-X bond. The fitted MSRD $\sigma^2$ are generally larger than that of the single-halide references, which we attribute to static disorder arising additionally from octahedral framework strain, owing to size differences between co-occupying halides on a shared site, assuming random site occupation. Within this framework, the smaller halide is stretched to longer bond distances, whereas the larger halide is compressed below its equilibrium bond length. The bond compression for the larger halide is likely accompanied by angular distortions within the octahedron, which translates to a wider range of bond distances and hence a larger fitted $\sigma^2$. By contrast, the smaller halide is stretched to longer bond lengths with comparatively less angular distortion. The dependence of MSRD $\sigma^2$ on halide size is thus consistent with both halide species mixing within a single-phase shared lattice. The EXAFS-determined Pb-X bond lengths also trend toward intermediate values compared to the single-halide compositions (**Figure 4e**). For $MAPb(Br_{0.6}Cl_{0.4})_3$, the Pb-Cl bond (2.844 Å) is expanded by 0.011 Å, whilst the Pb-Br bond (2.962 Å) is contracted by 0.016 Å with respect to those in $MAPbCl_3$ and $MAPbBr_3$, respectively. These changes are consistent with a strained mixed-halide octahedron, where even in a single-phase material, incorporation of halides with varying ionic radius and electronegativity leads us to expect varying Pb-X bond lengths. The single-halide Pb-Cl and Pb-Br bonds have a difference in length of 0.145 Å, which means that the Pb-Cl and Pb-Br bonds each compensate for approximately 7 and 11% of their difference in bond length, respectively, with the $Br^-$ bearing the greater strain. A similar, albeit weaker, trend is observed in $MAPb(Br_{0.6}I_{0.4})_3$ (1% expansion of the Pb-Br bond and 13% contraction of the Pb-I bond). Across both double-halide compositions, the convergence of Pb-X bond lengths towards intermediate values, together with the size-dependent increase in MSRD, provides consistent signatures of halide mixing, despite not providing conclusive evidence thereof.

After identifying these signatures of double-halide mixing, we likewise probe the nearest-neighbor environment of lead in the single-phase triple-halide perovskite, $MAPb(I_{0.2}Br_{0.6}Cl_{0.2})_3$ via Pb $L_3$-edge EXAFS fitting (**Figure 4c**). Because we observed two crystallographic phases for nominal $MAPb(I_{0.4}Br_{0.2}Cl_{0.4})_3$, EXAFS fits assuming a single phase would be misleading, so we only perform quantitative fits for the miscible composition. Like in the double-halide systems, we observe converging Pb-X bond lengths of 2.848 Å, 2.966 Å, and 3.174 Å for X = Cl, Br, and I respectively (**Figure 4e**). We note that for the triple-halide system, fitting coordination numbers directly is impractical owing to the large number of correlated fit parameters relative to the available data range. Consequently, we hold all path amplitudes $N$ fixed according to nominal stoichiometry and, like in the double-halide fits, $S_0^2$ is held fixed to the values obtained for single-halide perovskites. Although these compositional assumptions become more complicated for triple-halide perovskites – where the degree of halide mixing is an open question – the consistency between LCA, EXAFS, and input stoichiometry in the double-halide cases provides a validated benchmark for interpreting bond distances and disorder in fully miscible perovskites. Thus, because similar trends in fit results for $MAPb(I_{0.2}Br_{0.6}Cl_{0.2})_3$ are obtained compared to the double-halide case, it indicates a similar local structure, i.e., halide mixing. The high bromide content of $MAPb(I_{0.2}Br_{0.6}Cl_{0.2})_3$ is expected to favor miscibility, because the intermediate-radius $Br^-$ can readily accommodate both larger $I^-$ and smaller $Cl^-$ within the octahedral framework. All three Pb-X scattering paths were necessary to effectively reproduce the observed scattering (**Table S10**), providing further evidence for coordination of all three halides around Pb. To probe potential halide mixing, analysis of best-fit bond distances and disorder is required. In $MAPb(I_{0.2}Br_{0.6}Cl_{0.2})_3$, as for the double-halides, the longer Pb-X bonds have larger MSRDs than the smaller halides, consistent with halide miscibility in the lattice. Moreover, the Pb-X bond distances converge toward intermediate values (**Figure 4e**) with shrunken Pb-I and Pb-Br bonds and stretched Pb-Cl bonds. Thus, as for the double-halides, the overall trend is consistent with halide mixing and the changing MSRD is consistent with the proximity of all three halides in a single crystallographic phase.

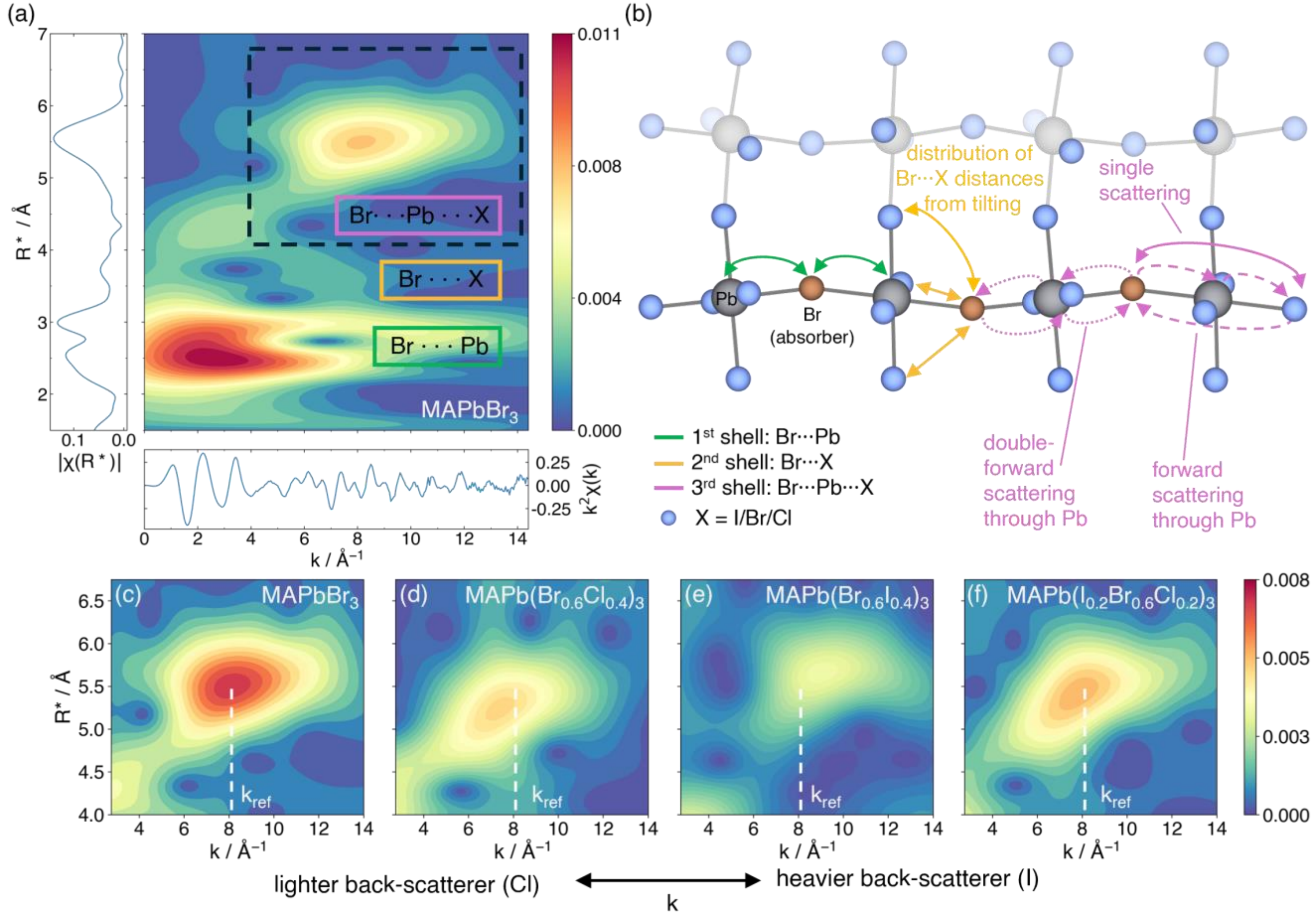

**Figure 5.** Cauchy wavelet transforms of the Br K-edge EXAFS for (a) $MAPbBr_3$, with three contributions labelled: the first, second, and third inorganic shell features indicated. The dashed box indicates the $R^*$-$k$ region shown in (c)-(f). (b) Schematic of representative scattering paths from an absorbing Br within the corner-sharing inorganic perovskite framework indicating the first, second, and distinct third shell scattering paths represented in the WT in (a). WT visualizations of the third shell region, all normalized to the same color scale, are shown for (c) $MAPbBr_3$ (d), $MAPb(Br_{0.6}Cl_{0.4})_3$ (e), $MAPb(Br_{0.6}I_{0.4})_3$ (f) and $MAPb(I_{0.2}Br_{0.6}Cl_{0.2})_3$. The dashed vertical line marks $k_{ref}$, the $k$-position of the Br···Pb···X maximum in pure $MAPbBr_3$.

Because Pb $L_3$-edge XAS probes an average over all Pb environments, it cannot distinguish between true mixed-halide coordination (i.e., multiple halide species within individual $PbX_6$ octahedra) and nanoscale halide clustering or domain segregation un-resolvable via XRD. To address this uncertainty, we employ Br K-edge EXAFS, which directly probes the local environment from the perspective of Br. The Br K-edge spectra (**Figure S16**) exhibit interference from Pb $L_3$-edge oscillations beyond $k = 10$ Å$^{-1}$ (**Figure S17**), complicating direct quantitative Br EXAFS fits. For a qualitative analysis, we instead apply the Cauchy WT (**Figure 5**) to resolve distinct scattering contributions in both distance and *k*-space. For $MAPbBr_3$ (**Figure 5a**), the WT reveals a strong intensity feature at $R^* \approx 2.5$-3 Å, appearing as the overlap of two closely spaced lobes, corresponding to Br $\cdots$ Pb scattering paths within the $PbBr_6$ octahedron. While there are two inequivalent Br-Pb distances in the low-temperature orthorhombic phase, this peak splitting can be reproduced in FEFF calculations just with a single Br $\cdots$ Pb scattering path (**Figure S18**). This split first shell feature is visible for all compositions measured at the Br K-edge (**Figures S19-S22**). Heavy backscattering atoms like Pb exhibit non-uniform, oscillatory *k*-dependencies in backscattering amplitude and phase because of relativistic effects.[75] The observed split can thus be rationalized by interference between the low-*k* and high-*k* amplitudes and phases of a single Br $\cdots$ Pb scattering path.

A second, weaker intensity wavelet feature is observed at intermediate $R^*$ (~ 3.5 - 5 Å) and distributed widely in *k*, corresponding to the second shell region. In the orthorhombic $MAPbBr_3$ lattice, octahedral tilting gives rise to six unique Br $\cdots$ Br single scattering paths that fall within this distance range (**Figure 5b**). Scattering amplitude calculations for these second shell Br $\cdots$ Br paths (**Figure S23,24**) show that the superposition of these contributions leads to partial destructive interference, yielding a weak second shell signal. Likewise, in the mixed-halide compositions (full WTs in **Figures S19-S22**), there is some destructive interference between Br-Br and Br-I scattering paths, as shown by phase shifts close to $\pi$ above $k = 8$ Å$^{-1}$ (**Figure S25**). However, the third coordination shell appears in a higher-intensity feature at $R^* \approx$ 5.5 Å and $k \approx 8$ Å$^{-1}$, corresponding to colinear Br ··· Pb ··· Br scattering within a single $PbX_6$ octahedron (**Figure 5b**, pink arrows). At each of the two inequivalent Br sites, three pathways contribute to the third shell: single scattering $Br_i$ ··· Br ···$Br_i$ (**Figure 5b**, solid pink arrow), forward-scattering $Br_i$ ···Pb ···Br ···$Br_i$ (**Figure 5b**, dashed pink arrows) and double-forward-scattering $Br_i$ ···Pb ···Br ···Pb ···$Br_i$ (**Figure 5b**, dotted pink arrows). In such nearly colinear geometries, forward scattering through the centric atom (here, Pb) significantly enhances the backscattering strength of the third atom (Br); however, the measured backscattered amplitude can be reduced, owing to out-of-phase destructive interference from the other forward scattering contribution (**Figures S23-24**).[75,76] The single scattering $Br_i$ ··· Br ···$Br_i$ contribution remains but is comparatively weaker; the overall intensity is dominated by the amplified forward scattering. For colinear multiple scattering paths, the *k*-dependence of the scattering amplitude remains dominated by the backscattering atom at a scattering angle of 180°, with distinct signatures predicted for each X (**Figures S26-S27**).[75] Thus, this third shell EXAFS feature is governed by contributions from halides on the opposite site of the same octahedron, as probed by the absorbing Br. The focusing effect from the forward scattering of the centric Pb amplifies the signal, making it an ideal probe for halide mixing on the single octahedron level. Of note, all contributing scattering paths to this shell have the same effective length and are independent of octahedral tilt in the orthorhombic structure at cryogenic temperatures.

For comparison between single-, double-, and triple-halide samples, the WTs of the third shell are shown in **Figure 5b-e**, with intensity normalized to that of $MAPbBr_3$. Systematic shifts in the $R^*$ and $k$ positions of the maxima depending on halide composition are apparent: In $MAPb(Br_{0.6}Cl_{0.4})_3$ (**Figure 5c**), this feature is shifted to lower $R^*$ and $k$ ($R^* \approx 5.25$ Å$^{-1}$, $k \approx 7$ Å$^{-}$

$^{1}$), whereas in $MAPb(Br_{0.6}I_{0.4})_3$ (**Figure 5d**) it appears at higher $R^*$ and $k$ ($R^* \approx 5.6$ Å$^{-1}$, $k \approx 9$ Å$^{-1}$) with respect to single halide $MAPbBr_3$. These shifts reflect additional scattering from lighter ($Cl^-$) and heavier ($I^-$) halides opposing the probing $Br^-$ in the lead octahedron, respectively. In $MAPb(I_{0.2}Br_{0.6}Cl_{0.2})_3$ (**Figure 5e**), the distribution along $k$ is broader than in $MAPbBr_3$ and the double-halide compositions, evidence of a wider range of backscattering contributions and therefore of halide intermixing on a local scale.

The most prominent difference, however, is not the position but the intensity of the focused third shell feature when comparing the single-halide with the mixed-halide systems. It is greatest in $MAPbBr_3$, consistent with a uniform Br-Br neighbor environment, in which all contributions scatter coherently at the same $R^*$ and $k$. In the mixed-halide compositions, this intensity is reduced relative to $MAPbBr_3$. Note that this reduction in amplitude is not caused by the reduced amount of Br in the structure, as the EXAFS is already normalized to the edge step of the Br K-absorption edge. Instead, it can be rationalized by the introduction of multiple different halide species at the opposite octahedral site: scattering contributions from different halides with significantly different phases lead to partial destructive interference, reducing the net WT intensity. Note that a phase-segregated $Br^-$-rich region, where no halide intermixing occurs on a single octahedron level, would be expected to exhibit the same third shell intensity as $MAPbBr_3$; the intensity reduction in all mixed-halide compositions thus is a direct consequence of halide mixing at the single octahedron level. The lowest observed WT intensity for $MAPb(Br_{0.6}I_{0.4})_3$ can be explained by the calculated scattering phases for X = Br and X = I (**Figures S26-27**), which differ by approximately $\pi$ or $3\pi$ for the two forward-scattering contributions in the region above $k = 7$ Å$^{-1}$ - both anti-phase - leading to strong destructive interference when both backscatterers are present at similar $R$. In $MAPb(Br_{0.6}Cl_{0.4})_3$, there is likewise destructive interference between the double forward scattering paths for X = Br and X = Cl due to a phase difference of approximately $3\pi$ above $k = 7$ Å$^{-1}$, but there is no such trend for the single forward scattering paths, leading to somewhat higher observed intensity. Meanwhile, in triple-halide $MAPb(I_{0.2}Br_{0.6}Cl_{0.2})_3$ the third shell WT intensity is intermediate between the double-halides and $MAPbBr_3$. This reflects the constructive interference between I and Cl double forward scattering contributions, which have the same phase at $k \sim 7.5$ Å$^{-1}$ and similar phase at greater $k$ (**Figure S27**). Thus, based both on the distribution of the third coordination shell in $k$ and $R^*$ and the moderately reduced intensity, we can conclude that bromide is intermixed both with iodide and with chloride on a single octahedron level in the triple-halide perovskite $MAPb(I_{0.2}Br_{0.6}Cl_{0.2})_3$. We note that the Br K-edge analysis directly probes only the halides neighboring Br and therefore cannot, by itself, establish Cl-I mixing within octahedra that contain no Br; a future I K-edge study could provide such direct evidence. However, knowing that Br and I intermix on the single octahedron level, coupled with results obtained at the Pb $L_3$-edge showing the direct impact of Cl on the Pb-I bond distance (see **Figure 4e**), strongly suggests all three halides in local proximity.

## Conclusions

We have demonstrated the formation of a solution-processed triple-halide perovskite in which all three halides coordinate within a single phase, with halide mixing evident down to the scale of individual $PbX_6$ octahedra. Using XRD, we establish that $Br^-$ content is a key determinant of phase stability in triple-halide compositions: a nominal 60% $Br^-$ composition yields a highly oriented and single-phase film, $MAPb(I_{0.2}Br_{0.6}Cl_{0.2})_3$, whereas a lower $Br^-$ content (20%) results in phase segregation into distinct $I^-$-rich and $Cl^-$-rich domains. We perform linear combination analysis of the Pb $L_3$-edge XANES, using experimental spectra for single-halide $MAPbX_3$ as standards, which demonstrates that in mixed compositions, all halides coordinate in a perovskite structure in ratios aligned with solution stoichiometry. WTs of the Pb $L_3$ EXAFS show that in the single-phase triple-halide $MAPb(I_{0.2}Br_{0.6}Cl_{0.2})_3$, more Pb-X bonds take on intermediate lengths compared to the phase-segregated composition, providing a local analogue to the trends observed in bulk-averaged XRD. Quantitative EXAFS fitting establishes these consistent signatures of halide mixing: for double-halides $MAPb(Br_{0.6}Cl_{0.4})_3$ and $MAPb(Br_{0.6}I_{0.4})_3$ and triple-halide $MAPb(I_{0.2}Br_{0.6}Cl_{0.2})_3$, we observe convergence of Pb-X bond lengths toward intermediate values relative to single-halide references, and systematically increased disorder (MSRD), consistent with strain from accommodating mixed ionic radii. Critically, Br K-edge WTs demonstrate third shell halide-to-halide backscattering, amplified by forward scattering through the centric Pb atom, forming a signature that is highly sensitive to the halides on the same octahedron as the absorbing Br atom. For the double-halide compositions $MAPb(Br_{0.6}Cl_{0.4})_3$ and $MAPb(Br_{0.6}I_{0.4})_3$, systematic shifts in the third shell WT intensity directly demonstrate that Br occupies the same $PbX_6$ octahedron with Cl or I, respectively. In triple-halide $MAPb(I_{0.2}Br_{0.6}Cl_{0.2})_3$, the *k*-dependence of this feature reveals a distribution of halide species at the octahedral sites opposite Br, providing evidence that halide mixing with $Br^-$ occurs within individual octahedra in this single-phase triple-halide perovskite. These results connect average crystallographic structure, Pb-centered coordination, and Br-centered local environment to establish mixed-halide incorporation from the bulk crystal structure down to the individual $PbX_6$ octahedron. These results highlight the remarkable structural flexibility of mixed-halide perovskites and highlight a robust spectroscopic framework for characterizing complex halide incorporation. As the compositional space of mixed halide perovskites continues to be explored, understanding and controlling halide coordination will be essential for the rational design of stable, high-performance wide-bandgap absorbers.

## ASSOCIATED CONTENT

**Supporting Information**.

Normalization and background subtraction of XAS data, quantitative fits of single-halide Pb EXAFS, Pb EXAFS fit values and plots in q for mixed-halide perovskites, full wavelet transform images, XRD on polyimide substrates, structural parameters from XRD fits, ab-initio amplitudes and phases of different EXAFS scattering paths (PDF)

## AUTHOR INFORMATION

**Corresponding Author**

* justus.just@maxiv.lu.se

* rbelisle@wellesley.edu

**Present Addresses**

† Department of Chemistry, University of Toronto, Toronto, Ontario M5S 3H6, Canada

**Author Contributions**

The manuscript was written through contributions of all authors. All authors have given approval to the final version of the manuscript.

## ACKNOWLEDGMENTS

This research was funded by the National Science Foundation (NSF) under award number DMR 2245435. We acknowledge the MAX IV Laboratory for beamtime on the Balder beamline under proposal 20240880. Research conducted at MAX IV, a Swedish national user facility, is supported by Vetenskapsrådet (Swedish Research Council, VR) under contract 2018-07152, Vinnova (Swedish Governmental Agency for Innovation Systems) under contract 2018-04969 and Formas under contract 2019-02496. Use of the Stanford Synchrotron Radiation Lightsource, SLAC National Accelerator Laboratory, is supported by the U.S. Department of Energy, Office of Science, Office of Basic Energy Sciences under contract no. DE-AC02-76SF00515. S.S.M. acknowledges support from the Jerome A. Schiff Fellowship.

REFERENCES

(1) Xiong, Z.; Zhang, Q.; Cai, K.; Zhou, H.; Song, Q.; Han, Z.; Kang, S.; Li, Y.; Jiang, Q.; Zhang, X.; You, J. Homogenized Chlorine Distribution for >27% Power Conversion Efficiency in Perovskite Solar Cells. *Science* **2025**, *390* (6773), 638–642. https://doi.org/10.1126/science.adw8780.

(2) Jia, L.; Xia, S.; Li, J.; Qin, Y.; Pei, B.; Ding, L.; Yin, J.; Du, T.; Fang, Z.; Yin, Y.; Liu, J.; Yang, Y.; Zhang, F.; Wu, X.; Li, Q.; Zhao, S.; Zhang, H.; Li, Q.; Jia, Q.; Liu, C.; Gu, X.; Liu, B.; Dong, X.; Liu, J.; Liu, T.; Gao, Y.; Yang, M.; Yin, S.; Ru, X.; Chen, H.; Yang, B.; Zheng, Z.; Zhou, W.; Dou, M.; Wang, S.; Gao, S.; Chen, L.; Qu, M.; Lu, J.; Fang, L.; Wang, Y.; Deng, H.; Yu, J.; Zhang, X.; Li, M.; Lang, X.; Xiao, C.; Hu, Q.; Xue, C.; Ning, L.; He, Y.; Li, Z.; Xu, X.; He, B. Efficient Perovskite/Silicon Tandem with Asymmetric Self-Assembly Molecule. *Nature* **2025**, *644* (8078), 912–919. https://doi.org/10.1038/s41586-025-09333-z.

(3) Noh, J. H.; Im, S. H.; Heo, J. H.; Mandal, T. N.; Seok, S. I. Chemical Management for Colorful, Efficient, and Stable Inorganic–Organic Hybrid Nanostructured Solar Cells. *Nano Lett.* **2013**, *13* (4), 1764–1769. https://doi.org/10.1021/nl400349b.

(4) Hörantner, M. T.; Leijtens, T.; Ziffer, M. E.; Eperon, G. E.; Christoforo, M. G.; McGehee, M. D.; Snaith, H. J. The Potential of Multijunction Perovskite Solar Cells. *ACS Energy Lett.* **2017**, *2* (10), 2506–2513. https://doi.org/10.1021/acsenergylett.7b00647.

(5) Xu, J.; Boyd, C. C.; Yu, Z. J.; Palmstrom, A. F.; Witter, D. J.; Larson, B. W.; France, R. M.; Werner, J.; Harvey, S. P.; Wolf, E. J.; Weigand, W.; Manzoor, S.; van Hest, M. F. A. M.; Berry, J. J.; Luther, J. M.; Holman, Z. C.; McGehee, M. D. Triple-Halide Wide–Band Gap Perovskites with Suppressed Phase Segregation for Efficient Tandems. *Science* **2020**, *367* (6482), 1097–1104. https://doi.org/10.1126/science.aaz5074.

(6) Shen, X.; Gallant, B. M.; Holzhey, P.; Smith, J. A.; Elmestekawy, K. A.; Yuan, Z.; Rathnayake, P. V. G. M.; Bernardi, S.; Dasgupta, A.; Kasparavicius, E.; Malinauskas, T.; Caprioglio, P.; Shargaieva, O.; Lin, Y.-H.; McCarthy, M. M.; Unger, E.; Getautis, V.; Widmer-Cooper, A.; Herz, L. M.; Snaith, H. J. Chloride-Based Additive Engineering for Efficient and Stable Wide-Bandgap Perovskite Solar Cells. *Adv. Mater.* **2023**, *35* (30), 2211742. https://doi.org/10.1002/adma.202211742.

(7) Castelli, I. E.; García-Lastra, J. M.; Thygesen, K. S.; Jacobsen, K. W. Bandgap Calculations and Trends of Organometal Halide Perovskites. *APL Mater.* **2014**, *2* (8), 081514. https://doi.org/10.1063/1.4893495.
(8) Wang, J.; Zeng, L.; Zhang, D.; Maxwell, A.; Chen, H.; Datta, K.; Caiazzo, A.; Remmerswaal, W. H. M.; Schipper, N. R. M.; Chen, Z.; Ho, K.; Dasgupta, A.; Kusch, G.; Ollearo, R.; Bellini, L.; Hu, S.; Wang, Z.; Li, C.; Teale, S.; Grater, L.; Chen, B.; Wienk, M. M.; Oliver, R. A.; Snaith, H. J.; Janssen, R. A. J.; Sargent, E. H. Halide Homogenization for Low Energy Loss in 2-eV-Bandgap Perovskites and Increased Efficiency in All-Perovskite Triple-Junction Solar Cells. *Nat. Energy* **2024**, *9* (1), 70–80. https://doi.org/10.1038/s41560-023-01406-5.
(9) Hu, S.; Wang, J.; Zhao, P.; Pascual, J.; Wang, J.; Rombach, F.; Dasgupta, A.; Liu, W.; Truong, M. A.; Zhu, H.; Kober-Czerny, M.; Drysdale, J. N.; Smith, J. A.; Yuan, Z.; Aalbers, G. J. W.; Schipper, N. R. M.; Yao, J.; Nakano, K.; Turren-Cruz, S.-H.; Dallmann, A.; Christoforo, M. G.; Ball, J. M.; McMeekin, D. P.; Zaininger, K.-A.; Liu, Z.; Noel, N. K.; Tajima, K.; Chen, W.; Ehara, M.; Janssen, R. A. J.; Wakamiya, A.; Snaith, H. J. Steering Perovskite Precursor Solutions for Multijunction Photovoltaics. *Nature* **2025**, *639* (8053), 93–101. https://doi.org/10.1038/s41586-024-08546-y.
(10) Frohna, K.; Anaya, M.; Macpherson, S.; Sung, J.; Doherty, T. A. S.; Chiang, Y.-H.; Winchester, A. J.; Orr, K. W. P.; Parker, J. E.; Quinn, P. D.; Dani, K. M.; Rao, A.; Stranks, S. D. Nanoscale Chemical Heterogeneity Dominates the Optoelectronic Response of Alloyed Perovskite Solar Cells. *Nat. Nanotechnol.* **2022**, *17* (2), 190–196. https://doi.org/10.1038/s41565-021-01019-7.
(11) Meggiolaro, D.; Motti, S. G.; Mosconi, E.; Barker, A. J.; Ball, J.; Perini, C. A. R.; Deschler, F.; Petrozza, A.; Angelis, F. D. Iodine Chemistry Determines the Defect Tolerance of Lead-Halide Perovskites. *Energy Environ. Sci.* **2018**, *11* (3), 702–713. https://doi.org/10.1039/C8EE00124C.
(12) Huang, T.; Tan, S.; Nuryyeva, S.; Yavuz, I.; Babbe, F.; Zhao, Y.; Abdelsamie, M.; Weber, M. H.; Wang, R.; Houk, K. N.; Sutter-Fella, C. M.; Yang, Y. Performance-Limiting Formation Dynamics in Mixed-Halide Perovskites. *Sci. Adv.* **2021**, *7* (46), eabj1799. https://doi.org/10.1126/sciadv.abj1799.
(13) Mahesh, S.; Ball, J. M.; Oliver, R. D. J.; McMeekin, D. P.; Nayak, P. K.; Johnston, M. B.; Snaith, H. J. Revealing the Origin of Voltage Loss in Mixed-Halide Perovskite Solar Cells. *Energy Environ. Sci.* **2020**, *13* (1), 258–267. https://doi.org/10.1039/C9EE02162K.
(14) Hoke, E. T.; Slotcavage, D. J.; Dohner, E. R.; Bowring, A. R.; Karunadasa, H. I.; McGehee, M. D. Reversible Photo-Induced Trap Formation in Mixed-Halide Hybrid Perovskites for Photovoltaics. *Chem. Sci.* **2014**, *6* (1), 613–617. https://doi.org/10.1039/C4SC03141E.
(15) Halford, G. C.; Deng, Q.; Gomez, A.; Green, T.; Mankoff, J. M.; Belisle, R. A. Structural Dynamics of Metal Halide Perovskites during Photoinduced Halide Segregation. *ACS Appl. Mater. Interfaces* **2022**, *14* (3), 4335–4343. https://doi.org/10.1021/acsami.1c22854.
(16) Ruth, A.; Okrepka, H.; Kamat, P.; Kuno, M. Thermodynamic Band Gap Model for Photoinduced Phase Segregation in Mixed-Halide Perovskites. *J. Phys. Chem. C* **2023**, *127* (37), 18547–18559. https://doi.org/10.1021/acs.jpcc.3c04708.
(17) Mao, W.; Hall, C. R.; Bernardi, S.; Cheng, Y.-B.; Widmer-Cooper, A.; Smith, T. A.; Bach, U. Light-Induced Reversal of Ion Segregation in Mixed-Halide Perovskites. *Nat. Mater.* **2021**, *20* (1), 55–61. https://doi.org/10.1038/s41563-020-00826-y.

(18) Bischak, C. G.; Hetherington, C. L.; Wu, H.; Aloni, S.; Ogletree, D. F.; Limmer, D. T.; Ginsberg, N. S. Origin of Reversible Photoinduced Phase Separation in Hybrid Perovskites. *Nano Lett.* **2017**, *17* (2), 1028–1033. https://doi.org/10.1021/acs.nanolett.6b04453.

(19) Kerner, R. A.; Xu, Z.; Larson, B. W.; Rand, B. P. The Role of Halide Oxidation in Perovskite Halide Phase Separation. *Joule* **2021**, *5* (9), 2273–2295. https://doi.org/10.1016/j.joule.2021.07.011.

(20) Slotcavage, D. J.; Karunadasa, H. I.; McGehee, M. D. Light-Induced Phase Segregation in Halide-Perovskite Absorbers. *ACS Energy Lett.* **2016**, *1* (6), 1199–1205. https://doi.org/10.1021/acsenergylett.6b00495.

(21) Brennan, M. C.; Ruth, A.; Kamat, P. V.; Kuno, M. Photoinduced Anion Segregation in Mixed Halide Perovskites. *Trends Chem.* **2020**, *2* (4), 282–301. https://doi.org/10.1016/j.trechm.2020.01.010.

(22) Yoon, S. J.; Draguta, S.; Manser, J. S.; Sharia, O.; Schneider, W. F.; Kuno, M.; Kamat, P. V. Tracking Iodide and Bromide Ion Segregation in Mixed Halide Lead Perovskites during Photoirradiation. *ACS Energy Lett.* **2016**, *1* (1), 290–296. https://doi.org/10.1021/acsenergylett.6b00158.

(23) Cho, J.; Kamat, P. V. How Chloride Suppresses Photoinduced Phase Segregation in Mixed Halide Perovskites. *Chem. Mater.* **2020**, *32* (14), 6206–6212. https://doi.org/10.1021/acs.chemmater.0c02100.

(24) McLeod, J. A.; Wu, Z.; Sun, B.; Liu, L. The Influence of the I/Cl Ratio on the Performance of $CH_3NH_3PbI_{3-x}Cl_x$-Based Solar Cells: Why Is $CH_3NH_3I$ : $PbCl_2$ = 3 : 1 the "Magic" Ratio? *Nanoscale* **2016**, *8* (12), 6361–6368. https://doi.org/10.1039/C5NR06217A.

(25) Stone, K. H.; Gold-Parker, A.; Pool, V. L.; Unger, E. L.; Bowring, A. R.; McGehee, M. D.; Toney, M. F.; Tassone, C. J. Transformation from Crystalline Precursor to Perovskite in PbCl2-Derived MAPbI3. *Nat. Commun.* **2018**, *9* (1), 3458. https://doi.org/10.1038/s41467-018-05937-4.

(26) Yang, B.; Keum, J.; Ovchinnikova, O. S.; Belianinov, A.; Chen, S.; Du, M.-H.; Ivanov, I. N.; Rouleau, C. M.; Geohegan, D. B.; Xiao, K. Deciphering Halogen Competition in Organometallic Halide Perovskite Growth. *J. Am. Chem. Soc.* **2016**, *138* (15), 5028–5035. https://doi.org/10.1021/jacs.5b13254.

(27) Wang, J.; Hu, S.; Gu, X.; Truong, M. A.; Yang, Y.; Liu, C.; Kusch, G.; Yuan, Z.; Kober-Czerny, M.; Zhang, Z.; Su, Z.; Nakano, K.; Dasgupta, A.; Zhang, X.; Shen, X.; Shioya, N.; Kurose, N.; Shirakura, D.; Wang, Z.; Zhou, W.; Li, M.; Hasegawa, T.; Gao, X.; Tajima, K.; Oliver, R. A.; Zhao, Y.; Ning, Z.; Wakamiya, A.; Snaith, H. J.; Chen, H. Homogenized Optoelectronic Properties in Perovskites: Achieving High-Efficiency Solar Cells with Common Chloride Additives. *J. Am. Chem. Soc.* **2026**, *148* (6), 6229–6237. https://doi.org/10.1021/jacs.5c18303.

(28) Chae, J.; Dong, Q.; Huang, J.; Centrone, A. Chloride Incorporation Process in CH3NH3PbI3−xClx Perovskites via Nanoscale Bandgap Maps. *Nano Lett.* **2015**, *15* (12), 8114–8121. https://doi.org/10.1021/acs.nanolett.5b03556.

(29) Kim, S.-Y.; Lee, H.-C.; Nam, Y.; Yun, Y.; Lee, S.-H.; Kim, D. H.; Noh, J. H.; Lee, J.-H.; Kim, D.-H.; Lee, S.; Heo, Y.-W. Ternary Diagrams of the Phase, Optical Bandgap Energy and Photoluminescence of Mixed-Halide Perovskites. *Acta Mater.* **2019**, *181*, 460–469. https://doi.org/10.1016/j.actamat.2019.10.008.

(30) Mariotti, S.; Köhnen, E.; Scheler, F.; Sveinbjörnsson, K.; Zimmermann, L.; Piot, M.; Yang, F.; Li, B.; Warby, J.; Musiienko, A.; Menzel, D.; Lang, F.; Keßler, S.; Levine, I.; Mantione,

D.; Al-Ashouri, A.; Härtel, M. S.; Xu, K.; Cruz, A.; Kurpiers, J.; Wagner, P.; Köbler, H.; Li, J.; Magomedov, A.; Mecerreyes, D.; Unger, E.; Abate, A.; Stolterfoht, M.; Stannowski, B.; Schlatmann, R.; Korte, L.; Albrecht, S. Interface Engineering for High-Performance, Triple-Halide Perovskite–Silicon Tandem Solar Cells. *Science* **2023**, *381* (6653), 63–69. https://doi.org/10.1126/science.adf5872.

(31) Yang, F.; Tockhorn, P.; Musiienko, A.; Lang, F.; Menzel, D.; Macqueen, R.; Köhnen, E.; Xu, K.; Mariotti, S.; Mantione, D.; Merten, L.; Hinderhofer, A.; Li, B.; Wargulski, D. R.; Harvey, S. P.; Zhang, J.; Scheler, F.; Berwig, S.; Roß, M.; Thiesbrummel, J.; Al-Ashouri, A.; Brinkmann, K. O.; Riedl, T.; Schreiber, F.; Abou-Ras, D.; Snaith, H.; Neher, D.; Korte, L.; Stolterfoht, M.; Albrecht, S. Minimizing Interfacial Recombination in 1.8 eV Triple-Halide Perovskites for 27.5% Efficient All-Perovskite Tandems. *Adv. Mater.* **2024**, *36* (6), 2307743. https://doi.org/10.1002/adma.202307743.

(32) Zheng, L.; Wei, M.; Eickemeyer, F. T.; Gao, J.; Huang, B.; Gunes, U.; Schouwink, P.; Bi, D. W.; Carnevali, V.; Mensi, M.; Biasoni, F.; Zhang, Y.; Agosta, L.; Slama, V.; Lempesis, N.; Hope, M. A.; Zakeeruddin, S. M.; Emsley, L.; Rothlisberger, U.; Pfeifer, L.; Xuan, Y.; Grätzel, M. Strain-Induced Rubidium Incorporation into Wide-Bandgap Perovskites Reduces Photovoltage Loss. *Science* **2025**, *388* (6742), 88–95. https://doi.org/10.1126/science.adt3417.

(33) Park, B.; Philippe, B.; Jain, S. M.; Zhang, X.; Edvinsson, T.; Rensmo, H.; Zietz, B.; Boschloo, G. Chemical Engineering of Methylammonium Lead Iodide/Bromide Perovskites: Tuning of Opto-Electronic Properties and Photovoltaic Performance. *J. Mater. Chem. A* **2015**, *3* (43), 21760–21771. https://doi.org/10.1039/C5TA05470B.

(34) Chen, Y.; Motti, S. G.; Oliver, R. D. J.; Wright, A. D.; Snaith, H. J.; Johnston, M. B.; Herz, L. M.; Filip, M. R. Optoelectronic Properties of Mixed Iodide–Bromide Perovskites from First-Principles Computational Modeling and Experiment. *J. Phys. Chem. Lett.* **2022**, *13* (18), 4184–4192. https://doi.org/10.1021/acs.jpclett.2c00938.

(35) Zhou, Y.; Wang, F.; Fang, H.-H.; Loi, M. A.; Xie, F.-Y.; Zhao, N.; Wong, C.-P. Distribution of Bromine in Mixed Iodide–Bromide Organolead Perovskites and Its Impact on Photovoltaic Performance. *J. Mater. Chem. A* **2016**, *4* (41), 16191–16197. https://doi.org/10.1039/C6TA07647E.

(36) Kamat, P. V.; Kuno, M. Halide Ion Migration in Perovskite Nanocrystals and Nanostructures. *Acc. Chem. Res.* **2021**, *54* (3), 520–531. https://doi.org/10.1021/acs.accounts.0c00749.

(37) Gratia, P.; Grancini, G.; Audinot, J.-N.; Jeanbourquin, X.; Mosconi, E.; Zimmermann, I.; Dowsett, D.; Lee, Y.; Grätzel, M.; De Angelis, F.; Sivula, K.; Wirtz, T.; Nazeeruddin, M. K. Intrinsic Halide Segregation at Nanometer Scale Determines the High Efficiency of Mixed Cation/Mixed Halide Perovskite Solar Cells. *J. Am. Chem. Soc.* **2016**, *138* (49), 15821–15824. https://doi.org/10.1021/jacs.6b10049.

(38) Lee, H.; Boonmongkolras, P.; Jun, S.; Kim, D.; Park, Y.; Koh, J.; Cho, Y.-H.; Shin, B.; Park, J. Y. In Situ Observation of Photoinduced Halide Segregation in Mixed Halide Perovskite. *ACS Appl. Energy Mater.* **2023**, *6* (3), 1565–1574. https://doi.org/10.1021/acsaem.2c03438.

(39) Tang, X.; van den Berg, M.; Gu, E.; Horneber, A.; Matt, G. J.; Osvet, A.; Meixner, A. J.; Zhang, D.; Brabec, C. J. Local Observation of Phase Segregation in Mixed-Halide Perovskite. *Nano Lett.* **2018**, *18* (3), 2172–2178. https://doi.org/10.1021/acs.nanolett.8b00505.

(40) Suchan, K.; Just, J.; Beblo, P.; Rehermann, C.; Merdasa, A.; Mainz, R.; Scheblykin, I. G.; Unger, E. Multi-Stage Phase-Segregation of Mixed Halide Perovskites under Illumination: A Quantitative Comparison of Experimental Observations and Thermodynamic Models. *Adv. Funct. Mater.* **2023**, *33* (3), 2206047. https://doi.org/10.1002/adfm.202206047.

(41) Suchan, K.; Just, J.; Becker, P.; Unger, E. L.; Unold, T. Optical in Situ Monitoring during the Synthesis of Halide Perovskite Solar Cells Reveals Formation Kinetics and Evolution of Optoelectronic Properties. *J. Mater. Chem. A* **2020**, *8* (20), 10439–10449. https://doi.org/10.1039/D0TA01237H.

(42) Goldschmidt, V. M. Die Gesetze der Krystallochemie. *Naturwissenschaften* **1926**, *14* (21), 477–485. https://doi.org/10.1007/BF01507527.

(43) Varadwaj, P. R.; Varadwaj, A.; Marques, H. M.; Yamashita, K. Significance of Hydrogen Bonding and Other Noncovalent Interactions in Determining Octahedral Tilting in the CH3NH3PbI3 Hybrid Organic-Inorganic Halide Perovskite Solar Cell Semiconductor. *Sci. Rep.* **2019**, *9* (1), 50. https://doi.org/10.1038/s41598-018-36218-1.

(44) Weadock, N. J.; MacKeen, C.; Qin, X.; Waquier, L.; Rakita, Y.; Vigil, J. A.; Karunadasa, H. I.; Blum, V.; Toney, M. F.; Bridges, F. Thermal Contributions to the Local and Long-Range Structural Disorder in CH 3 NH 3 Pb Br 3. *PRX Energy* **2023**, *2* (3), 033004. https://doi.org/10.1103/PRXEnergy.2.033004.

(45) Schuck, G.; Többens, D. M.; Wallacher, D.; Grimm, N.; Tien, T. S.; Schorr, S. Temperature-Dependent EXAFS Measurements of the Pb L3-Edge Allow Quantification of the Anharmonicity of the Lead–Halide Bond of Chlorine-Substituted Methylammonium (MA) Lead Triiodide. *J. Phys. Chem. C* **2022**, *126* (12), 5388–5402. https://doi.org/10.1021/acs.jpcc.1c05750.

(46) Schuck, G.; Többens, D. M.; Schorr, S. On the Thermal Expansion of the Tetragonal Phase of $MAPbI_3$ and $MAPbBr_3$. *Microstructures* **2024**, *4* (4). https://doi.org/10.20517/microstructures.2024.33.

(47) Singh, H.; Fei, R.; Rakita, Y.; Kulbak, M.; Cahen, D.; Rappe, A. M.; Frenkel, A. I. Origin of the Anomalous Pb-Br Bond Dynamics in Formamidinium Lead Bromide Perovskites. *Phys. Rev. B* **2020**, *101* (5), 054302. https://doi.org/10.1103/PhysRevB.101.054302.

(48) Bridges, F.; Gruzdas, J.; MacKeen, C.; Mayford, K.; Weadock, N. J.; Baltazar, V. U.; Rakita, Y.; Waquier, L.; Vigil, J. A.; Karunadasa, H. I.; Toney, M. F. Local Structure, Bonding, and Asymmetry of ((NH2)2⊠CH)⊠PbBr3, CsPbBr3, and (CH3⊠NH3)⊠PbBr3. *Phys. Rev. B* **2023**, *108* (21), 214102. https://doi.org/10.1103/PhysRevB.108.214102.

(49) Nandi, P.; Mahana, S.; Welter, E.; Topwal, D. Probing the Role of Local Structure in Driving the Stability of Halide Perovskites $CH_3NH_3PbX_3$. *J. Phys. Chem. C* **2021**, *125* (44), 24655–24662. https://doi.org/10.1021/acs.jpcc.1c07225.

(50) Zuo, S.; Niu, W.; Chu, S.; An, P.; Huang, H.; Zheng, L.; Zhao, L.; Zhang, J. Water-Regulated Lead Halide Perovskites Precursor Solution: Perovskite Structure Making and Breaking. *J. Phys. Chem. Lett.* **2023**, *14* (20), 4876–4885. https://doi.org/10.1021/acs.jpclett.3c00683.

(51) Sharenko, A.; Mackeen, C.; Jewell, L.; Bridges, F.; Toney, M. F. Evolution of Iodoplumbate Complexes in Methylammonium Lead Iodide Perovskite Precursor Solutions. *Chem. Mater.* **2017**, *29* (3), 1315–1320. https://doi.org/10.1021/acs.chemmater.6b04917.

(52) Flatken, M. A.; Radicchi, E.; Wendt, R.; Buzanich, A. G.; Härk, E.; Pascual, J.; Mathies, F.; Shargaieva, O.; Prause, A.; Dallmann, A.; De Angelis, F.; Hoell, A.; Abate, A. Role of the Alkali Metal Cation in the Early Stages of Crystallization of Halide Perovskites. *Chem. Mater.* **2022**, *34* (3), 1121–1131. https://doi.org/10.1021/acs.chemmater.1c03563.

(53) Kim, J.; Park, B.; Baek, J.; Yun, J. S.; Kwon, H.-W.; Seidel, J.; Min, H.; Coelho, S.; Lim, S.; Huang, S.; Gaus, K.; Green, M. A.; Shin, T. J.; Ho-baillie, A. W. Y.; Kim, M. G.; Seok, S. I. Unveiling the Relationship between the Perovskite Precursor Solution and the Resulting Device Performance. *J. Am. Chem. Soc.* **2020**, *142* (13), 6251–6260. https://doi.org/10.1021/jacs.0c00411.

(54) Intan, N. N.; Sorenson, B. A.; Choi, Y. K.; Choi, J. J.; Fulton, J. L.; Govind, N.; Kelly, S. D.; Schenter, G. K.; Clancy, P.; Mundy, C. J. Smaller Is Better: The Case for Lower-Order Iodoplumbate Species Dominating MAPbI3/Dimethylformamide Solutions. *Chem. Mater.* **2024**, *36* (17), 8424–8436. https://doi.org/10.1021/acs.chemmater.4c01523.

(55) Drisdell, W. S.; Leppert, L.; Sutter-Fella, C. M.; Liang, Y.; Li, Y.; Ngo, Q. P.; Wan, L. F.; Gul, S.; Kroll, T.; Sokaras, D.; Javey, A.; Yano, J.; Neaton, J. B.; Toma, F. M.; Prendergast, D.; Sharp, I. D. Determining Atomic-Scale Structure and Composition of Organo-Lead Halide Perovskites by Combining High-Resolution X-Ray Absorption Spectroscopy and First-Principles Calculations. *ACS Energy Lett.* **2017**, *2* (5), 1183–1189. https://doi.org/10.1021/acsenergylett.7b00182.

(56) Kang, D.-H.; Park, Y.-J.; Jeon, Y.-S.; Park, N.-G. Extended X-Ray Absorption Fine Structure (EXAFS) of FAPbI3 for Understanding Local Structure-Stability Relation in Perovskite Solar Cells. *J. Energy Chem.* **2022**, *67*, 549–554. https://doi.org/10.1016/j.jechem.2021.10.028.

(57) Liu, B.; Cui, R.; Huang, H.; Guo, X.; Dong, J.; Yao, H.; Li, Y.; Zhao, D.; Wang, J.; Zhang, J.; Chen, Y.; Sun, B. Elucidating the Mechanisms Underlying PCBM Enhancement of CH3NH3PbI3 Perovskite Solar Cells Using GIXRD and XAFS. *J. Mater. Chem. A* **2020**, *8* (6), 3145–3153. https://doi.org/10.1039/C9TA10763K.

(58) Ishida, H.; Maeda, H.; Hirano, A.; Kubozono, Y.; Furukawa, Y. Local Structures around Pb(II) and Sn(II) in CH3NH3PbX3 (X = Cl, Br, I) and CH3NH3SnX3 (X = Br, I) Studied by Pb LIII-Edge and Sn K-Edge EXAFS. *Phys. Status Solidi A* **1997**, *159* (2), 277–282. https://doi.org/10.1002/1521-396X(199702)159:2%3C277::AID-PSSA277%3E3.0.CO;2-C.

(59) Pradeep, K. R.; Jain, P.; Suhas, K. T.; Murzin, V.; Narayana, C.; Viswanatha, R. Structure of Mixed Halide Perovskite Nanocrystals at Various Length Scales. *J. Phys. Chem. C* **2024**, *128* (39), 16781–16790. https://doi.org/10.1021/acs.jpcc.4c05095.

(60) Chen, Y.-C.; Chou, H.-L.; Lin, J.-C.; Lee, Y.-C.; Pao, C.-W.; Chen, J.-L.; Chang, C.-C.; Chi, R.-Y.; Kuo, T.-R.; Lu, C.-W.; Wang, D.-Y. Enhanced Luminescence and Stability of Cesium Lead Halide Perovskite CsPbX3 Nanocrystals by Cu2+-Assisted Anion Exchange Reactions. *J. Phys. Chem. C* **2019**, *123* (4), 2353–2360. https://doi.org/10.1021/acs.jpcc.8b11535.

(61) Toby, B. H.; Von Dreele, R. B. GSAS-II: The Genesis of a Modern Open-Source All Purpose Crystallography Software Package. *J. Appl. Crystallogr.* **2013**, *46* (2), 544–549. https://doi.org/10.1107/S0021889813003531.

(62) Ashiotis, G.; Deschildre, A.; Nawaz, Z.; Wright, J. P.; Karkoulis, D.; Picca, F. E.; Kieffer, J. The Fast Azimuthal Integration Python Library: pyFAI. *J. Appl. Crystallogr.* **2015**, *48* (2), 510–519. https://doi.org/10.1107/S1600576715004306.

(63) Dane, T. Tgdane/Pygix, 2024. https://github.com/tgdane/pygix (accessed 2026-04-12).

(64) Ravel, B.; Newville, M. ATHENA and ARTEMIS: Interactive Graphical Data Analysis Using IFEFFIT. *Phys. Scr.* **2005**, *2005* (T115), 1007. https://doi.org/10.1238/Physica.Topical.115a01007.

(65) Newville, M. *IFEFFIT* : Interactive XAFS Analysis and *FEFF* Fitting. *J. Synchrotron Radiat.* **2001**, *8* (2), 322–324. https://doi.org/10.1107/S0909049500016964.

(66) Muñoz, M.; Argoul, P.; Farges, F. Continuous Cauchy Wavelet Transform Analyses of EXAFS Spectra: A Qualitative Approach. *Am. Mineral.* **2003**, *88* (4), 694–700. https://doi.org/10.2138/am-2003-0423.

(67) Newville, M. Larch: An Analysis Package for XAFS and Related Spectroscopies. *J. Phys. Conf. Ser.* **2013**, *430* (1), 012007. https://doi.org/10.1088/1742-6596/430/1/012007.

(68) Sadhanala, A.; Ahmad, S.; Zhao, B.; Giesbrecht, N.; Pearce, P. M.; Deschler, F.; Hoye, R. L. Z.; Gödel, K. C.; Bein, T.; Docampo, P.; Dutton, S. E.; De Volder, M. F. L.; Friend, R. H. Blue-Green Color Tunable Solution Processable Organolead Chloride–Bromide Mixed Halide Perovskites for Optoelectronic Applications. *Nano Lett.* **2015**, *15* (9), 6095–6101. https://doi.org/10.1021/acs.nanolett.5b02369.

(69) Franz, A.; Többens, D. M.; Steckhan, J.; Schorr, S. Determination of the Miscibility Gap in the Solid Solutions Series of Methyl-ammonium Lead Iodide/Chloride. *Acta Crystallogr. Sect. B Struct. Sci. Cryst. Eng. Mater.* **2018**, *74* (5), 445–449. https://doi.org/10.1107/S2052520618010764.

(70) Lehmann, F.; Franz, A.; Többens, D. M.; Levcenco, S.; Unold, T.; Taubert, A.; Schorr, S. The Phase Diagram of a Mixed Halide (Br, I) Hybrid Perovskite Obtained by Synchrotron X-Ray Diffraction. *RSC Adv.* **2019**, *9* (20), 11151–11159. https://doi.org/10.1039/C8RA09398A.

(71) Pols, M.; van Duin, A. C. T.; Calero, S.; Tao, S. Mixing I and Br in Inorganic Perovskites: Atomistic Insights from Reactive Molecular Dynamics Simulations. *J. Phys. Chem. C* **2024**, *128* (9), 4111–4118. https://doi.org/10.1021/acs.jpcc.4c00563.

(72) Drużbicki, K.; Pinna, R. S.; Rudić, S.; Jura, M.; Gorini, G.; Fernandez-Alonso, F. Unexpected Cation Dynamics in the Low-Temperature Phase of Methylammonium Lead Iodide: The Need for Improved Models. *J. Phys. Chem. Lett.* **2016**, *7* (22), 4701–4709. https://doi.org/10.1021/acs.jpclett.6b01822.

(73) Alvarez-Galván, M. C.; Alonso, J. A.; López, C. A.; López-Linares, E.; Contreras, C.; Lázaro, M. J.; Fauth, F.; Martínez-Huerta, M. V. Crystal Growth, Structural Phase Transitions, and Optical Gap Evolution of CH3NH3Pb(Br1−xClx)3 Perovskites. *Cryst. Growth Des.* **2019**, *19* (2), 918–924. https://doi.org/10.1021/acs.cgd.8b01463.

(74) Iqbal, A. N.; Orr, K. W. P.; Nagane, S.; Ferrer Orri, J.; Doherty, T. A. S.; Jung, Y.-K.; Chiang, Y.-H.; Selby, T. A.; Lu, Y.; Mirabelli, A. J.; Baldwin, A.; Ooi, Z. Y.; Gu, Q.; Anaya, M.; Stranks, S. D. Composition Dictates Octahedral Tilt and Photostability in Halide Perovskites. *Adv. Mater.* **2024**, *36* (28), 2307508. https://doi.org/10.1002/adma.202307508.

(75) Rehr, J. J.; Albers, R. C. Theoretical Approaches to X-Ray Absorption Fine Structure. *Rev. Mod. Phys.* **2000**, *72* (3), 621–654. https://doi.org/10.1103/RevModPhys.72.621.

(76) Vila, F. D.; Spencer, J. W.; Kas, J. J.; Rehr, J. J.; Bridges, F. Extended X-Ray Absorption Fine Structure of ZrW2O8: Theory vs. Experiment. *Front. Chem.* **2018**, *6*. https://doi.org/10.3389/fchem.2018.00356.

**TOC Graphic**

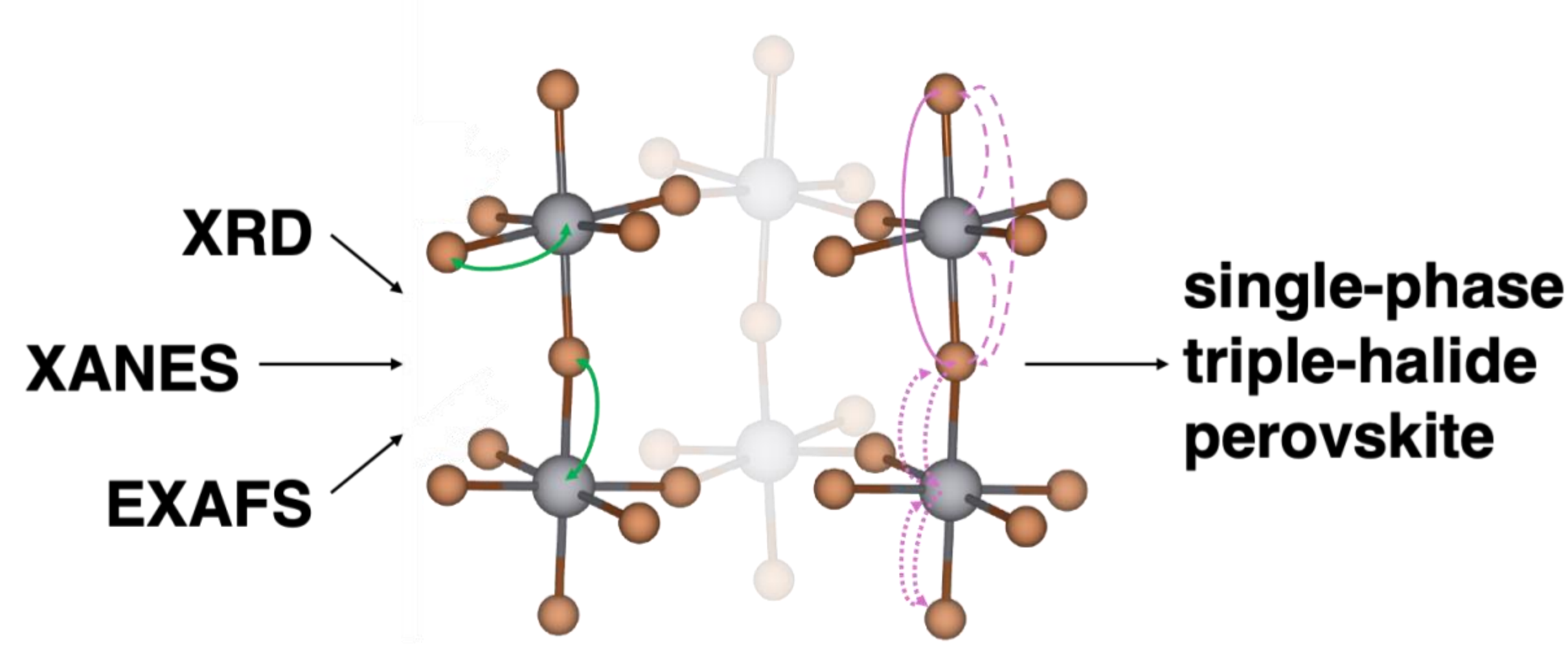

# Supporting Information: Probing local coordination and halide miscibility in single-, double-, and triple-halide perovskites using EXAFS

*Sonia S. Mulgund,[1†] Esther Y.-H. Hung,[2] Leslie Bostwick,[3] Ashley Galbraith,[1] Owen M. Romberg,[4] Justus Just,[2*] Rebecca A. Belisle[1*]*

[1] Department of Physics and Astronomy, Wellesley College, Wellesley, Massachusetts 02481, United States

[2] MAX IV Laboratory, Lund University, 22484 Lund, Sweden

[3] Olin College of Engineering, Needham, Massachusetts 02492, United States

[4] Department of Chemistry, Wellesley College, Wellesley, Massachusetts 02471 United States

**X-ray Diffraction Analysis**

All X-ray diffraction images were collected at grazing incidence of 3° and approximate detector distance of 175 mm. Image calibration was done using a $LaB_6$ standard in pyFAI,[1] followed by grazing incidence geometry correction using pygix.[2] Images and diffraction patterns represent the average diffraction from three subsequent scans normalized for beam intensity. Full azimuthal integrations (chi width of 170°) were completed for all samples. Relevant structural parameters (e.g. FWHM of diffracted peaks, peak area, etc.) were determined by curve fitting background subtracted integrated data to a pseudo-Voight peak model using a non-linear least squares method. Lattice parameters were determined from Pawley fit performed using GSAS-II over the Q≈1.0–3.3 $Å^{-1}$ data range for all apparent mixed compositions.[3] Lattice parameters and other structural parameters are presented in **Table S1**. Data and fits from Pawley fit assuming $Pm\bar{3}m$ symmetry shown in **Figures S32-34.**

**EXAFS Analysis**

Br K-edge EXAFS data were processed and Cauchy wavelet transforms W($k$,$R$*) computed using a background removal parameter of $r_{bkg}$ ~1.6 Å, a $k$-range of 2.5-14.45 Å$^{-1}$ with $k^2$ weighting. For a low $r_{bkg}$ (~1.0 Å), a low intensity feature which is broad in $k$ appears at low $R$* in all Br $k$-edge WTs (**Figure S30**). This is attributed to residual background rather than a structural coordination shell. The higher $r_{bkg}$ value suppresses this without affecting the first shell Br-Pb feature. FEFF6 (implemented in Artemis) was used to calculate scattering paths for the quantitative Pb $L_3$-edge fits and FEFF8 (implemented in Larch) was used to calculate scattering amplitudes and phases as shown in **Figure S10** and **Figures S25-S27**.

<table>
<tr><td>Chemistry</td><td colspan="2">(100) Pseudocubic lattice parameter (Å)</td><td colspan="2">(100) Pseudo cubic FWHM (Å $^{-1}$)</td></tr>
<tr><td>$MAPb(Br_{0.6}I_{0.4})_3$</td><td colspan="2">6.04*</td><td colspan="2">0.0162</td></tr>
<tr><td>$MAPb(Br_{0.6}Cl_{0.4})_3$</td><td colspan="2">5.81*</td><td colspan="2">0.0135</td></tr>
<tr><td>$MAPb(Br_{0.6}I_{0.2}\ Cl_{0.2})_3$</td><td colspan="2">5.91*</td><td colspan="2">0.0160</td></tr>
<tr><td>$MAPb(Br_{0.2}I_{0.4}\ Cl_{0.4})_3$</td><td>6.16</td><td>5.76</td><td>0.0167</td><td>0.0170</td></tr>
</table>

**Table S1.** *Cubic lattice parameter determined from Pawley fit. All other Structural parameters from pseudo-Voigt peak fitting of diffraction data.

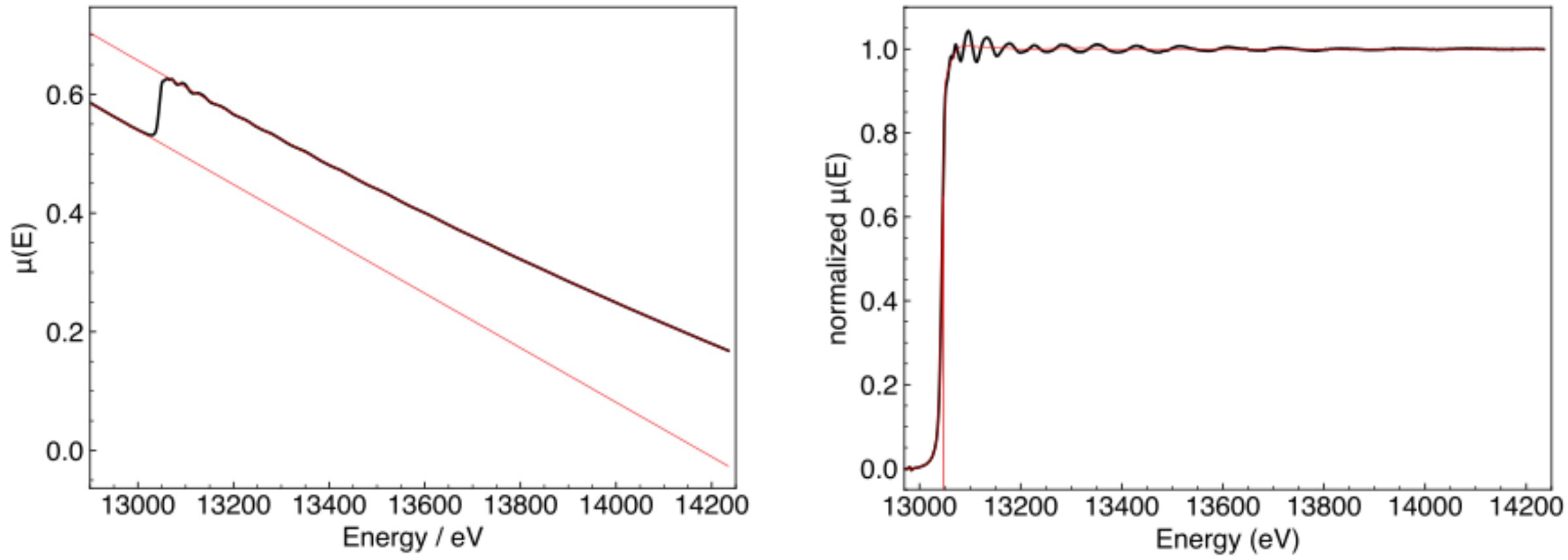

**Figure S1**. Pre-and post-edge polynomials and normalization of Pb L3 EXAFS for $MAPbI_3$. Data in black.

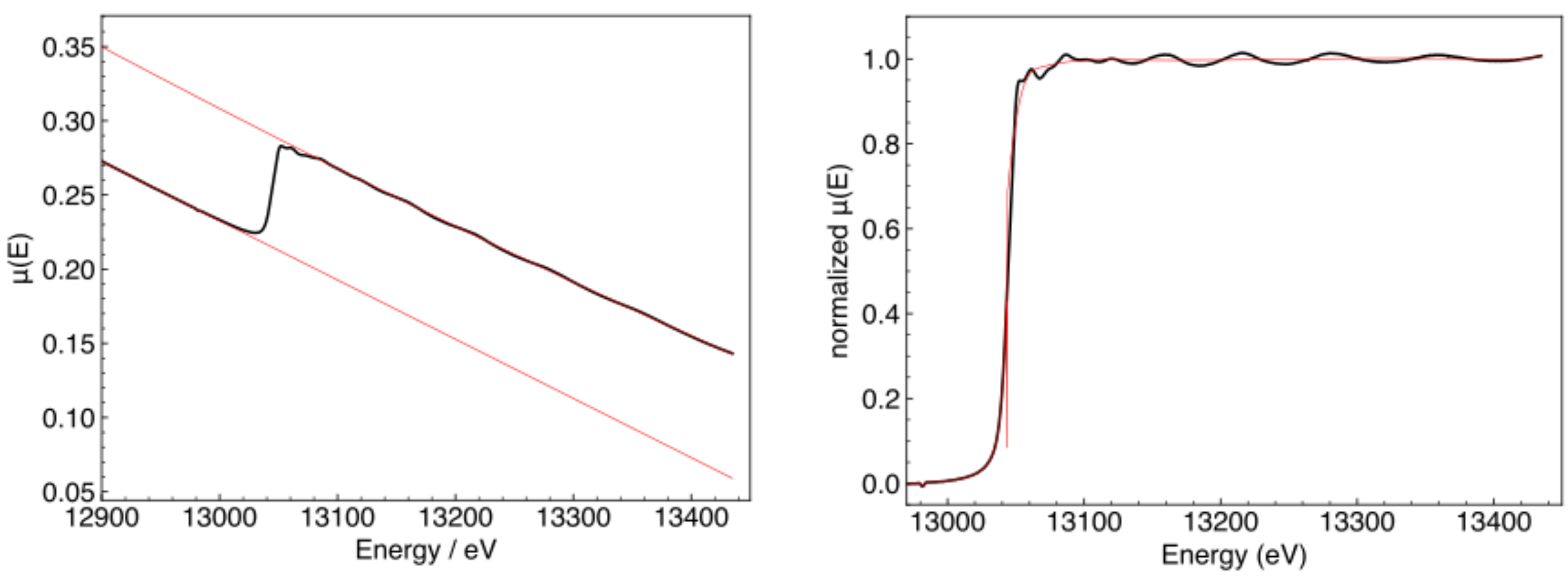

**Figure S2**. Pre-and post-edge polynomials and normalization of Pb L3 EXAFS for $MAPbBr_3$. Data in black.

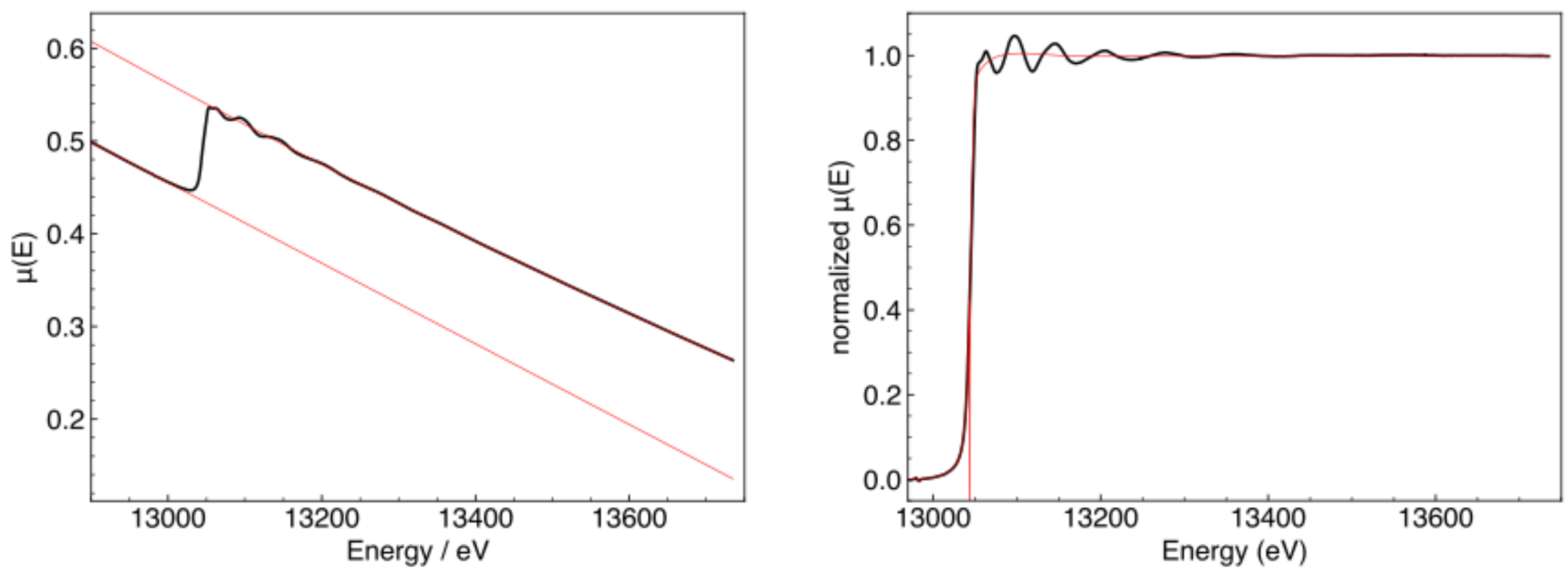

**Figure S3**. Pre-and post-edge polynomials and normalization of Pb L3 EXAFS for $MAPbCl_3$. Data in black.

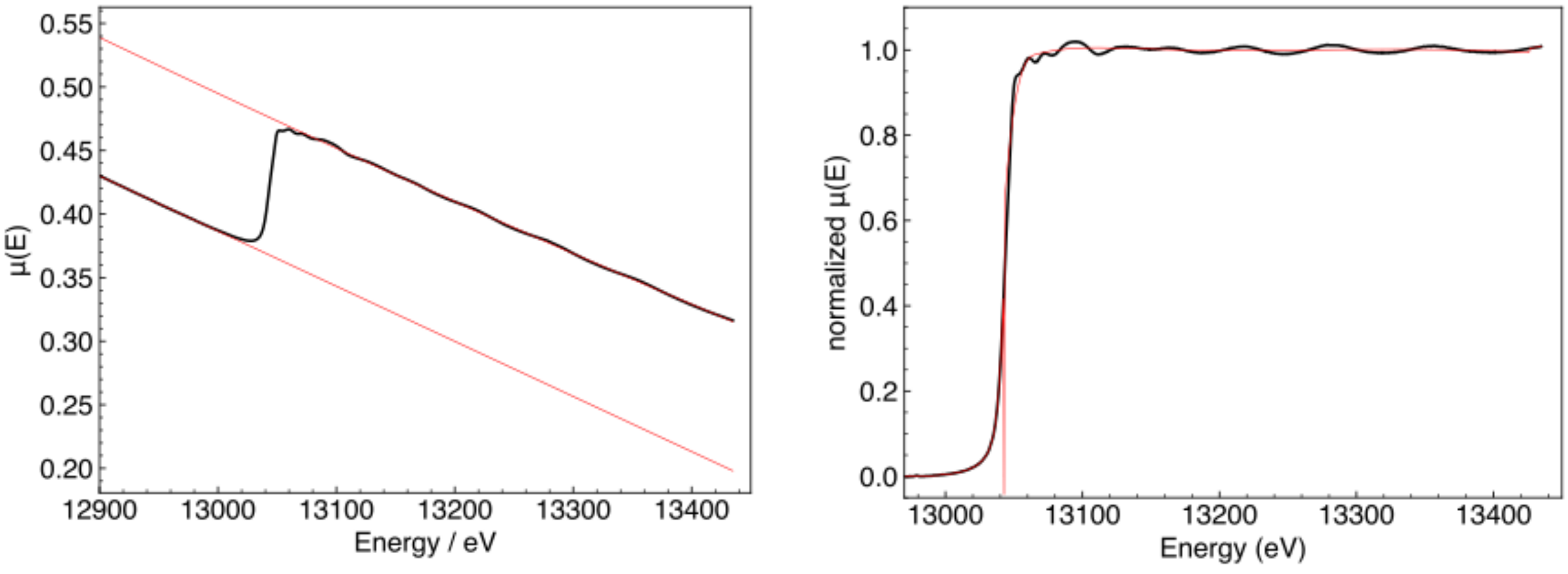

**Figure S4**. Pre-and post-edge polynomials and normalization of Pb L3 EXAFS for $MAPb(Br_{0.6}I_{0.4})_3$. Data in black.

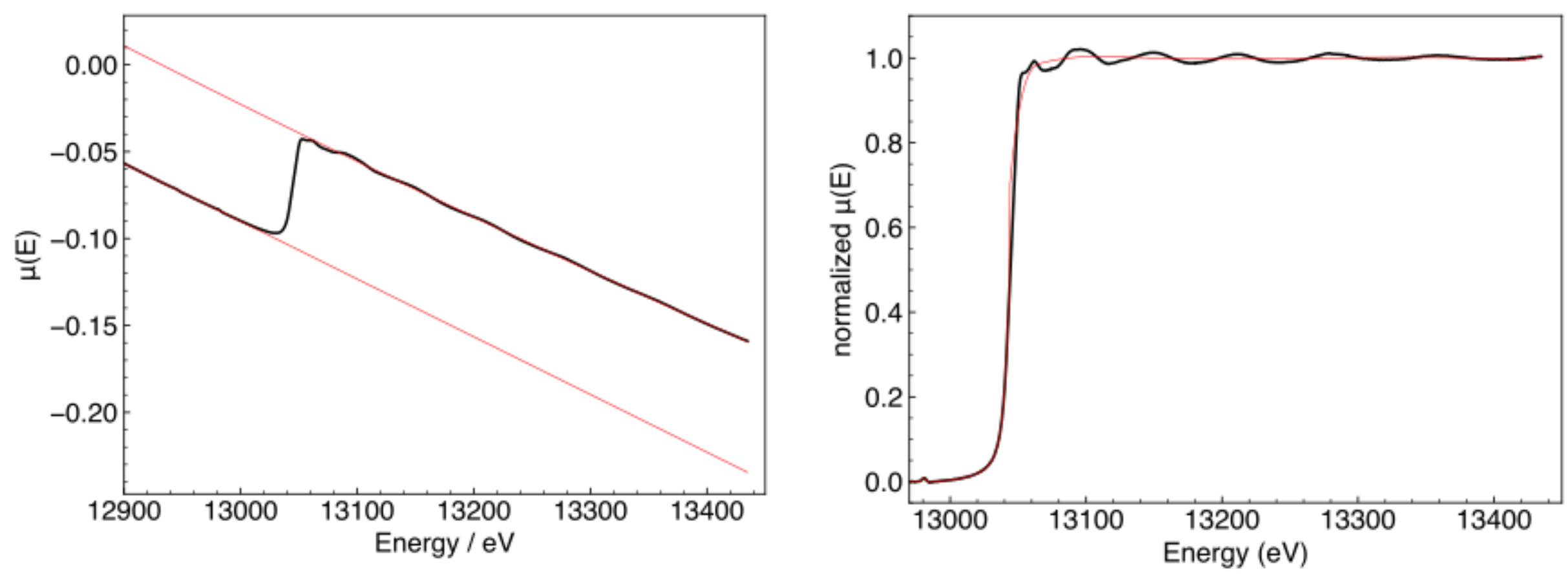

**Figure S5**. Pre-and post-edge polynomials and normalization of Pb L3 EXAFS for $MAPb(Br_{0.6}Cl_{0.4})_3$. Data in black.

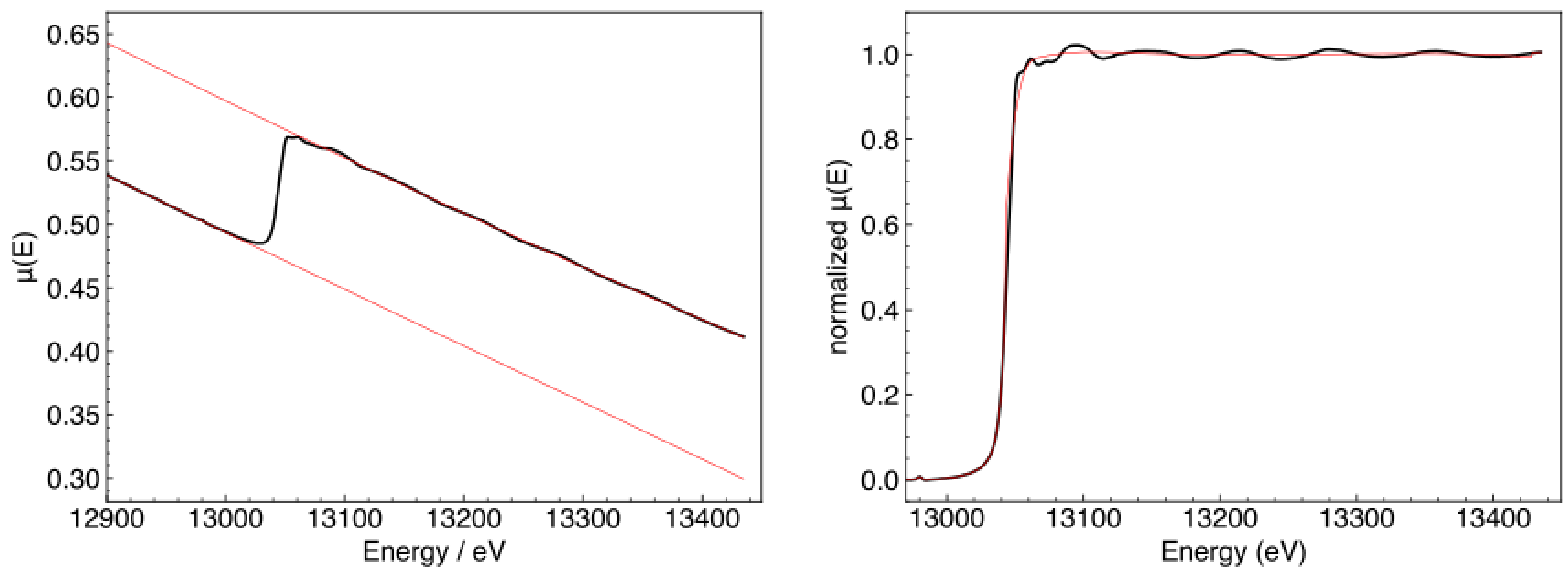

**Figure S6**. Pre-and post-edge polynomials and normalization of Pb L3 EXAFS for $MAPb(I_{0.2}Br_{0.6}Cl_{0.2})_3$. Data in black.

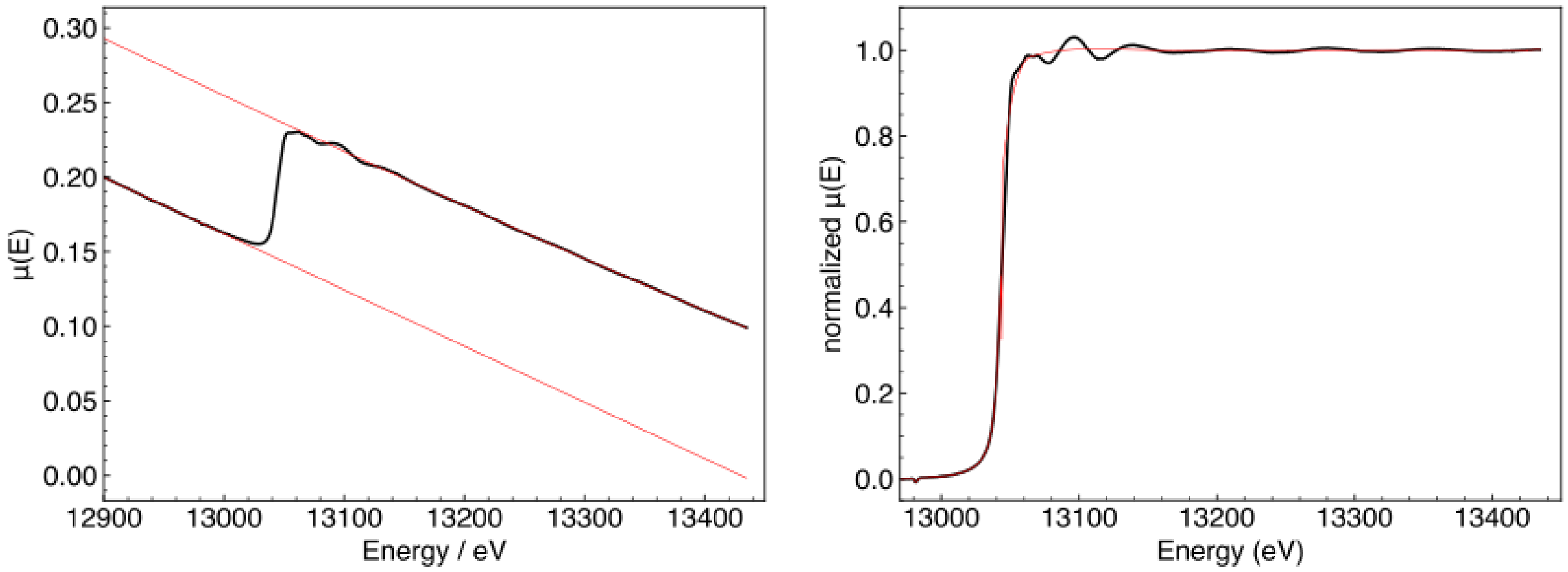

**Figure S7**. Pre-and post-edge polynomials and normalization of Pb L3 EXAFS for $MAPb(I_{0.4}Br_{0.2}Cl_{0.4})_3$. Data in black.

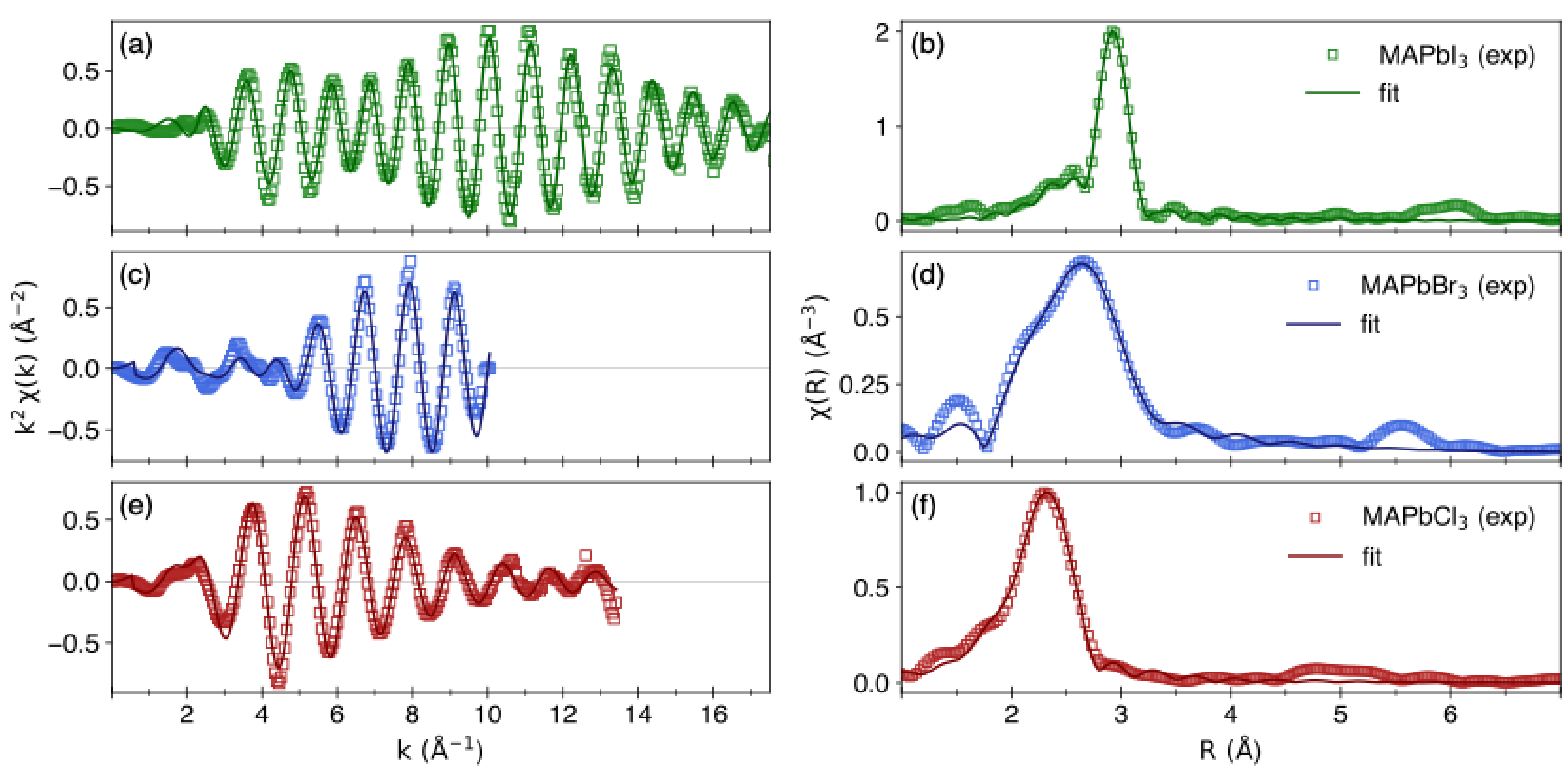

**Figure S8.** Pb $L_3$ EXAFS data with first shell fits for single-halide samples. (b), (d), and (f) show the Fourier transform in R, while (a), (c), and (e) show the q-space fit overlaid on the χ(k) data. (a-b): $MAPbI_3$. (c-d): $MAPbBr_3$. (e-f): $MAPbCl_3$.

| Material | Authors | Pb-X distance | Temperature |
|---|---|---|---|
| $MAPbI_3$ | Weller et al.[4] | 3.19 Å | 100 K |
| $MAPbBr_3$ | Swainson et al.[5] | 3.04 Å | 11 K |
| $MAPbCl_3$ | Chi et al.[6] | 2.73-3.02 Å | 80 K |

**Table S2.** Information from CIFs for single-halide perovskites (all *Pnma*) used as inputs to ATOMS. $MAPbCl_3$ has a lower symmetry $PbX_6$ octahedron, resulting in a spread of Pb-X bond distances.

| Sample | k-weight | R-range (Å) | $\Delta E_0$ (eV) | $S_0^2$ | $R_{Pb-X}$ (Å) | $\sigma^2$ (Å$^2$) |
|---|---|---|---|---|---|---|
| $MAPbI_3$ | 3 | 2-3.3 | -0.71 | 0.85 | 3.181 | 0.004 |
| $MAPbBr_3$ | 3 | 1.75-3.4 | 1.51 | 0.96 | 2.978 | 0.0071 |
| $MAPbCl_3$ | 2,3 | 1.35-3 | 1.30 | 1.09 | 2.723-3.017 | 0.0045 |

**Table S3.** First-shell EXAFS fits (performed in back-Fourier transformed q-space) from each single-halide perovskite composition, as shown in **Figure S1**. The reduced symmetry in the low-temperature orthorhombic phase of $MAPbCl_3$ results in multiple Pb-Cl bond distances, which were all constrained to change by the same fractional amount.

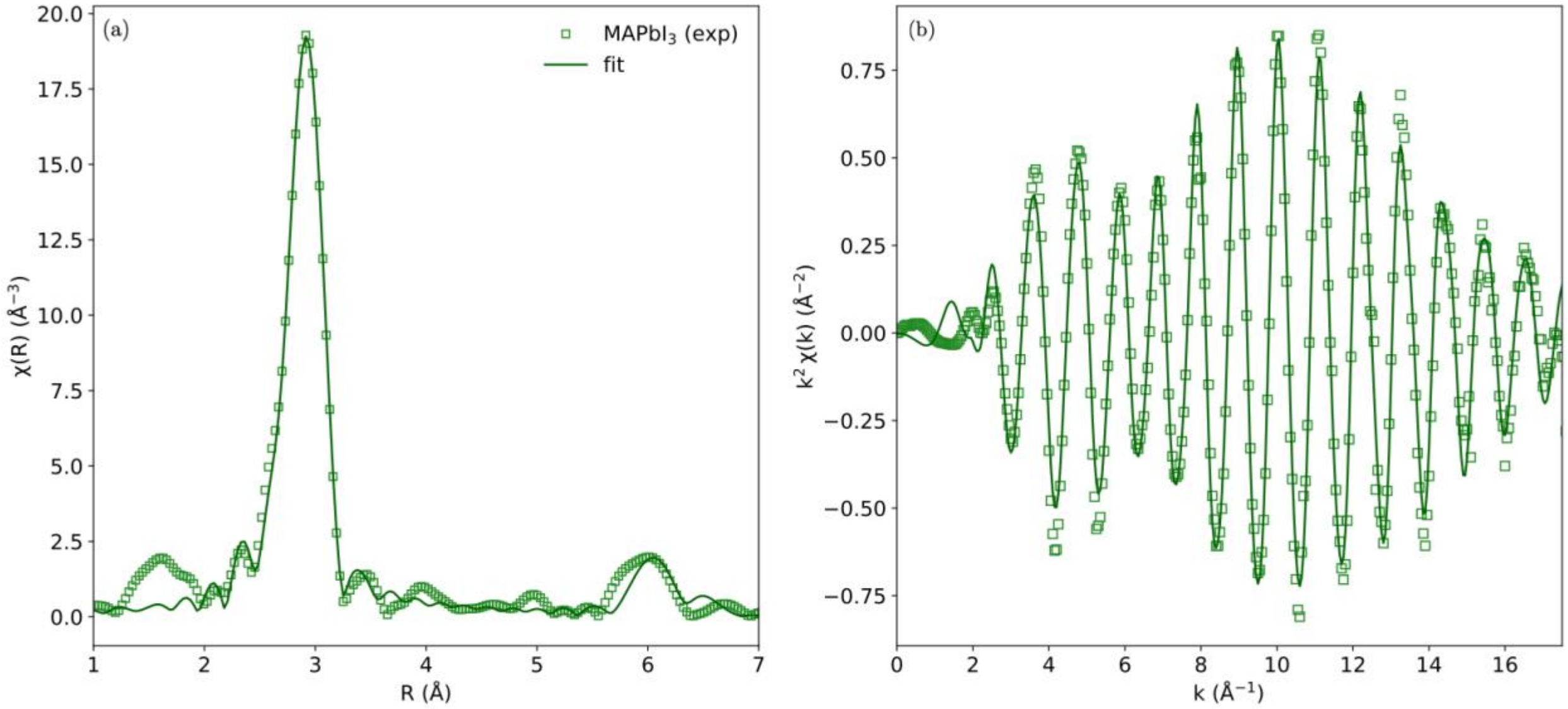

**Figure S9**. Pb L3 EXAFS data for $MAPbI_3$ with second shell fit, performed in back-Fourier transformed q-space. (a) Fit to χ(R). (b) Fit to χ(k). All paths generated by ATOMS containing Pb and I atoms with an effective path length less than 6.5 Å were incorporated into the fit. Values for $S_0^2$ and $\Delta E_0$ were fixed to the first-shell values (**Table S2**). The first shell Pb-I-Pb scattering path was fit to one set of values for ΔR and $\sigma^2$, and all other paths were fit to a second set of values for ΔR and $\sigma^2$. For the first shell R = 3.181 Å and $\sigma^2$ = 0.0041 Å$^2$, and for the second shell ΔR = 0.018 Å and $\sigma^2$ = 0.0048 Å$^2$.

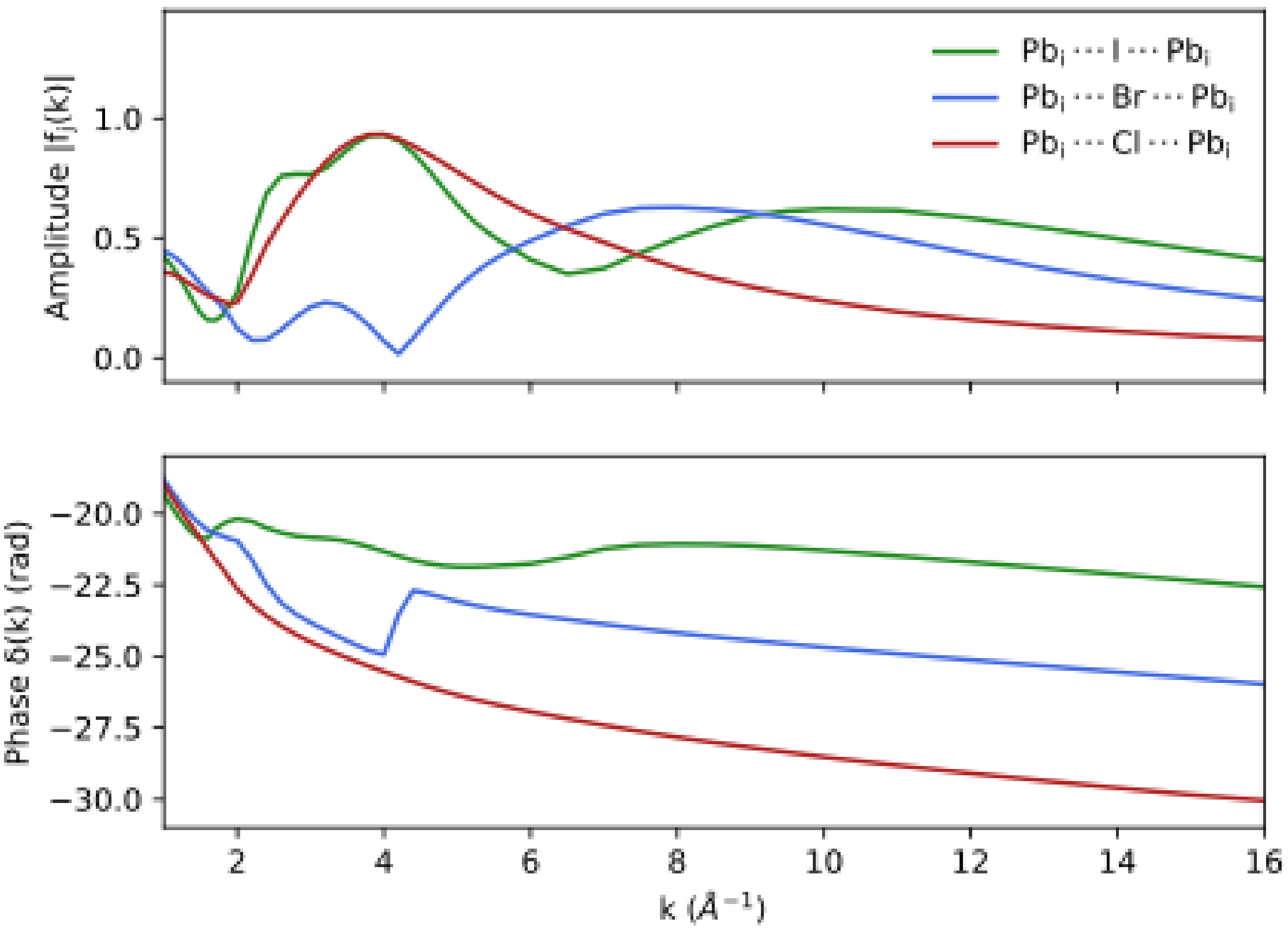

**Figure S10**. Calculated scattering amplitudes and phase shifts for Pb-X single scattering paths as calculated in FEFF from the structures in **Table S1** using the $\sigma^2$ from **Tables S4-S6**.

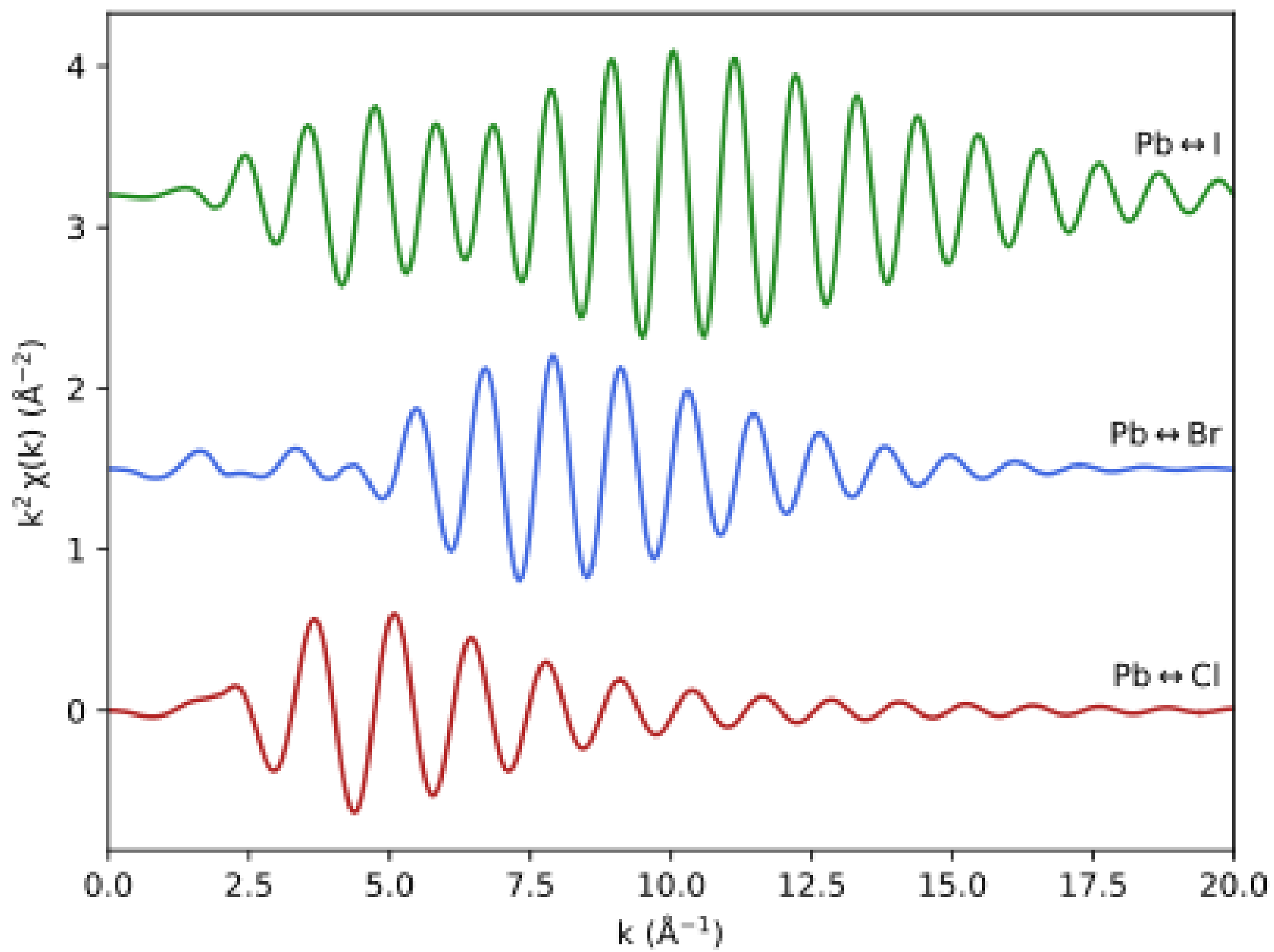

**Figure S11**. Calculated χ(k) for the three single-halide scattering paths using the fit parameters described in **Tables S4-S6**.

| Sample | k-weight | R-range (Å) | $\Delta E_0$ (eV) | $\sigma^2_{Cl}$ (Å$^2$) | $\sigma^2_{Br}$ (Å$^2$) | $\sigma^2_{I}$ (Å$^2$) |
|---|---|---|---|---|---|---|
| $MAPb(Br_{0.6}Cl_{0.4})_3$ | 3 | 1.75-3.4 | 0.46 | 0.0068 | 0.0098 | --- |
| $MAPb(Br_{0.6}I_{0.4})_3$ | 3 | 1.8-3.4 | 1.12 | --- | 0.0047 | 0.0093 |
| $MAPb(I_{0.2}Br_{0.6}Cl_{0.2})_3$ | 2,3 | 1.85-3.5 | 2.63 | 0.0058 | 0.0068 | 0.0086 |
| Sample | $R_{Pb-Cl}$(Å) | $R_{Pb-Br}$(Å) | $R_{Pb-I}$(Å) | $n_{Cl}$ | $n_{Br}$ | $n_{I}$ |
| $MAPb(Br_{0.6}Cl_{0.4})_3$ | 2.846 | 2.963 | --- | 2.07 | 3.93 | --- |
| $MAPb(Br_{0.6}I_{0.4})_3$ | --- | 2.979 | 3.156 | --- | 2.93 | 3.07 |
| $MAPb(I_{0.2}Br_{0.6}Cl_{0.2})_3$ | 2.869 | 2.974 | 3.175 | 1.2 | 3.6 | 1.2 |

**Table S4**. First-shell EXAFS fits (performed in back-Fourier transformed q-space) from each mixed-halide perovskite composition, as shown in **Figure 4a-d**.

| fit | amp | enot | del_r | ss |
|---|---|---|---|---|
| kw = 3 | | | | |
| kmin = 1.5 | 0.847203 | -0.77373 | -0.00918 | 0.004078 |
| kmin = 1.75 | 0.8472 | -0.77099 | -0.00917 | 0.004078 |
| kmin = 2 | 0.847224 | -0.76649 | -0.00916 | 0.004078 |
| 2.25 | 0.847368 | -0.76189 | -0.00914 | 0.004079 |
| 2.5 | 0.847564 | -0.75828 | -0.00913 | 0.00408 |
| kw = 2,3 | | | | |
| 1.5 | 0.867005 | -0.67038 | -0.00873 | 0.004158 |
| 1.75 | 0.867641 | -0.6542 | -0.00867 | 0.004161 |
| 2 | 0.86837 | -0.63325 | -0.0086 | 0.004164 |
| 2.25 | 0.870031 | -0.61761 | -0.00854 | 0.004171 |
| 2.5 | 0.872055 | -0.60376 | -0.00849 | 0.004179 |
| | | | | |
| AVGS | 0.858166 | -0.70106 | -0.00888 | 0.004123 |
| STDS | 0.011522 | 0.071174 | 0.000298 | 4.66E-05 |

**Table S5.** First-shell EXAFS fits (performed in back-Fourier transformed q-space) for $MAPbI_3$ for a range of k-weights and minimum k-values, which generated the error bar for this composition in **Figure 4e**.

| fit | amp | enot | del_r | ss |
|---|---|---|---|---|
| kw = 3 | | | | |
| kmin = 1.5 | 0.955872 | 1.506708 | -0.06277 | 0.007097 |
| kmin = 1.75 | 0.956511 | 1.510931 | -0.06274 | 0.007103 |
| kmin = 2 | 0.955914 | 1.505223 | -0.06277 | 0.007097 |

| 2.25 | 0.95646 | 1.50862 | -0.06275 | 0.007102 |
|---|---|---|---|---|
| 2.5 | 0.959319 | 1.541919 | -0.06252 | 0.00713 |
| kw = 2,3 | | | | |
| 1.5 | 0.957248 | 1.725591 | -0.06141 | 0.007106 |
| 1.75 | 0.963817 | 1.764445 | -0.06111 | 0.007164 |
| 2 | 0.960667 | 1.696386 | -0.06153 | 0.007134 |
| 2.25 | 0.958456 | 1.653289 | -0.06181 | 0.007114 |
| 2.5 | 0.962555 | 1.726037 | -0.06133 | 0.007155 |
| | | | | |
| AVGS | 0.958682 | 1.613915 | -0.06208 | 0.00712 |
| STDS | 0.002849 | 0.108647 | 0.000694 | 2.45E-05 |

**Table S6.** First-shell EXAFS fits (performed in back-Fourier transformed q-space) for $MAPbBr_3$ for a range of k-weights and minimum k-values, which generated the error bar for this composition in **Figure 4e**.

| fit | amp | enot | frac | ss | delr_1 | delr_2 | delr_3 | delr_4 |
|---|---|---|---|---|---|---|---|---|
| kw = 3 | | | | | | | | |
| kmin = 1.5 | 1.118937 | 0.901241 | 0.996622 | 0.00473 | -0.00922 | -0.00958 | -0.00969 | -0.01022 |
| kmin = 1.75 | 1.118273 | 0.895801 | 0.996606 | 0.004725 | -0.00926 | -0.00962 | -0.00974 | -0.01026 |
| kmin = 2 | 1.11875 | 0.886771 | 0.996586 | 0.004728 | -0.00932 | -0.00968 | -0.0098 | -0.01033 |
| 2.25 | 1.121972 | 0.888445 | 0.996595 | 0.004751 | -0.00929 | -0.00965 | -0.00977 | -0.0103 |
| 2.5 | 1.127111 | 0.902246 | 0.996635 | 0.004788 | -0.00918 | -0.00954 | -0.00965 | -0.01017 |
| kw = 2,3 | | | | | | | | |
| 1.5 | 1.072634 | 1.277381 | 0.997659 | 0.004361 | -0.00639 | -0.00664 | -0.00672 | -0.00708 |
| 1.75 | 1.069779 | 1.265478 | 0.997614 | 0.004336 | -0.00651 | -0.00676 | -0.00684 | -0.00721 |
| 2 | 1.069648 | 1.242589 | 0.997546 | 0.004334 | -0.0067 | -0.00696 | -0.00704 | -0.00742 |

| 2.25 | 1.077024 | 1.2453 | 0.997562 | 0.004396 | -0.00665 | -0.00691 | -0.00699 | -0.00737 |
|---|---|---|---|---|---|---|---|---|
| 2.5 | 1.089191 | 1.301203 | 0.997729 | 0.004498 | -0.0062 | -0.00644 | -0.00652 | -0.00687 |
| | | | | | | | | |
| AVGS | 1.098332 | 1.080645 | 0.997115 | 0.004565 | -0.00787 | -0.00818 | -0.00828 | -0.00872 |
| STDS | 0.024636 | 0.196514 | 0.000536 | 0.000196 | 0.001464 | 0.001521 | 0.001539 | 0.001622 |

**Table S7.** First-shell EXAFS fits (performed in back-Fourier transformed q-space) for $MAPbCl_3$ for a range of k-weights and minimum k-values, which generated the error bar for this composition in **Figure 4e**.

| fit | enot | delr_br | delr_i | ss_br | ss_i | n_i | n_br |
|---|---|---|---|---|---|---|---|
| kw = 3 | | | | | | | |
| kmin = 1.5 | 1.179582 | -0.06099 | -0.03394 | 0.004803 | 0.009245 | 3.045457 | 2.954543 |
| kmin = 1.75 | 1.163692 | -0.06111 | -0.03395 | 0.004782 | 0.009275 | 3.052434 | 2.947566 |
| kmin = 2 | 1.118133 | -0.0614 | -0.03411 | 0.004738 | 0.00933 | 3.068721 | 2.931279 |
| 2.25 | 1.038902 | -0.06198 | -0.0343 | 0.004622 | 0.009491 | 3.107443 | 2.892557 |
| 2.5 | 1.032948 | -0.06257 | -0.03349 | 0.004335 | 0.009963 | 3.189856 | 2.810144 |
| kw = 2,3 | | | | | | | |
| 1.5 | 1.278106 | -0.05827 | -0.03552 | 0.00588 | 0.007344 | 2.742385 | 3.257615 |
| 1.75 | 1.309714 | -0.05821 | -0.03519 | 0.00583 | 0.007442 | 2.754837 | 3.245163 |
| 2 | 1.236263 | -0.05869 | -0.03556 | 0.005721 | 0.007602 | 2.791512 | 3.208488 |
| 2.25 | 1.036507 | -0.06005 | -0.03651 | 0.005446 | 0.007995 | 2.882759 | 3.117241 |
| 2.5 | 0.946705 | -0.06163 | -0.03594 | 0.00483 | 0.008985 | 3.061896 | 2.938104 |
| | | | | | | | |
| AVGS | 1.134055 | -0.06049 | -0.03485 | 0.005099 | 0.008667 | 2.96973 | 3.03027 |
| STDS | 0.119792 | 0.001593 | 0.001018 | 0.000563 | 0.000968 | 0.161754 | 0.161754 |

**Table S8**. First-shell EXAFS fits (performed in back-Fourier transformed q-space) for $MAPb(Br_{0.6}I_{0.4})_3$ for a range of k-weights and minimum k-values, which generated the error bar for this composition in **Figure 4e**.

| fit | enot | delr_br | delr_cl | ss_br | ss_cl | n_cl | n_br |
|---|---|---|---|---|---|---|---|
| kw = 3 | | | | | | | |
| kmin = 1.5 | 0.515593 | -0.07738 | 0.004276 | 0.009828 | 0.006678 | 2.061879 | 3.938121 |
| kmin = 1.75 | 0.456591 | -0.07741 | 0.003964 | 0.009779 | 0.006788 | 2.068427 | 3.931573 |
| kmin = 2 | 0.376799 | -0.0774 | 0.003592 | 0.009686 | 0.006994 | 2.084304 | 3.915696 |
| 2.25 | 0.372137 | -0.07719 | 0.003717 | 0.00959 | 0.007194 | 2.109091 | 3.890909 |
| 2.5 | 0.519706 | -0.07682 | 0.004645 | 0.009604 | 0.007143 | 2.121358 | 3.878642 |
| kw = 2,3 | | | | | | | |
| 1.5 | 0.644961 | -0.08009 | 0.002129 | 0.010985 | 0.004487 | 1.837344 | 4.162656 |
| 1.75 | 0.548625 | -0.08011 | 0.001549 | 0.010938 | 0.004587 | 1.838473 | 4.161527 |
| 2 | 0.415875 | -0.08011 | 0.000796 | 0.010843 | 0.004784 | 1.84773 | 4.15227 |
| 2.25 | 0.407914 | -0.07969 | 0.001179 | 0.010661 | 0.005122 | 1.884738 | 4.115262 |
| 2.5 | 0.624267 | -0.07888 | 0.003019 | 0.010538 | 0.005328 | 1.928566 | 4.071434 |
| | | | | | | | |
| AVGS | 0.488247 | -0.07851 | 0.002887 | 0.010245 | 0.005911 | 1.978191 | 4.021809 |
| STDS | 0.098341 | 0.001393 | 0.001376 | 0.000595 | 0.001141 | 0.120928 | 0.120928 |

**Table S9**. First-shell EXAFS fits (performed in back-Fourier transformed q-space) for $MAPb(Br_{0.6}Cl_{0.4})_3$ for a range of k-weights and minimum k-values, which generated the error bar for this composition in **Figure 4e**.

| fit | enot | delr_cl | delr_br | delr_i | ss_cl | ss_br | ss_i |
|---|---|---|---|---|---|---|---|
| kw = 3 | | | | | | | |
| kmin = 1.5 | 1.4459168 | 0.00232891 | -0.07673137 | -0.01258876 | 0.00372107 | 0.00856729 | 0.00513212 |
| kmin = 1.75 | 1.41373272 | 0.00202601 | -0.07684082 | -0.01280089 | 0.00373347 | 0.00857563 | 0.00511026 |
| kmin = 2 | 1.28172063 | 0.00024242 | -0.07733305 | -0.01367189 | 0.00372018 | 0.00867957 | 0.00491284 |
| 2.25 | 1.06427275 | -0.00318504 | -0.07843807 | -0.01478913 | 0.00356897 | 0.00893906 | 0.00451393 |
| 2.5 | 1.01443497 | -0.00477677 | -0.0792632 | -0.01449042 | 0.00331344 | 0.00914684 | 0.00430529 |
| kw = 2,3 | | | | | | | |
| 1.5 | 2.08502917 | 0.01788753 | -0.06964993 | -0.01681804 | 0.00557795 | 0.00727148 | 0.00714118 |
| 1.75 | 2.31410011 | 0.02242141 | -0.06870423 | -0.01602519 | 0.00568153 | 0.00703214 | 0.00786916 |
| 2 | 1.98080244 | 0.01729762 | -0.06950865 | -0.01866363 | 0.00583674 | 0.00723601 | 0.00712295 |
| 2.25 | 1.42811921 | 0.00679478 | -0.07166264 | -0.02084915 | 0.0055702 | 0.0078003 | 0.00573637 |
| 2.5 | 1.25759033 | 0.00127307 | -0.0733836 | -0.01959332 | 0.00489999 | 0.00828811 | 0.00502115 |
| | | | | | | | |
| AVGS | 1.52857191 | 0.00623099 | -0.07415156 | -0.01602904 | 0.00456235 | 0.00815364 | 0.00568653 |
| STDS | 0.44417207 | 0.00956833 | 0.00403859 | 0.00289095 | 0.00103754 | 0.00076377 | 0.00124317 |

**Table S10**. First-shell EXAFS fits (performed in back-Fourier transformed q-space) for $MAPb(I_{0.2}Br_{0.6}Cl_{0.2})_3$ for a range of k-weights and minimum k-values, which generated the error bar for this composition in **Figure 4e**.

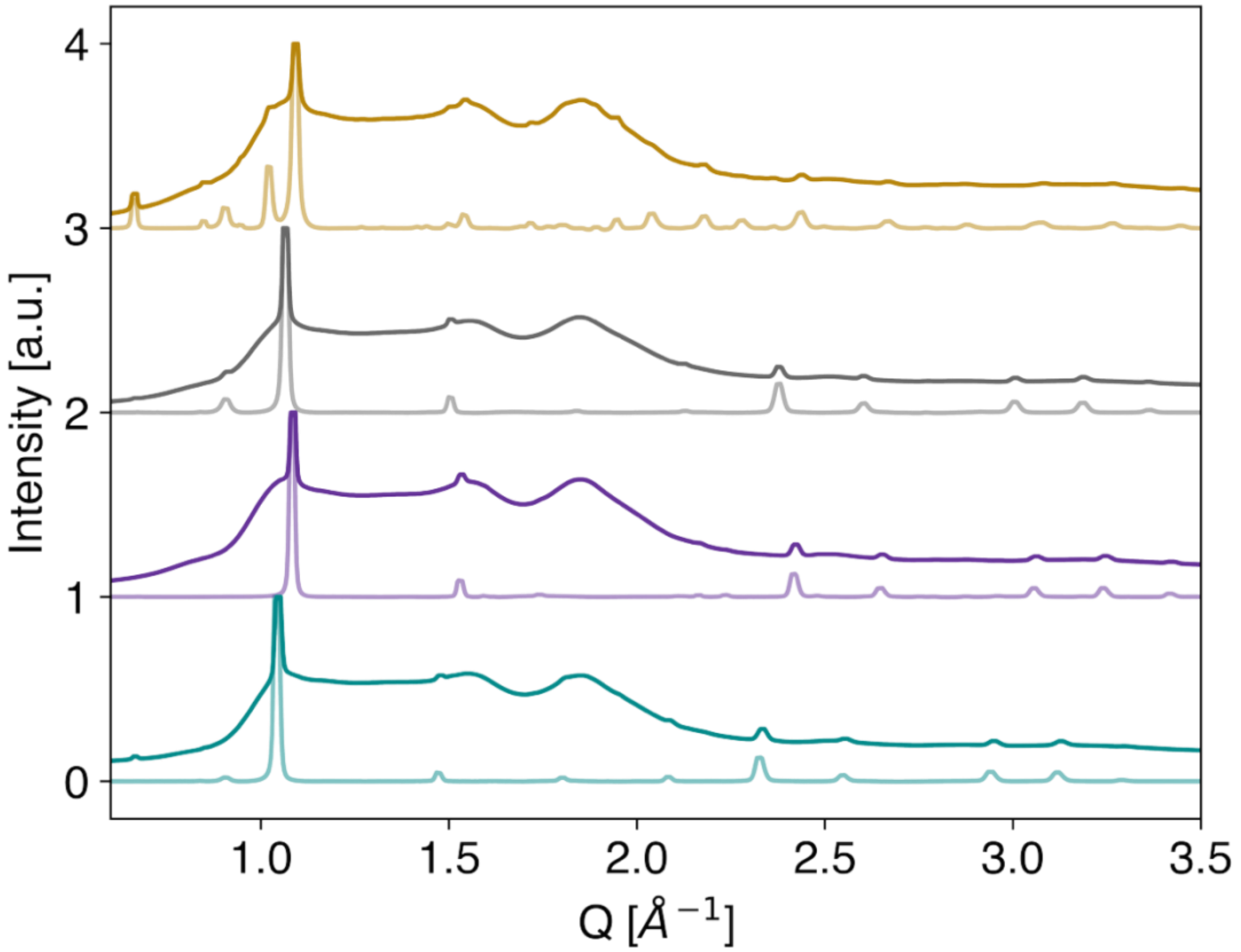

**Figure S12**. X-ray diffraction of mixed-halide perovskites on polyimide substrate (used for EXAFS measurements, dark) vs. glass (light; same data as **Figure 1**). The broad amorphous background results from the polyimide substrate.

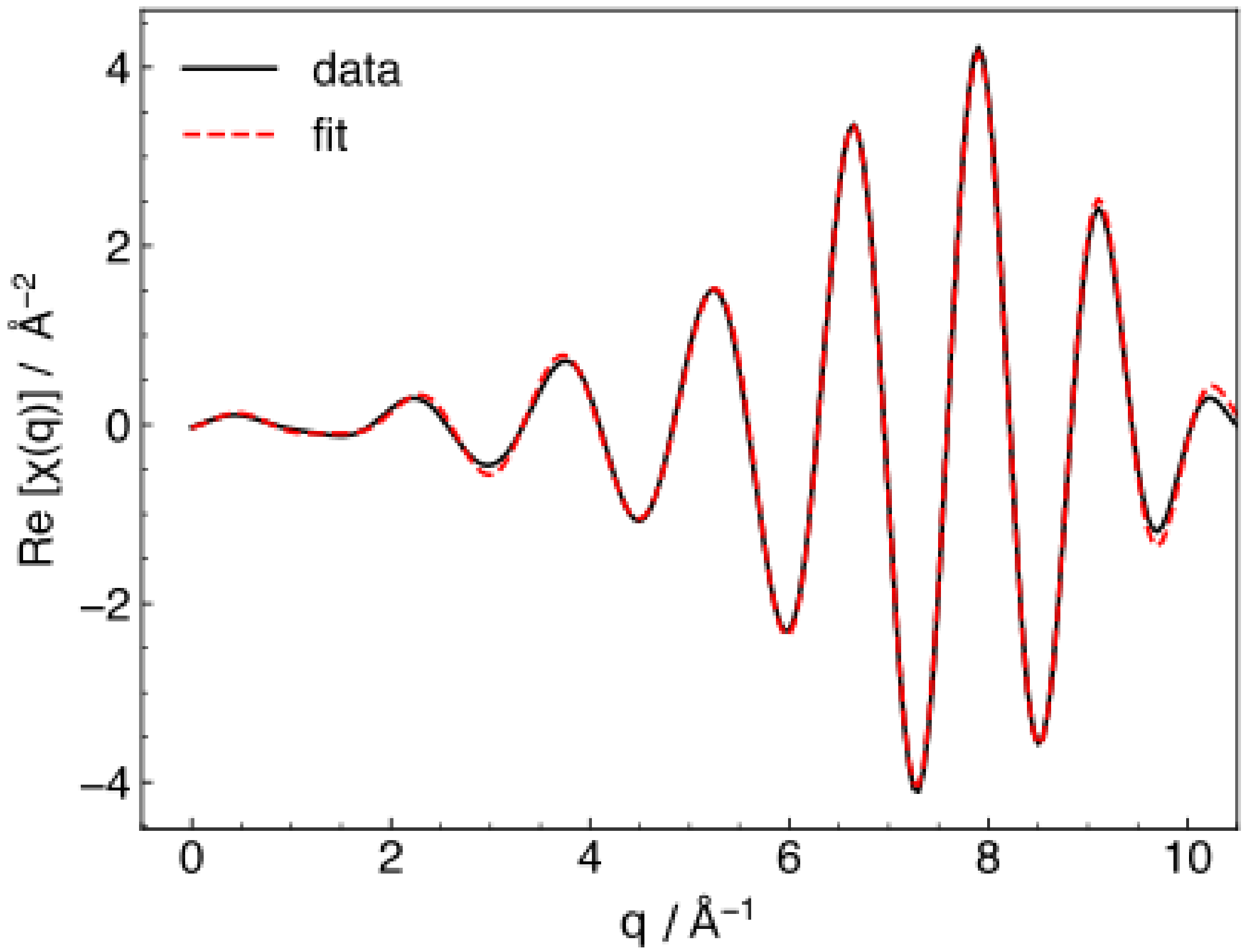

**Figure S13.** First-shell EXAFS fit (performed in back Fourier-transformed q-space) of $MAPb(Br_{0.6}Cl_{0.4})_3$ overlaid with the q-space data for $MAPb(Br_{0.6}Cl_{0.4})_3$. Fit values are in **Table S3**.

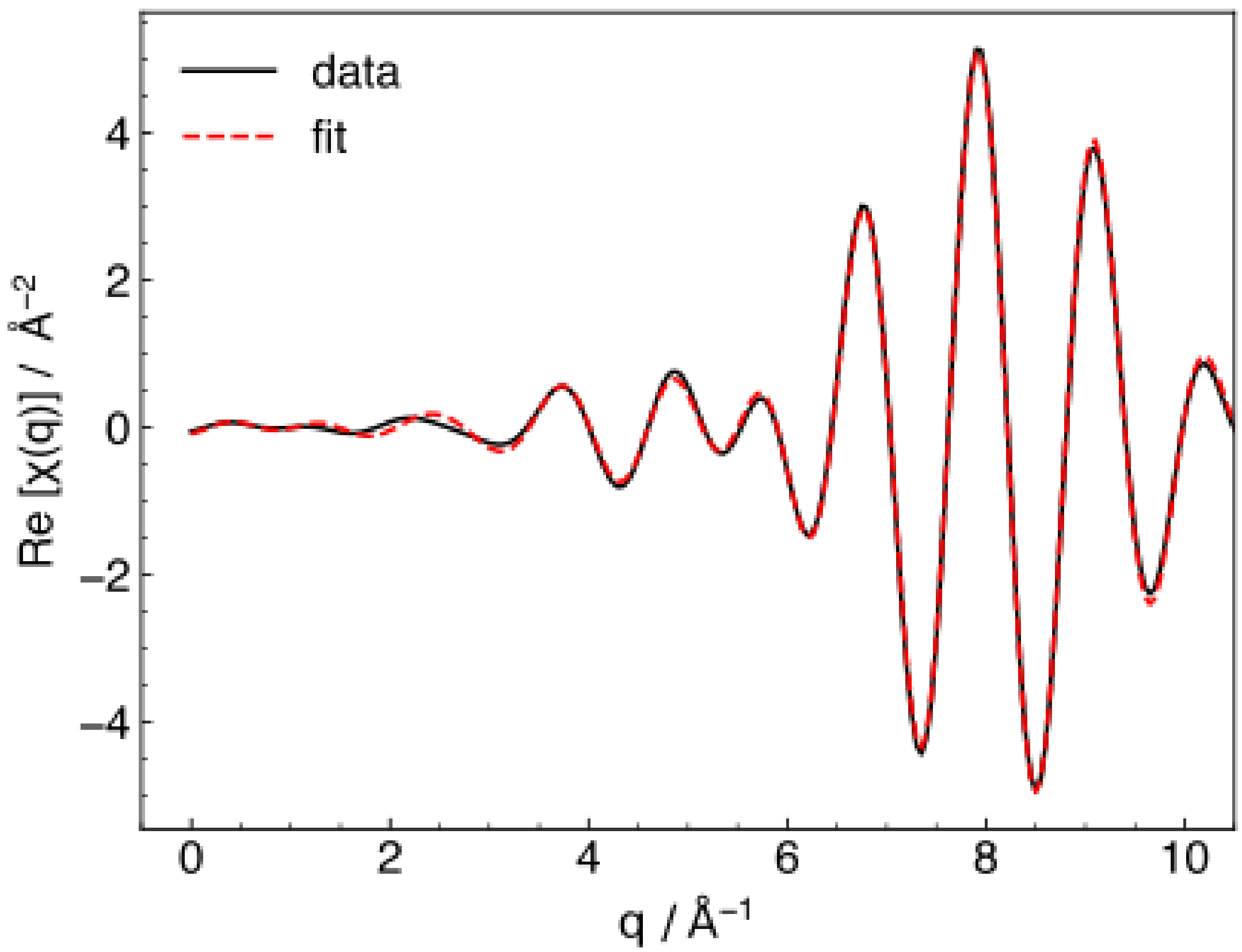

**Figure S14.** First-shell EXAFS fit (performed in back Fourier-transformed q-space) of $MAPb(Br_{0.6}I_{0.4})_3$ overlaid with the q-space data for $MAPb(Br_{0.6}I_{0.4})_3$. Fit values are in **Table S3**.

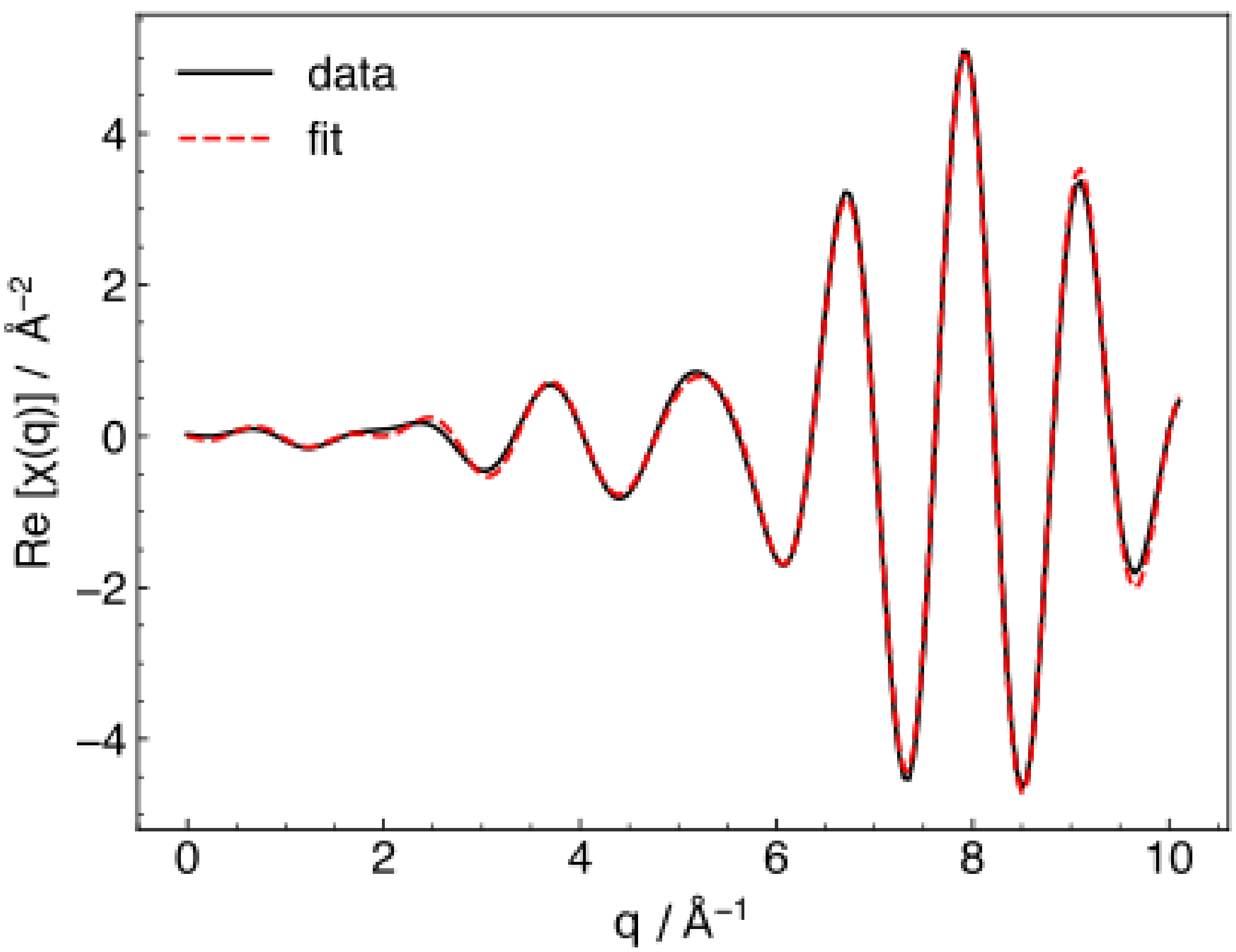

**Figure S15.** First-shell EXAFS fit (performed in back Fourier-transformed q-space) of $MAPb(I_{0.2}Br_{0.6}Cl_{0.2})_3$ overlaid with the q-space data for $MAPb(I_{0.2}Br_{0.6}Cl_{0.2})_3$. Fit values are in **Table S3**.

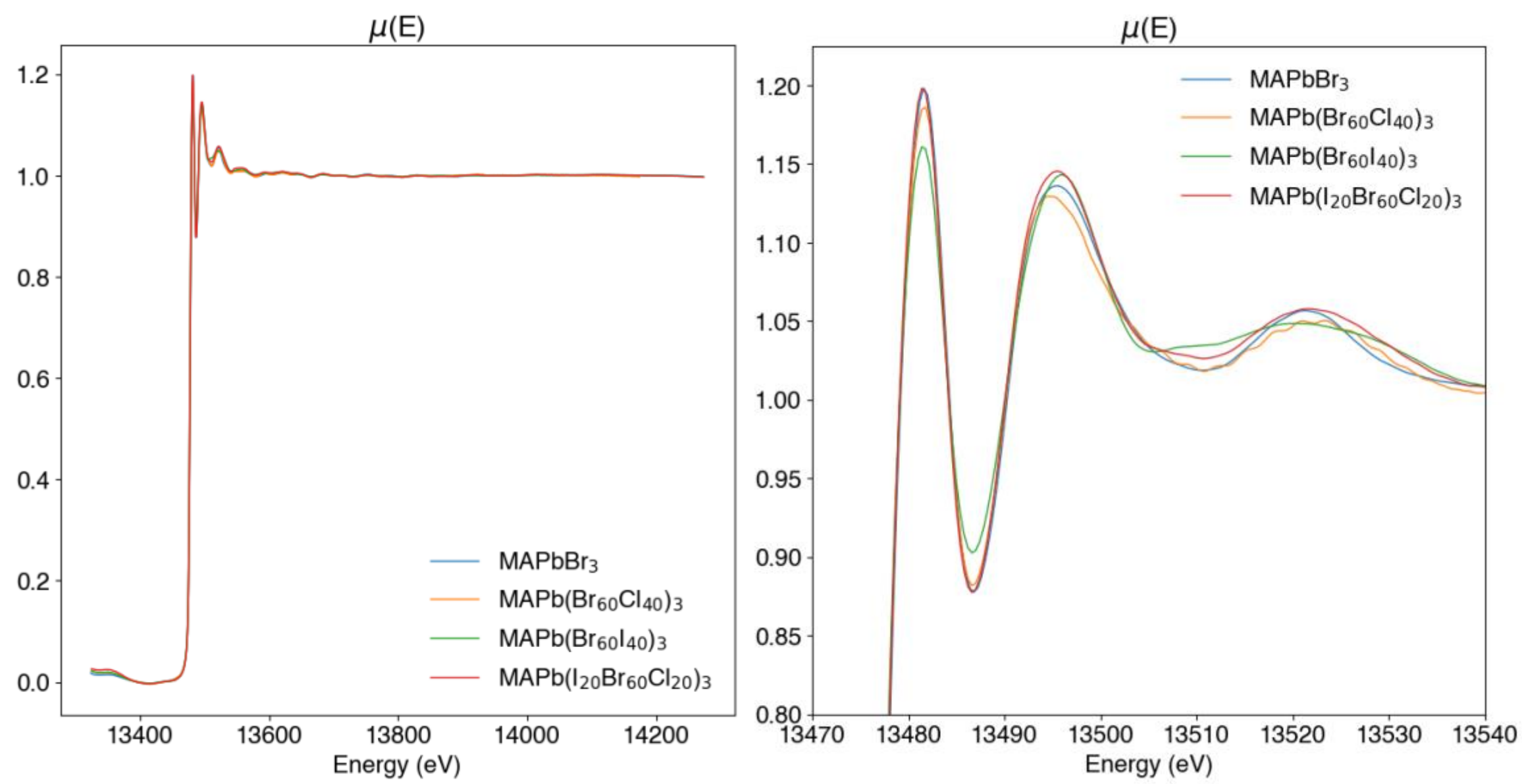

**Figure S16:** Full μ(E) (left) and XANES portion (right) taken at the Br K edge for the mixed-halide compositions.

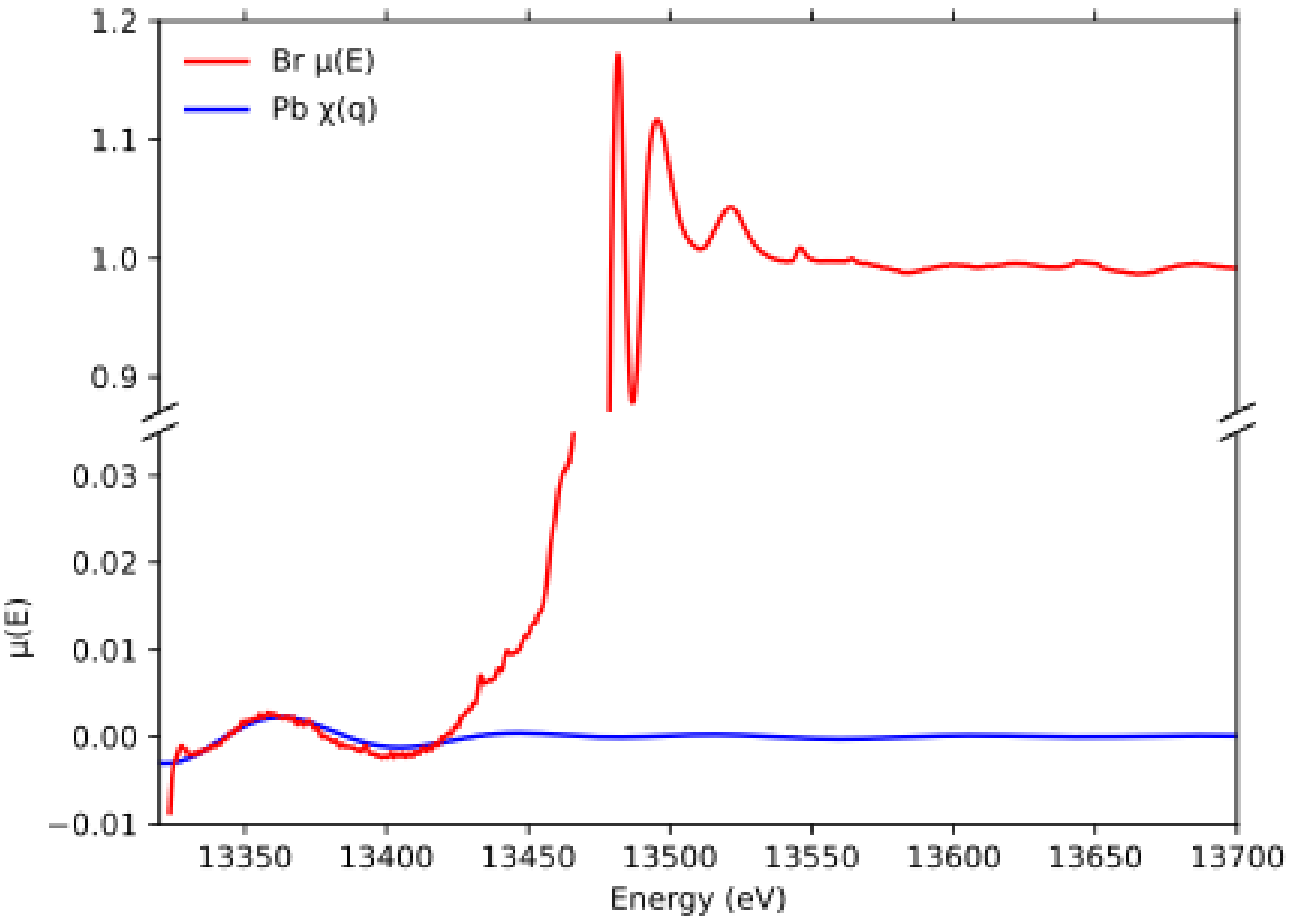

**Figure S17.** Overlap of the Pb $L_3$-edge oscillations with the Br K-edge for $MAPbBr_3$.

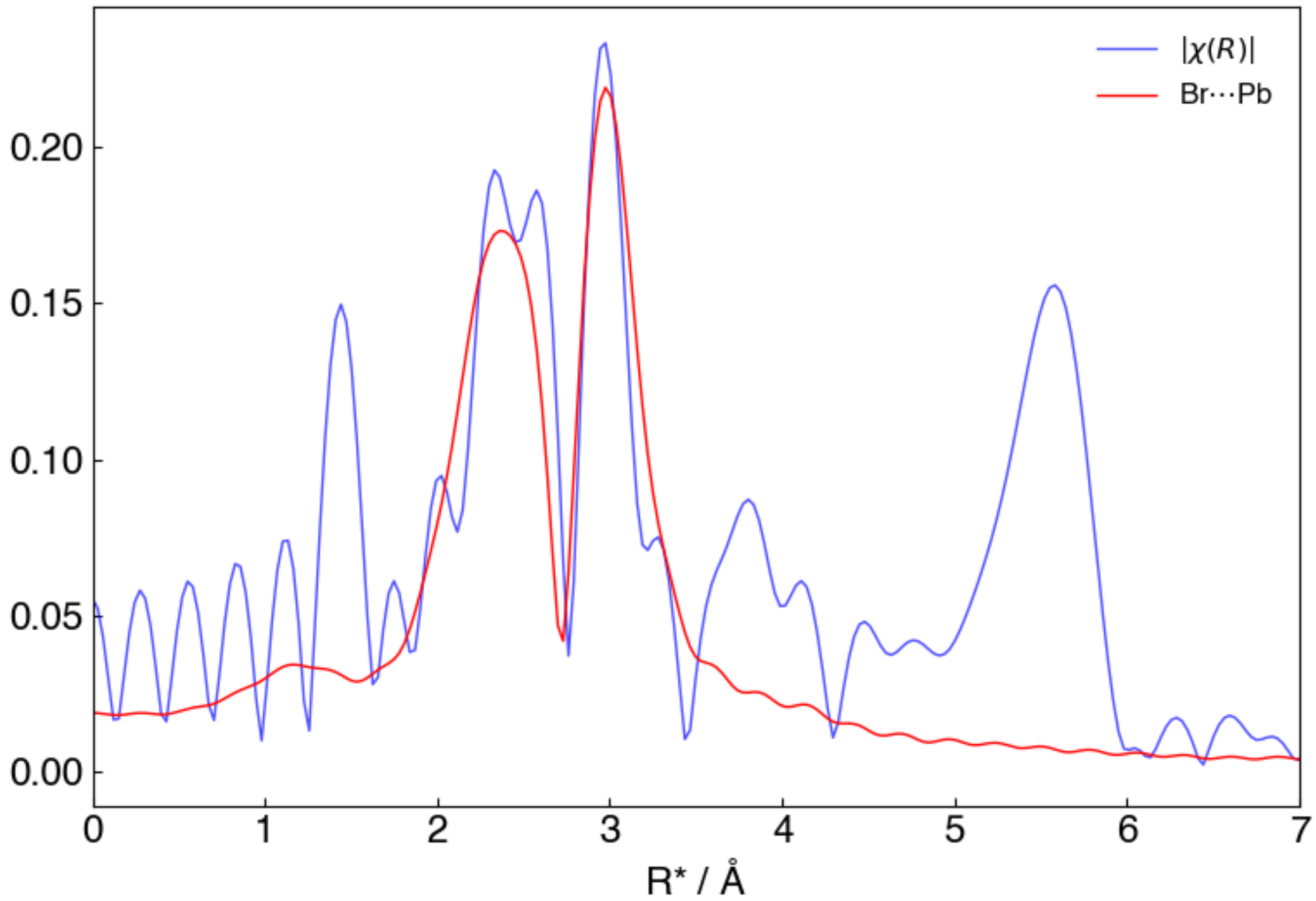

**Figure S18:** Magnitude of the Fourier-transformed Br K-edge EXAFS, ($|\chi(R)|$), for $MAPbBr_3$ (blue), overlaid with the FEFF-calculated first-shell, single-scattering path Br ···Pb path (red) with bond distance fixed to the orthorhombic $MAPbBr_3$ crystal structure and $\sigma^2$ fitted to the first-shell peak. The double-peaked structure at R*~2-3.5 Å is reproduced by this path alone.

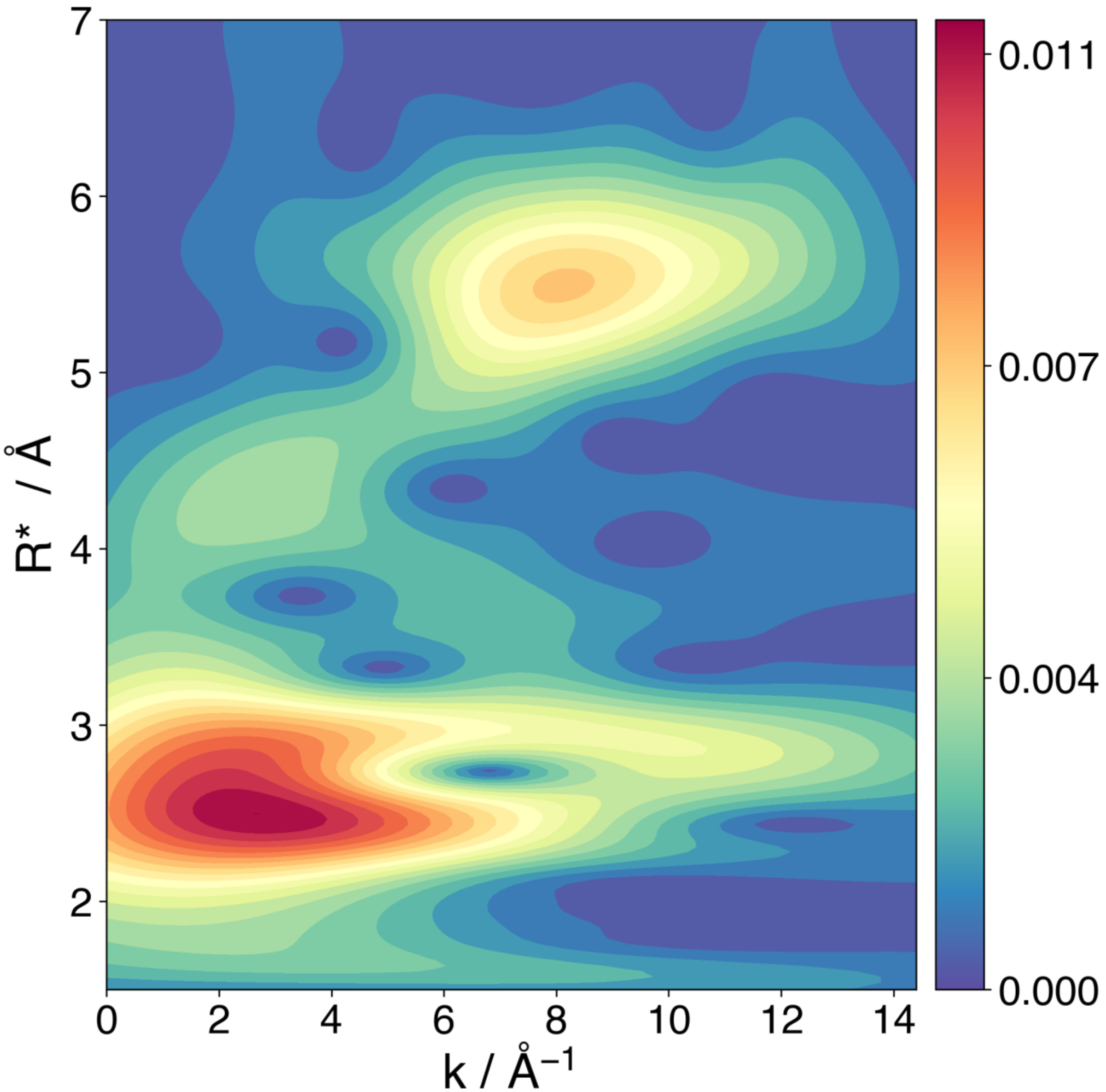

**Figure S19.** Full Cauchy wavelet transform (WT) for $MAPbBr_3$ measured at Br K edge.

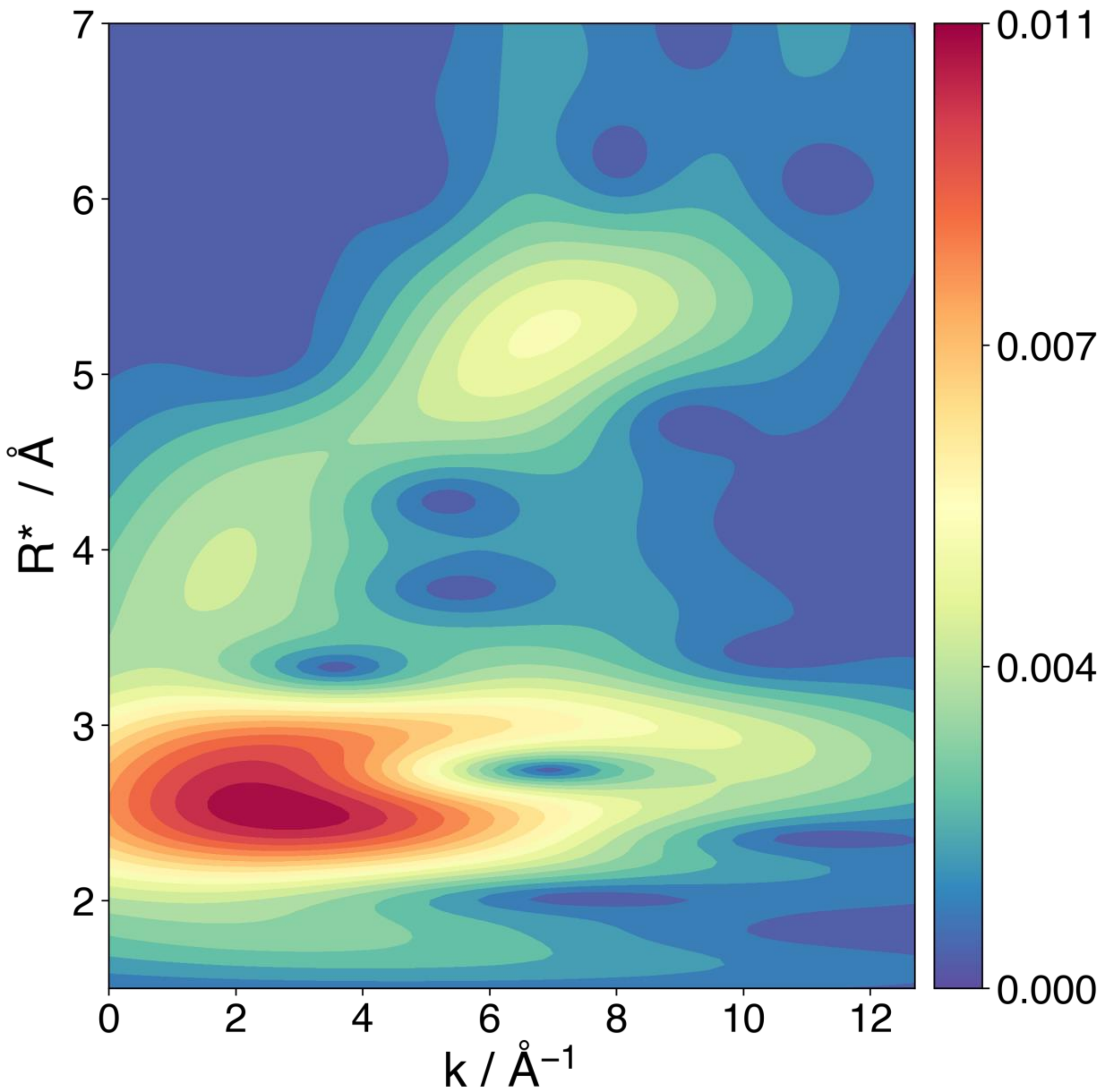

**Figure S20.** Full Cauchy wavelet transform for $MAPb(Br_{0.6}Cl_{0.4})_3$ measured at Br K edge.

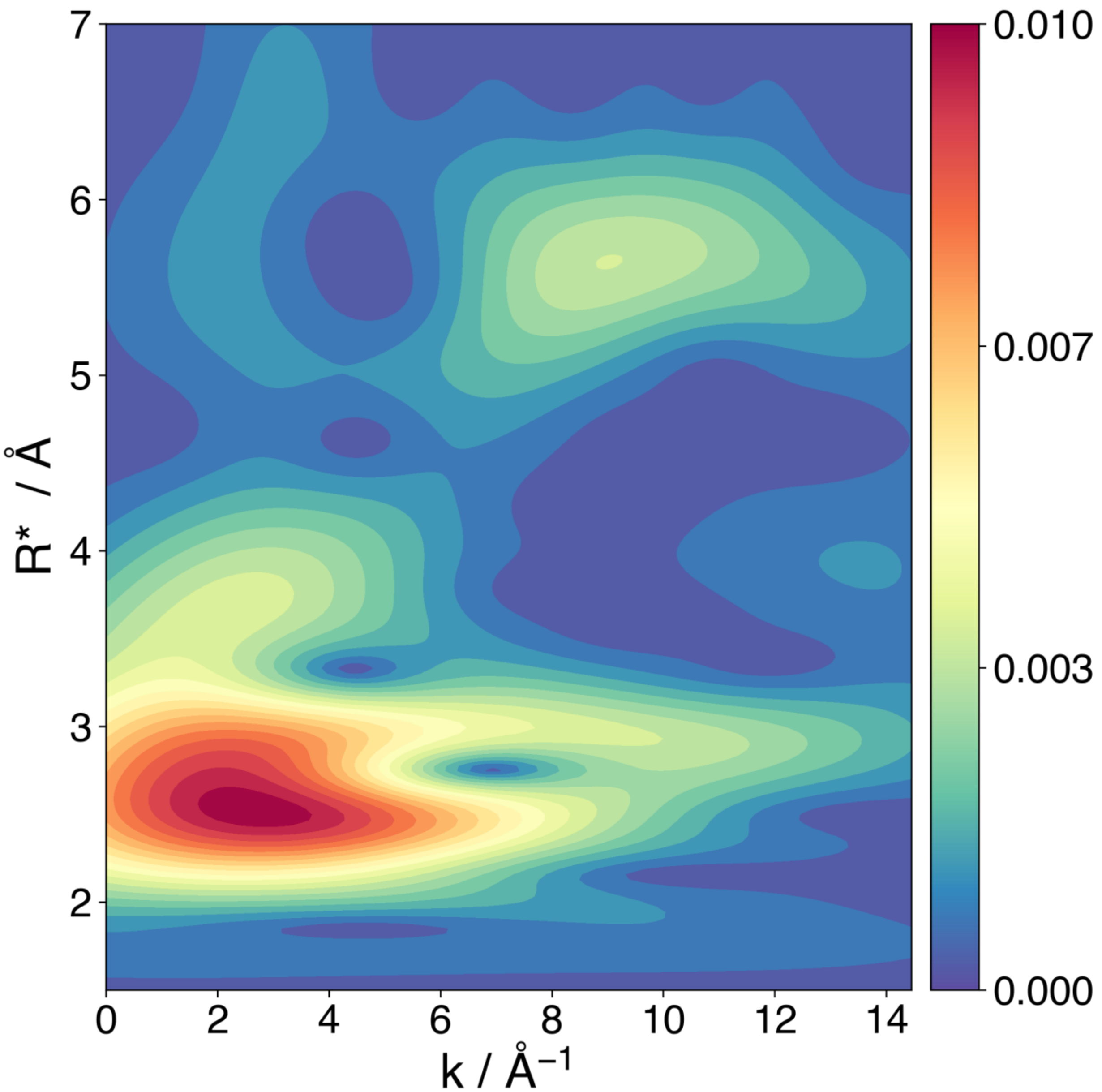

**Figure S21.** Full Cauchy wavelet transform for $MAPb(Br_{0.6}I_{0.4})_3$ measured at Br K edge.

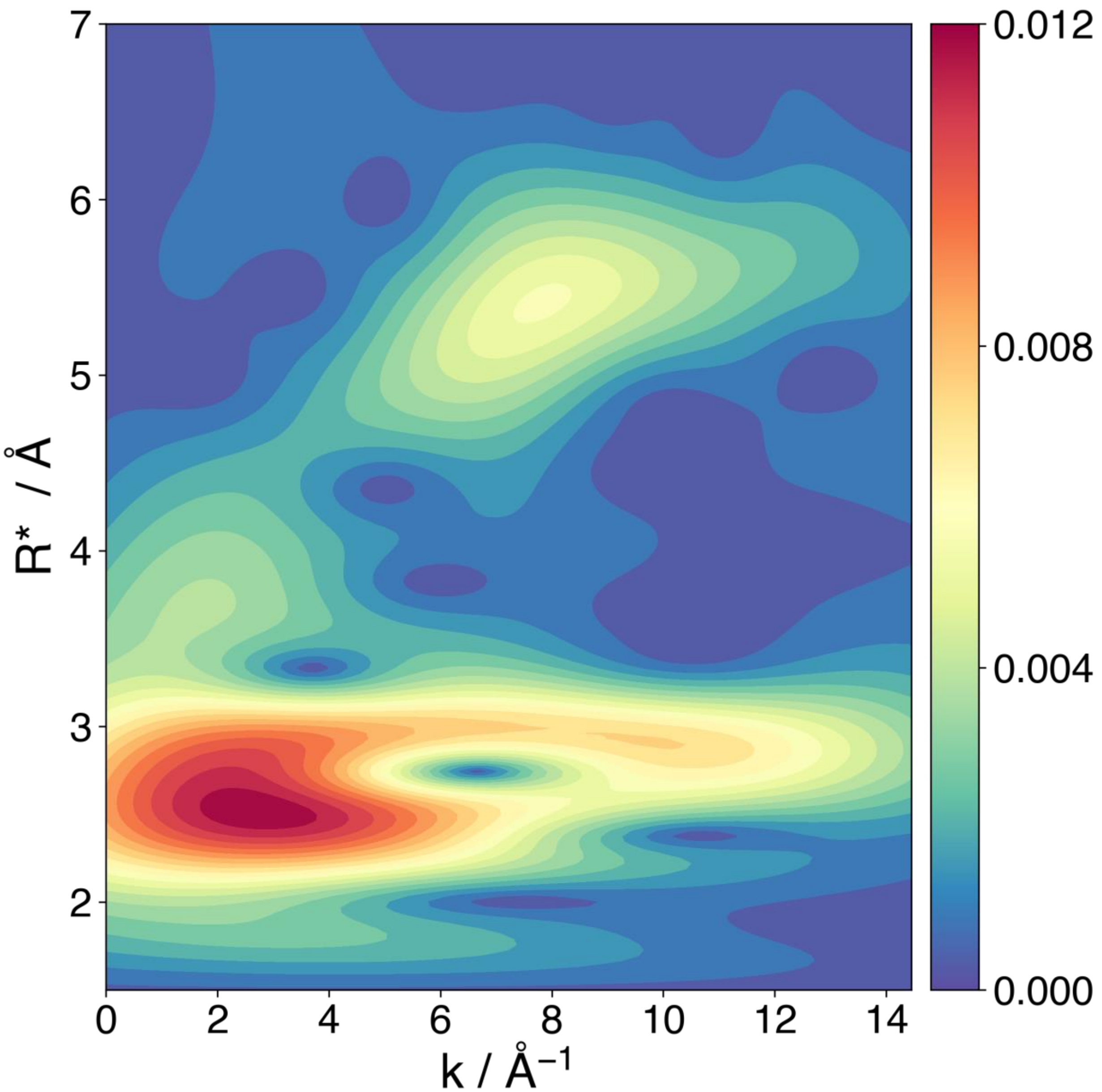

**Figure S22.** Full Cauchy wavelet transform for $MAPb(I_{0.2}Br_{0.6}Cl_{0.2})_3$ measured at Br K edge.

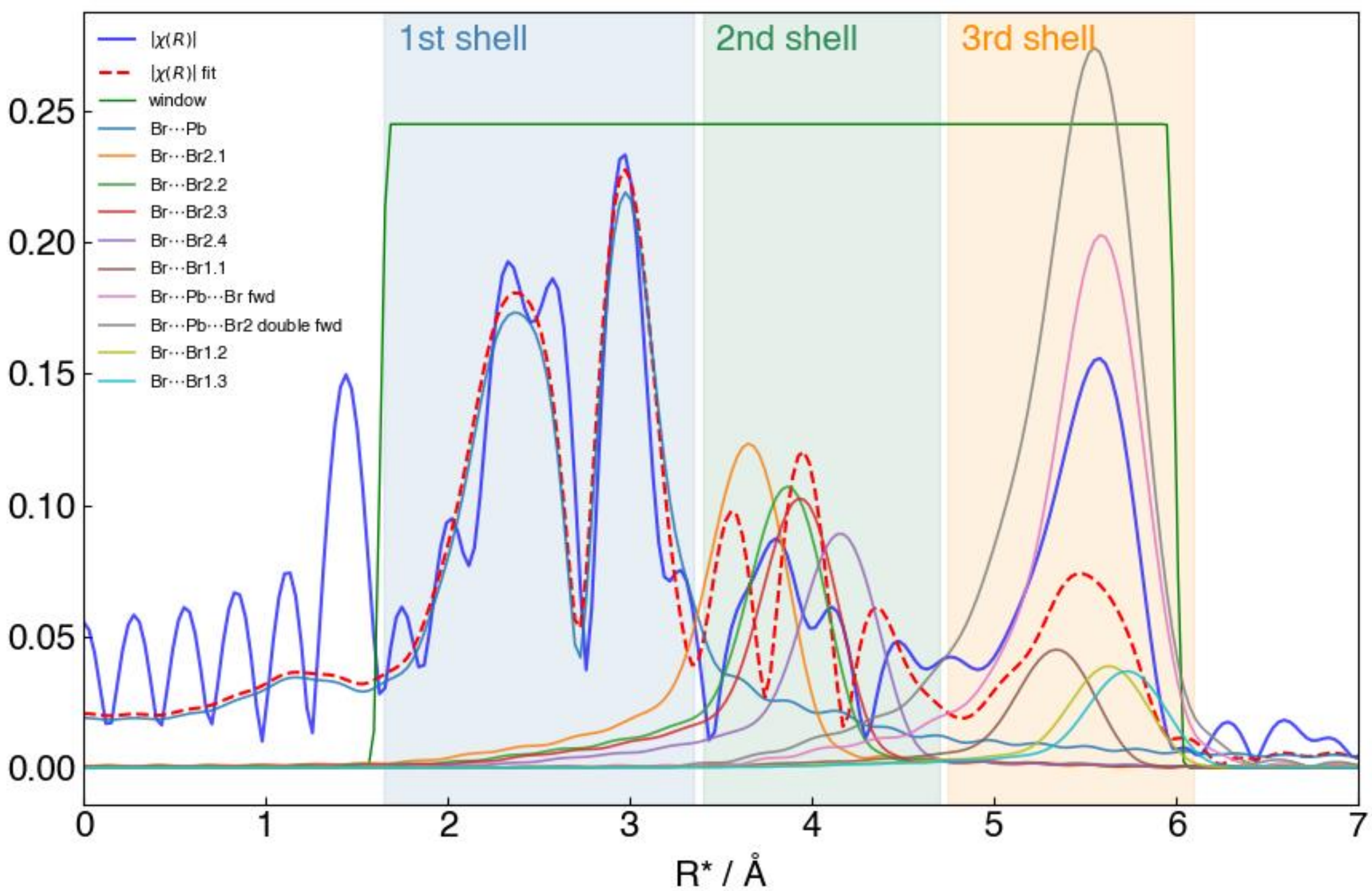

**Figure S23:** Magnitude of the Fourier-transformed Br K-edge EXAFS, $|\chi(R)|$, for $MAPbBr_3$ (blue), overlaid with the total fit (red dashed) and individual FEFF-calculated scattering path contributions shown. $\sigma^2$ for all paths was fixed to the same value used for the first-shell path (in Fig. S18).

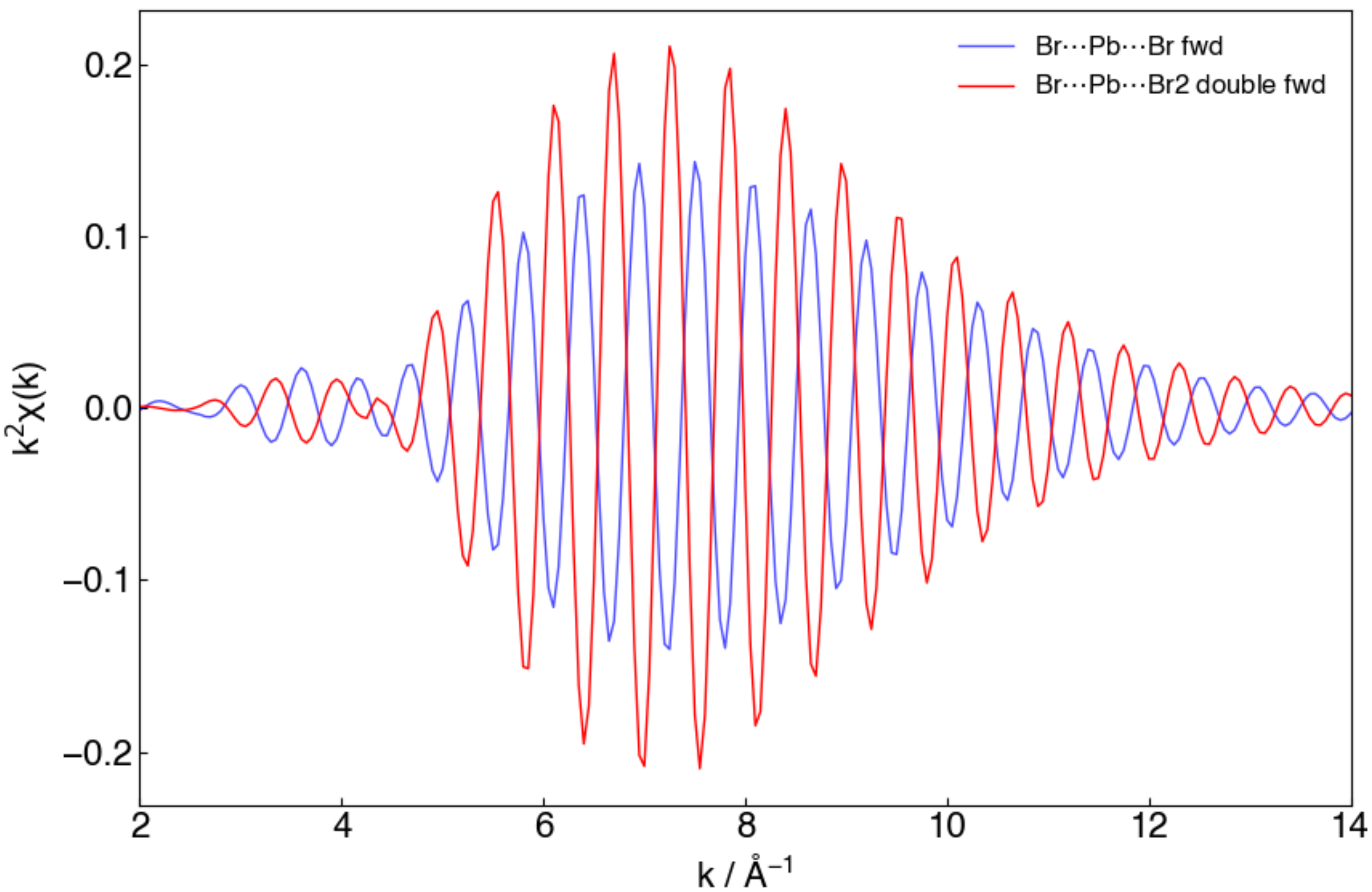

**Figure S24:** FEFF-calculated $k^2$-weighted EXAFS contributions in $k$ -space for the two dominant third-shell multiple-scattering paths in orthorhombic $MAPbBr_3$: the single forward-scattering path Br ··· Pb ··· Br (blue) and the double forward-scattering path Br ··· Pb ··· Br ··· Pb ··· Br (red).

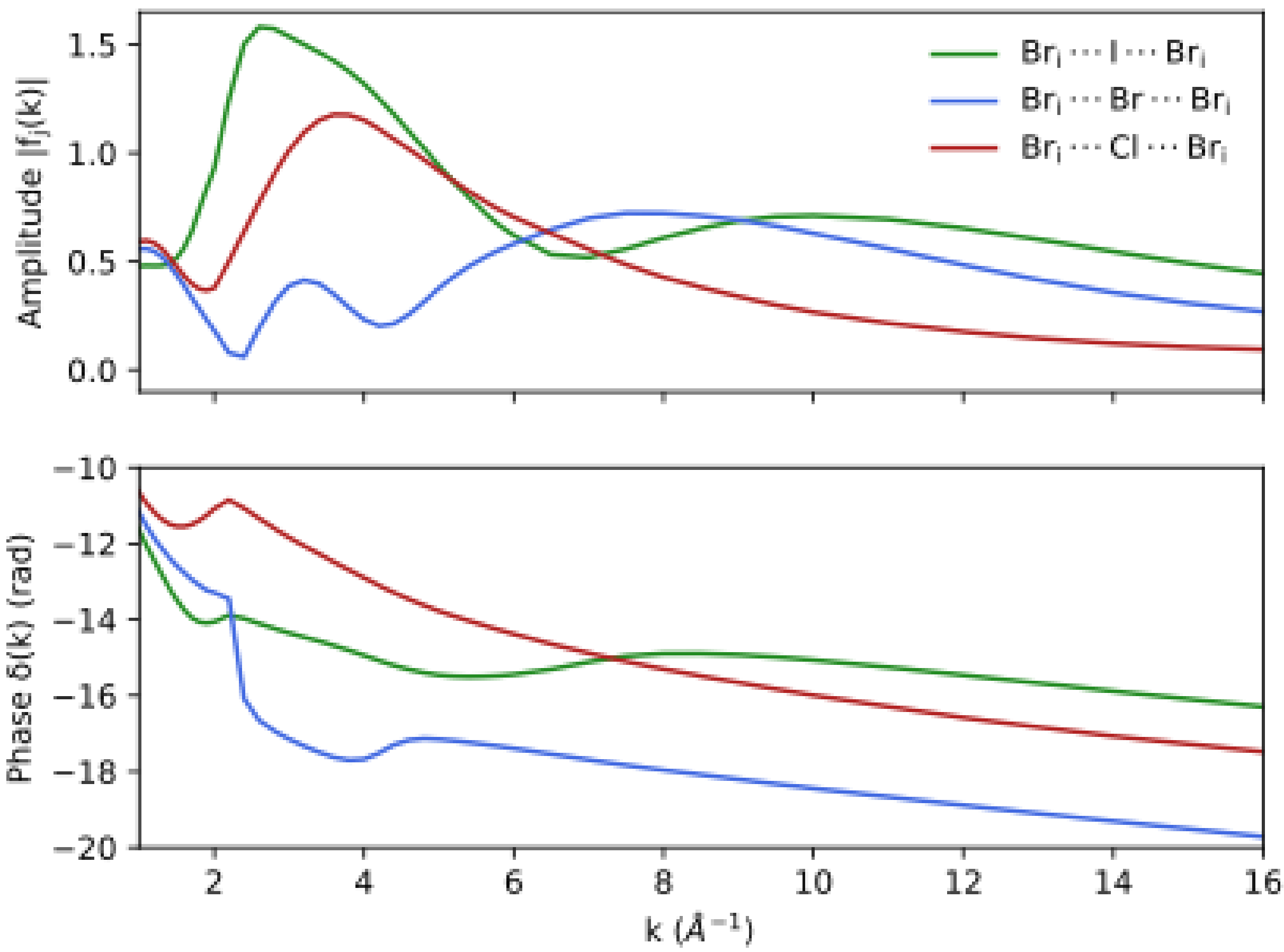

**Figure S25.** Calculated scattering amplitudes and phase shifts for Br-X single scattering paths as calculated in FEFF, using a Br-X distance of 3.96 Å corresponding to a $MAPbBr_3$ orthorhombic structure and changing the identity of X.

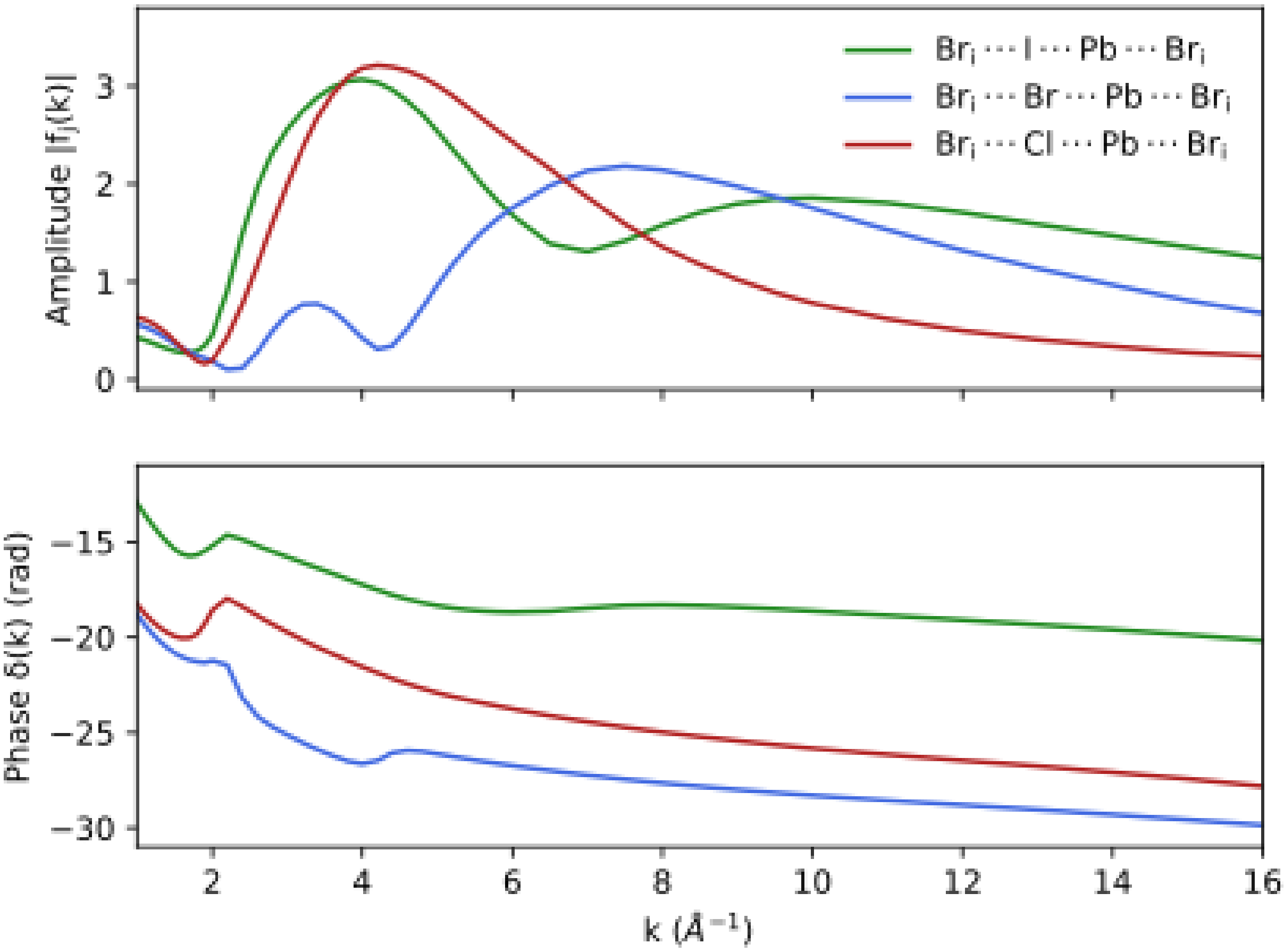

**Figure S26.** Calculated scattering amplitudes and phase shifts for $Br_i$-X-Pb-$Br_i$ single forward scattering paths as calculated in FEFF, using a $MAPbBr_3$ orthorhombic structure and changing the identity of X. Above approximately $k = 8$ $Å^{-1}$, the paths for X = Br and X = I differ in phase by approximately $3\pi$. Near $k = 7$ $Å^{-1}$, the paths for X = I and X = Cl differ in phase by approximately $2\pi$.

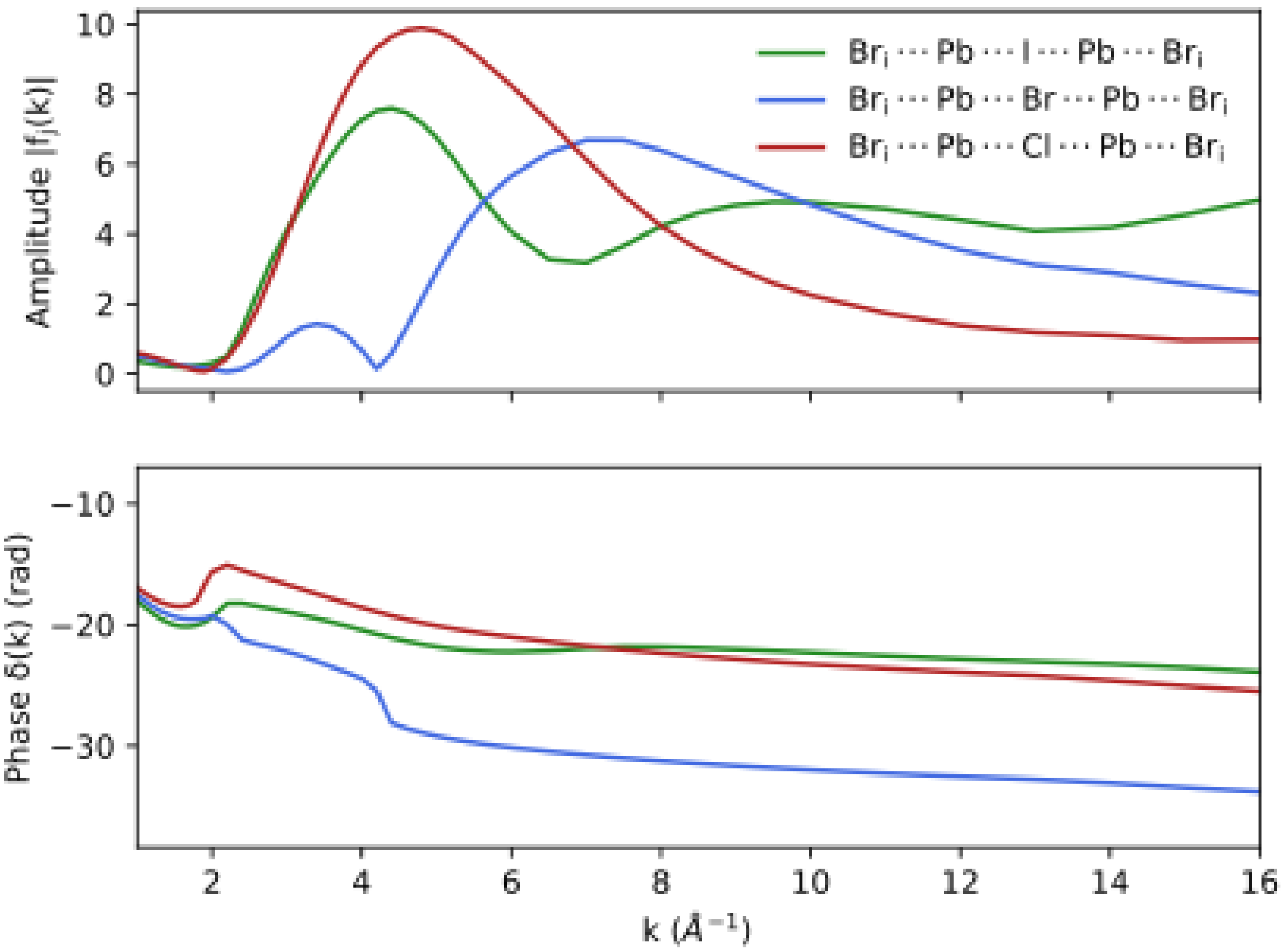

**Figure S27.** Calculated scattering amplitudes and phase shifts for $Br_i$-Pb-X-Pb-$Br_i$ double forward scattering paths as calculated in FEFF, using a $MAPbBr_3$ orthorhombic structure and changing the identity of X. Above approximately $k = 7$ Å$^{-1}$, there is a phase difference between the X = Br path and the paths for X = Cl and X = I of approximately $3\pi$.

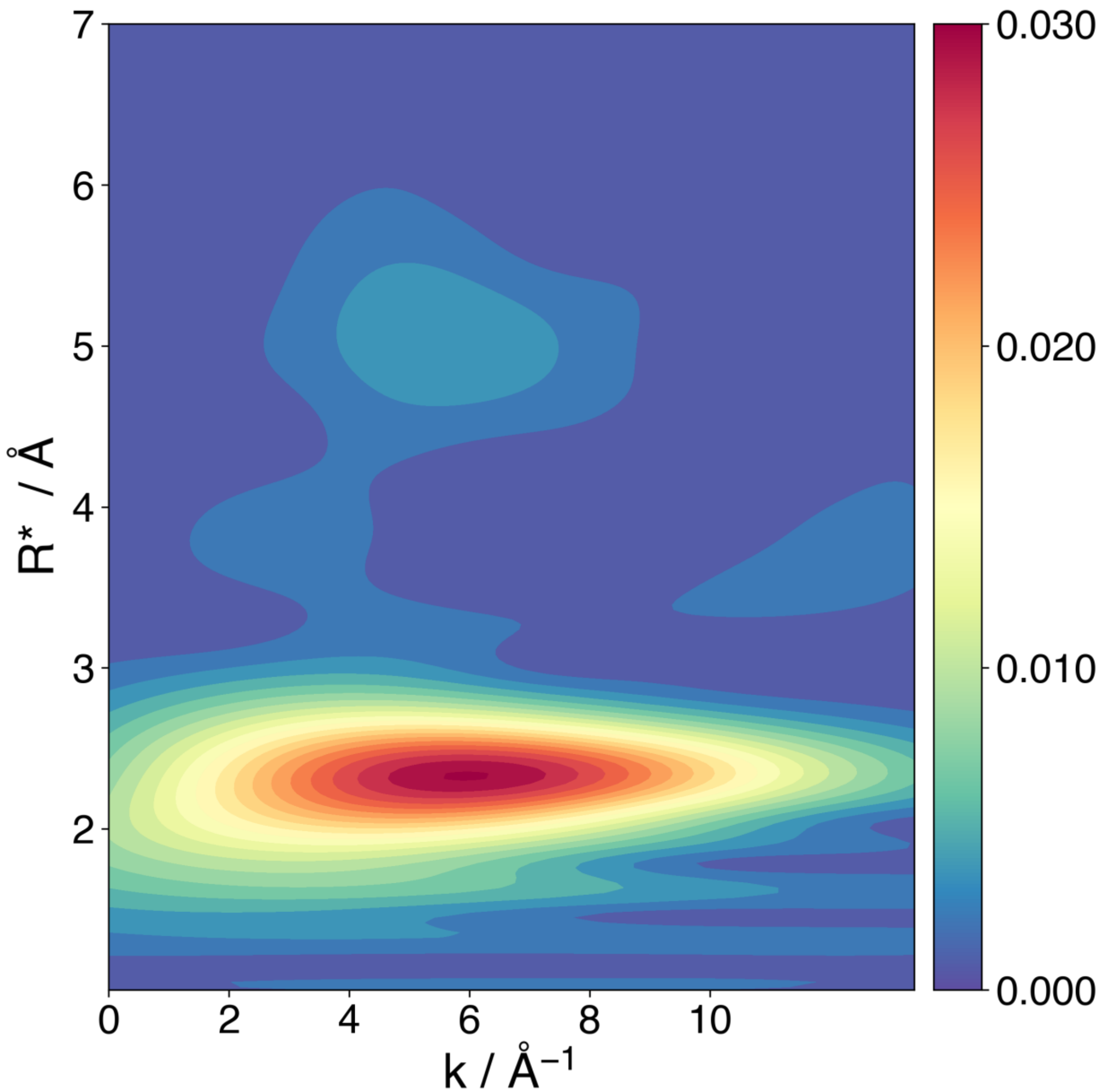

**Figure S28.** Full Cauchy WT of Pb $L_3$-edge EXAFS for $MAPbCl_3$.

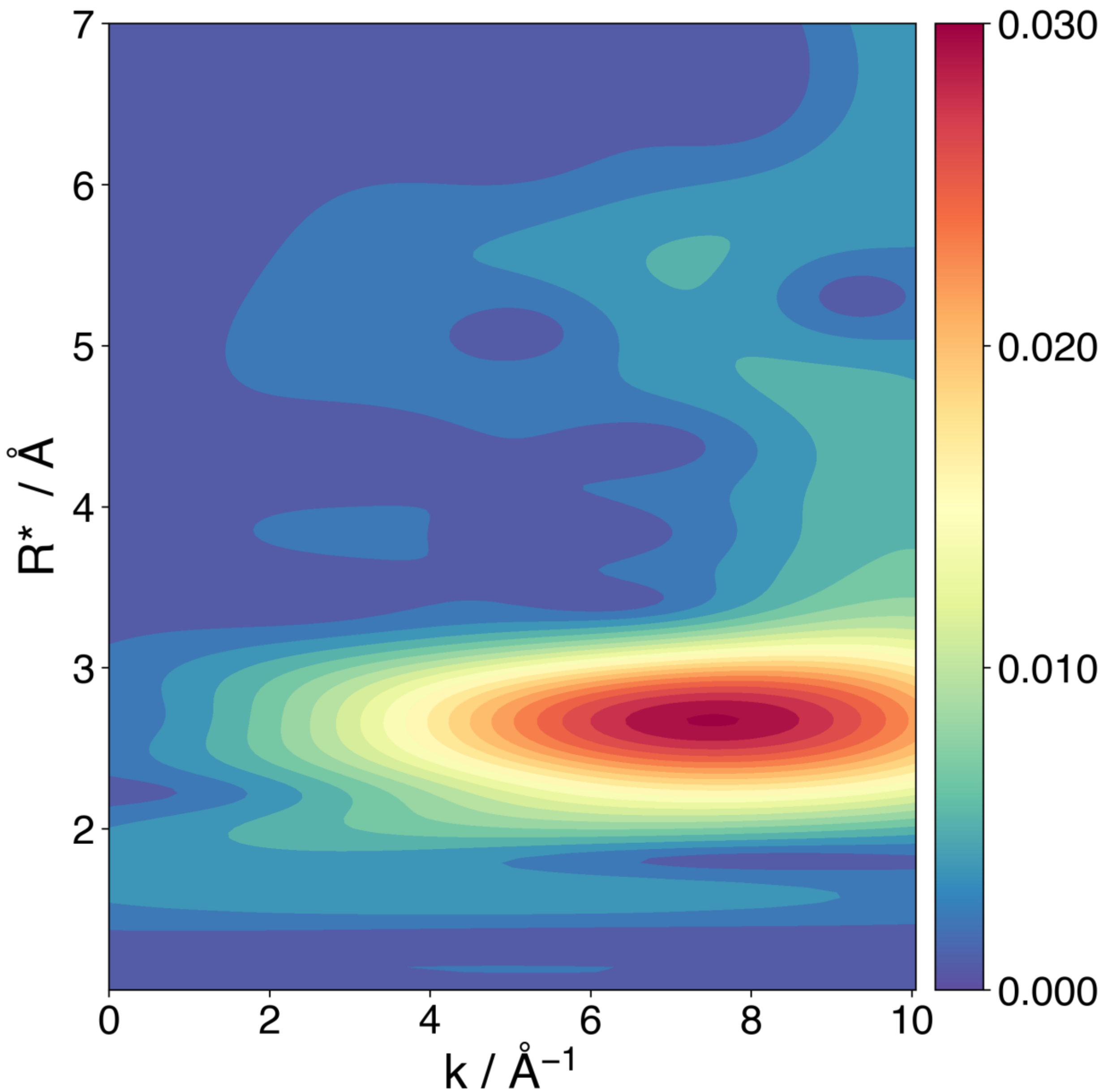

**Figure S29.** Full Cauchy WT of Pb $L_3$-edge EXAFS for $MAPbBr_3$.

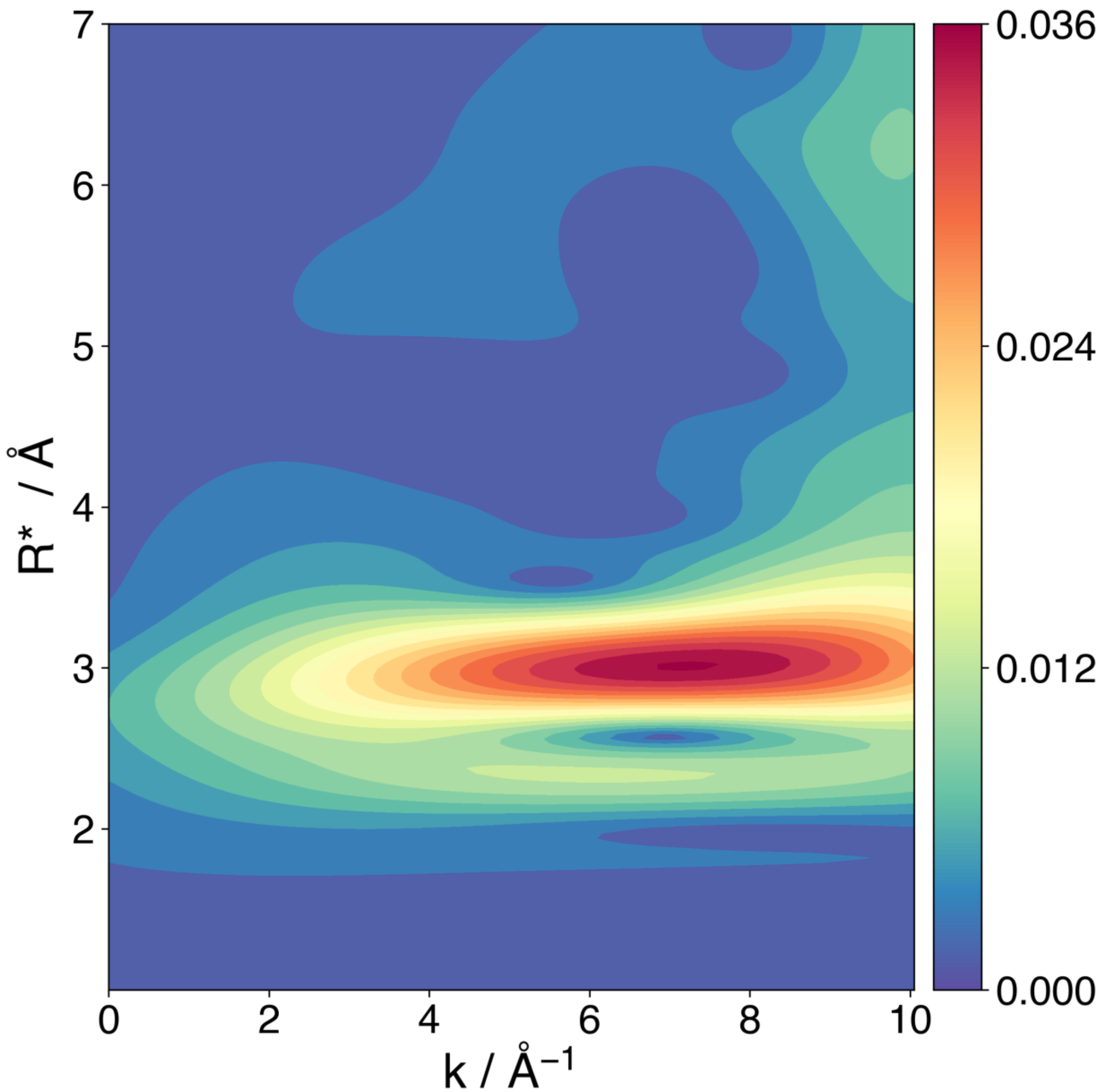

**Figure S30.** Full Cauchy WT of Pb $L_3$-edge EXAFS for $MAPbI_3$.

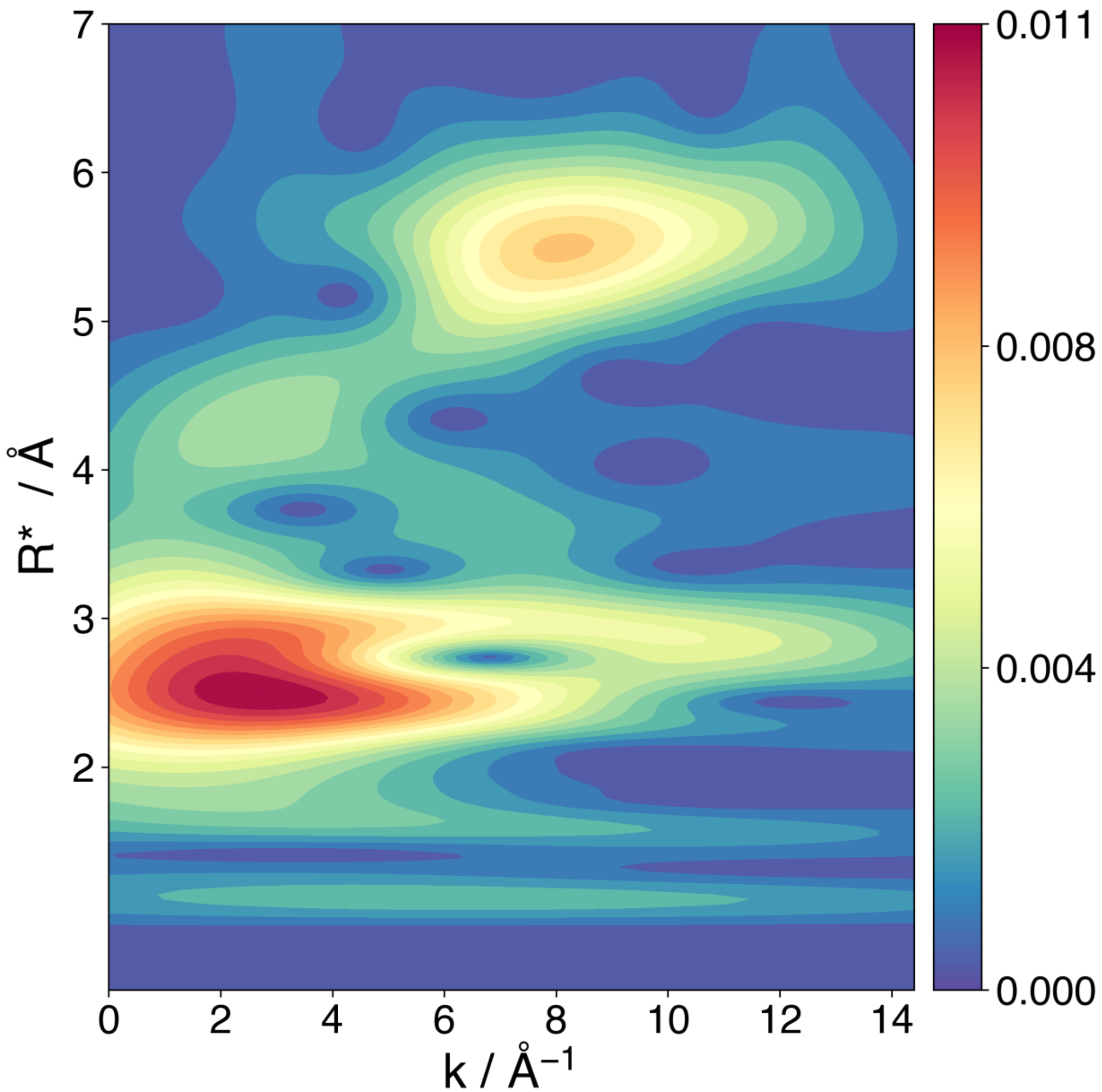

**Figure S31.** Full Cauchy WT of Br K-edge EXAFS for $MAPbBr_3$, computed using $r_{bkg}$ = 1.0 Å.

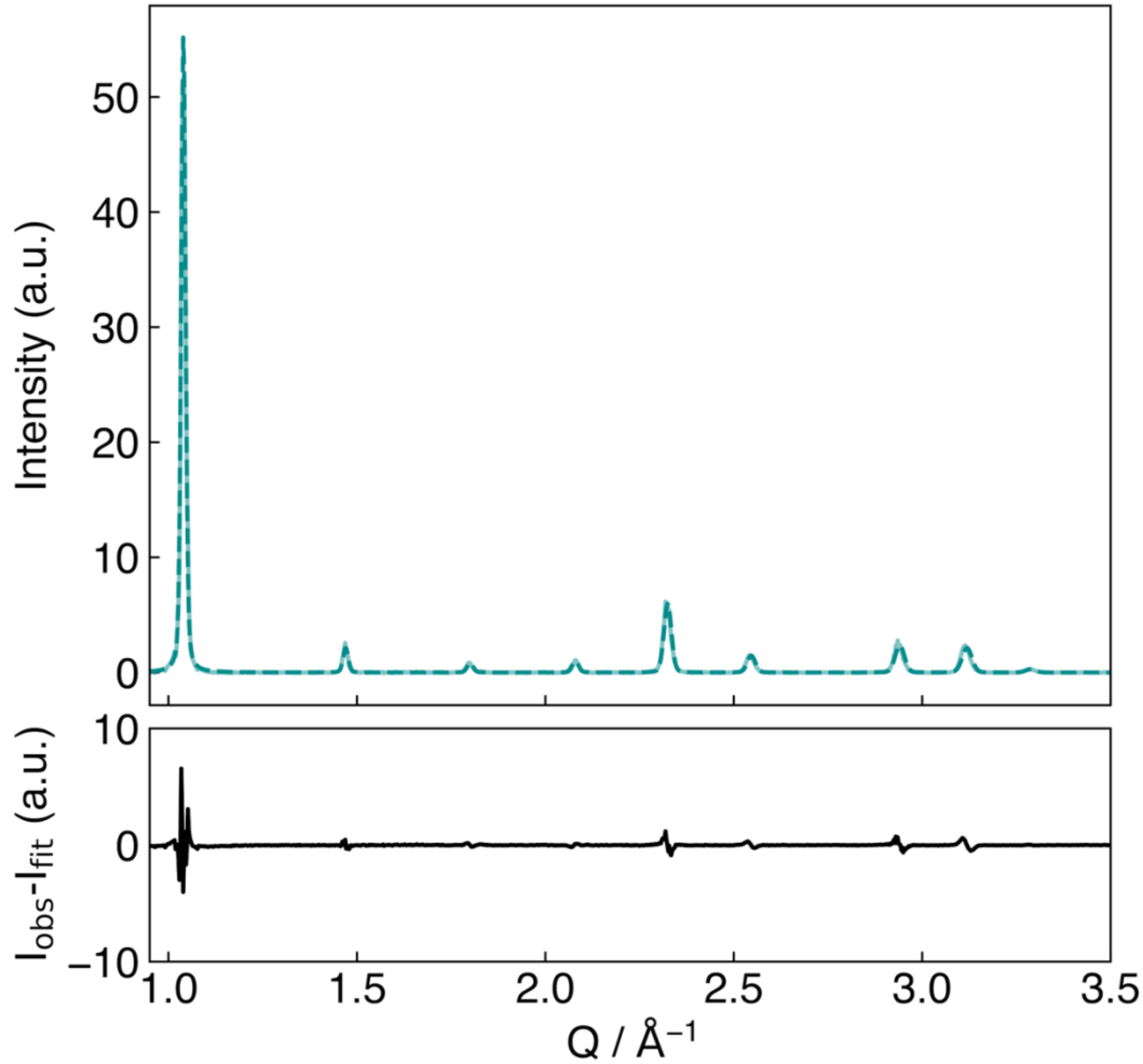

**Figure S32.** Integrated X-ray scattering of $MAPb(Br_{0.6}I_{0.4})_3$ (solid line) and calculated scattering from Pawley fit (dashed line). Peak positions are well described by a cubic $Pm\overline{3}m$ model, with $a = 6.04$ Å .

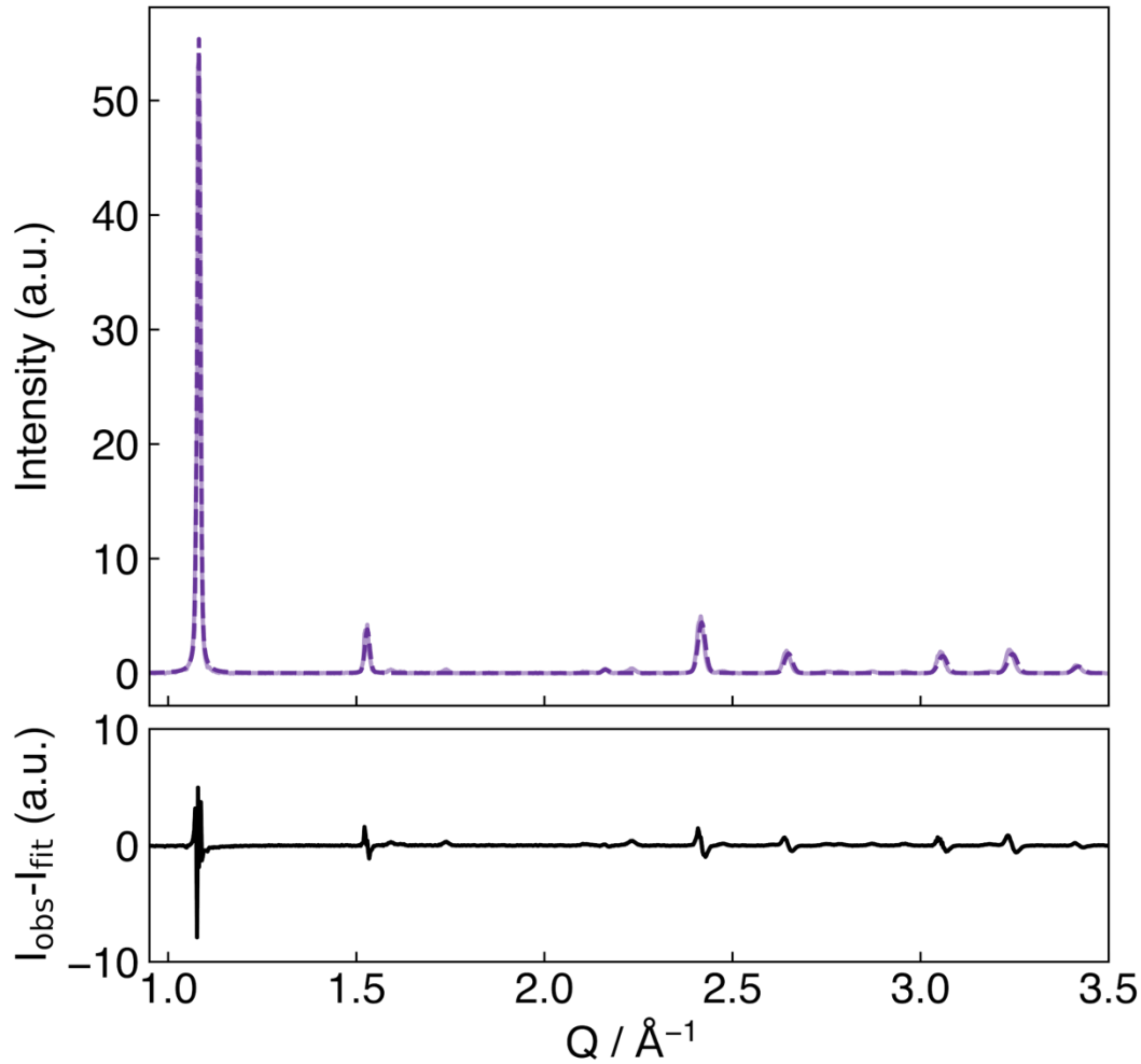

**Figure S32.** Integrated X-ray scattering of $MAPb(Br_{0.6}Cl_{0.4})_3$ (solid line) and calculated scattering from Pawley fit (dashed line). Peak positions are well described by a cubic $Pm\bar{3}m$ model, with $a = 5.81$ Å. Reflections not captured by fit consistent with trace crystalline $PbCl_2$

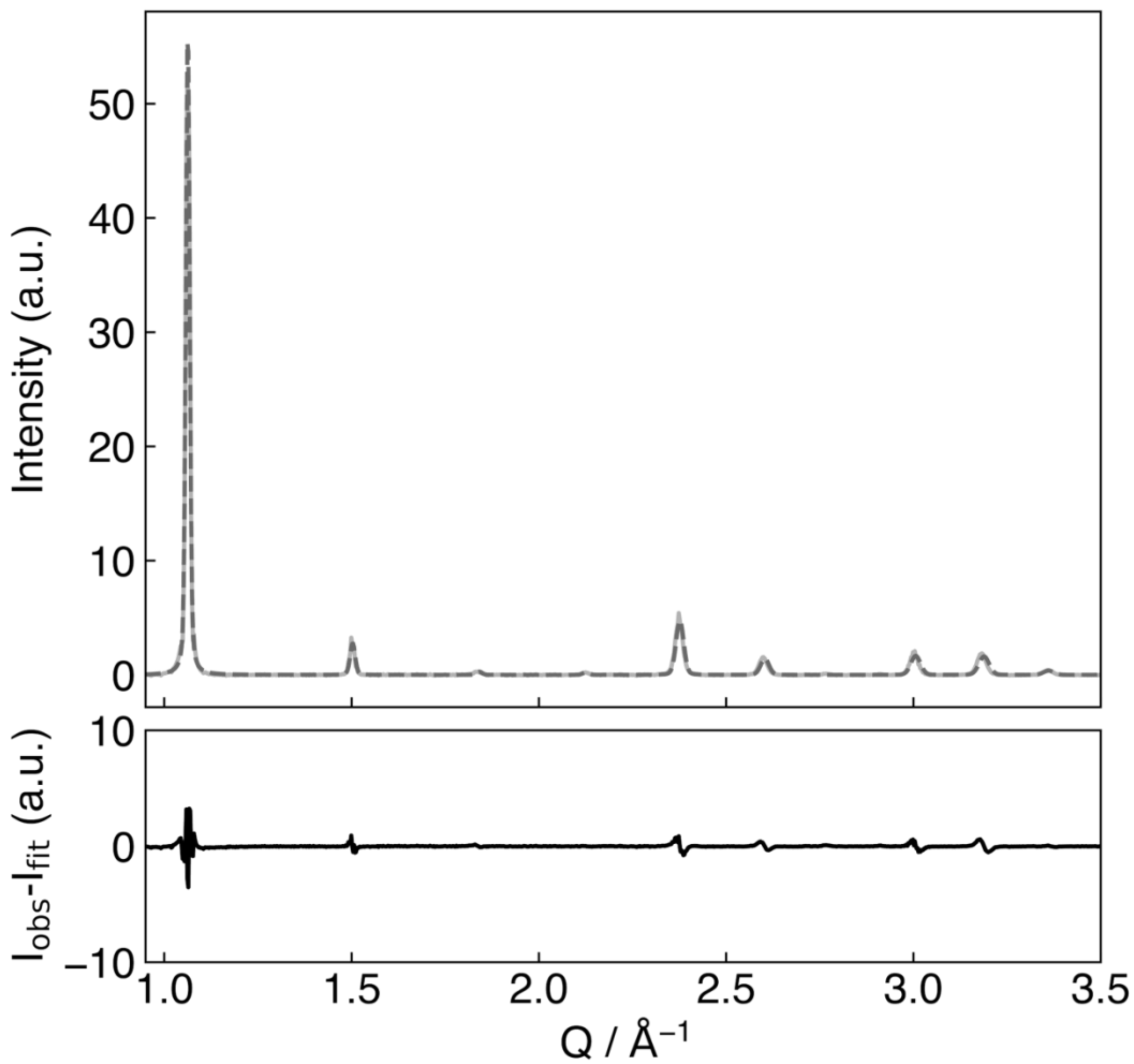

**Figure S33.** Integrated X-ray scattering of $MAPb(Br_{0.6}I_{0.2}Cl_{0.2})_3$ (solid line) and calculated scattering from Pawley fit (dashed line). Peak positions are well described by a cubic $Pm\overline{3}m$ model, with $a = 5.91$ Å.

| Nominal composition | Goldschmidt tolerance factor, $t$ | Octahedral factor, $\mu$ | Ionic Pb-X distance (Å) | Fitted EXAFS Pb-X (Å) |
| --- | --- | --- | --- | --- |
| $MAPbI_3$ | 0.912 | 0.541 | 3.39 | 3.18 |
| $MAPbBr_3$ | 0.927 | 0.607 | 3.15 | 2.98 |
| $MAPbCl_3$ | 0.938 | 0.657 | 3.00 | 2.78-3.017 |
| $MAPb(Br_{0.6}Cl_{0.4})_3$ | 0.931 | 0.626 | N/A | 2.844, 2.962 |
| $MAPb(Br_{0.6}I_{0.4})_3$ | 0.921 | 0.579 | N/A | 2.979, 3.156 |

| $MAPb(I_{0.2}Br_{0.6}Cl_{0.2})_3$ | 0.926 | 0.602 | N/A | 2.847- 3.174 |
| --- | --- | --- | --- | --- |
| $MAPb(I_{0.4}Br_{0.2}Cl_{0.4})_3$ | 0.925 | 0.596 | N/A | N/A |

**Table S11**. Calculated Goldschmidt tolerance factor and octahedral factors for the nominal compositions studied in this work, calculated using the Shannon ionic radii ($Pb^{2+}$: 1.19 Å; $I^-$: 2.20 Å; $Br^-$: 1.96 Å; $Cl^-$: 1.81 Å)[7] and effective $R_{MA}$ = 2.17 Å,[8] along with comparison with ionic Pb-X distance and fitted Pb-X bond lengths.

## References

(1) Ashiotis, G.; Deschildre, A.; Nawaz, Z.; Wright, J. P.; Karkoulis, D.; Picca, F. E.; Kieffer, J. The Fast Azimuthal Integration Python Library: pyFAI. *J. Appl. Crystallogr.* **2015**, *48* (2), 510–519. https://doi.org/10.1107/S1600576715004306.

(2) Dane, T. Tgdane/Pygix, 2024. https://github.com/tgdane/pygix (accessed 2026-04-12).

(3) Toby, B. H.; Von Dreele, R. B. GSAS-II: The Genesis of a Modern Open-Source All Purpose Crystallography Software Package. *J. Appl. Crystallogr.* **2013**, *46* (2), 544–549. https://doi.org/10.1107/S0021889813003531.

(4) Weller, M. T.; Weber, O. J.; Henry, P. F.; Pumpo, A. M. D.; Hansen, T. C. Complete Structure and Cation Orientation in the Perovskite Photovoltaic Methylammonium Lead Iodide between 100 and 352 K. *Chem. Commun.* **2015**, *51* (20), 4180–4183. https://doi.org/10.1039/C4CC09944C.

(5) Swainson, I. P.; Hammond, R. P.; Soullière, C.; Knop, O.; Massa, W. Phase Transitions in the Perovskite Methylammonium Lead Bromide, CH3ND3PbBr3. *J. Solid State Chem.* **2003**, *176* (1), 97–104. https://doi.org/10.1016/S0022-4596(03)00352-9.

(6) Chi, L.; Swainson, I.; Cranswick, L.; Her, J.-H.; Stephens, P.; Knop, O. The Ordered Phase of Methylammonium Lead Chloride CH3ND3PbCl3. *J. Solid State Chem.* **2005**, *178* (5), 1376–1385. https://doi.org/10.1016/j.jssc.2004.12.037.

(7) Shannon, R. D. Revised Effective Ionic Radii and Systematic Studies of Interatomic Distances in Halides and Chalcogenides. *Acta Crystallogr. Sect. A* **1976**, *32* (5), 751–767. https://doi.org/10.1107/S0567739476001551.

(8) Kieslich, G.; Sun, S.; Cheetham, A. K. An Extended Tolerance Factor Approach for Organic–Inorganic Perovskites. *Chem. Sci.* **2015**, *6* (6), 3430–3433. https://doi.org/10.1039/c5sc00961h.